\documentclass[useAMS,usenatbib]{mn2e}

\bibpunct[,]{(}{)}{;}{a}{}{,}

\usepackage{graphicx}                                                           
\usepackage{amssymb}
\usepackage{epsfig}
\usepackage{epstopdf}
\usepackage{verbatim}
\usepackage{subfigure}
\usepackage{lscape}
\usepackage{longtable}
\usepackage{deluxetable}
\usepackage{rotating}
\usepackage{lscape}
\usepackage{color}
\definecolor{dkcyan}{rgb}{0.000,0.600,0.600}
\usepackage[totalwidth=17.8cm, totalheight=24.0cm]{geometry}

\newcommand{\apj}{ApJ}
\newcommand{\apjl}{ApJ}
\newcommand{\apjs}{ApJS}
\newcommand{\aj}{AJ}
\newcommand{\mnras}{MNRAS}
\newcommand{\aaps}{A\&AS}
\newcommand{\aap}{A\&A}
\newcommand{\pasp}{PASP}

\newcommand{\araa}{ARA\&A}
\newcommand{\nat}{Nature}

\def\arcsec{$^{\prime\prime}$}

\definecolor{dkcyan}{rgb}{0.000,0.600,0.600}
\definecolor{navy}{RGB}{0,0,102}
\definecolor{royal}{RGB}{0,0,204}

 \newcommand{\atlas}{ATLAS$^{\rm 3D}$ }

\title[Angular momentum and nuclear light profiles]{The ATLAS$^{\rm3D}$ Project -- XXIII.  Angular momentum and nuclear surface brightness profiles}
\author[Davor Krajnovi\'c et al. ]{Davor Krajnovi\'c$^{1}$\thanks{E-mail: dkrajnovic@aip.de}, A. M. Karick$^{2}$, Roger L. Davies$^{2}$, Thorsten Naab$^{3}$, Marc Sarzi$^{4}$, \newauthor Eric Emsellem$^{5,6}$, Michele Cappellari$^{2}$, Paolo Serra$^{7}$, P.T. de Zeeuw$^{5,8}$, \newauthor Nicholas Scott$^{9}$, Richard M. McDermid$^{10}$, Anne-Marie Weijmans$^{11}$\thanks{Dunlop Fellow}, \newauthor Timothy A. Davis$^{5}$, Katherine Alatalo$^{12,13}$, Leo Blitz$^{12}$, Maxime Bois$^{14}$, \newauthor Martin Bureau$^2$, Frederic Bournaud$^{15}$, Alison Crocker$^{15}$, Pierre-Alain Duc$^{16}$, \newauthor Sadegh Khochfar$^{17,18}$, Harald Kuntschner$^{5}$,  Raffaella Morganti$^{6,19}$, Tom Oosterloo$^{6,19}$,   \newauthor Lisa M. Young$^{20}$\\ 
$^{1}$Leibniz-Institut f\"ur Astrophysik Potsdam (AIP), An der Sternwarte 16, D-14482 Potsdam, Germany\\
$^{2}$Sub-department of Astrophysics, Department of Physics, University of Oxford, Denys Wilkinson Building, Keble Road, Oxford OX1 3RH, UK\\
$^{3}$Max-Planck-Institut f\"ur Astrophysik, Karl-Schwarzschild-Str. 1, 85741 Garching, Germany\\ 
$^{4}$Centre for Astrophysics Research, University of Hertfordshire, Hatfield, Herts AL1 9AB, UK\\
$^5$European Southern Observatory, Karl-Schwarzschild-Str. 2, 85748 Garching, Germany\\
$^6$Universit\'e Lyon 1, Observatoire de Lyon, Centre de Recherche Astrophysique de Lyon and Ecole Normale Sup\'erieure de Lyon,  9 avenue \\ Charles Andr\'e, F-69230 Saint-Genis Laval, France\\
$^{7}$Netherlands Institute for Radio Astronomy (ASTRON), Postbus 2, 7990 AA Dwingeloo, The Netherlands\\
$^8$Sterrewacht Leiden, Leiden University, Postbus 9513, 2300 RA Leiden, the Netherlands\\
$^{9}$Centre for Astrophysics and Supercomputing, Swinburne University of Technology, Hawthorn, Victoria 3122, Australia\\
$^{10}$Gemini Observatory, Northern Operations Centre, 670 N. A`ohoku Place, Hilo, HI 96720, USA\\
$^{11}$Dunlap Institute for Astronomy \& Astrophysics, University of Toronto, 50 St. George Street, Toronto, ON M5S 3H4, Canada\\ 
$^{12}$Department of Astronomy, Campbell Hall, University of California, Berkeley, CA 94720, USA\\
$^{13}$Infrared Processing and Analysis Center, California Institute of Technology, Pasadena, CA 91125, USA\\
$^{14}$Observatoire de Paris, LERMA and CNRS, 61 Av. de l'Observatoire, F-75014 Paris, France\\
$^{15}$Ritter Astrophysical Observatory, University of Toledo, Toledo, OH 43606, USA\\ 
$^{16}$Laboratoire AIM Paris-Saclay, CEA/IRFU/SAp Ð CNRS Ð Universit\'e Paris Diderot, 91191 Gif-sur-Yvette Cedex, France\\
$^{17}$Max Planck Institut f\"ur extraterrestrische Physik, PO Box 1312, D-85478 Garching, Germany\\
$^{18}$Institute for Astronomy, University of Edinburgh, Royal Observatory, Blackford Hill, Edinburgh EH9 3HJ, UK\\
$^{19}$Kapteyn Astronomical Institute, University of Groningen, Postbus 800, 9700 AV Groningen, The Netherlands\\
$^{20}$Physics Department, New Mexico Institute of Mining and Technology, Socorro, NM 87801, USA\\
}

\pagerange{\pageref{firstpage}--\pageref{lastpage}} \pubyear{2012}

\begin{document}
\date{Draft version, \today}
\date{Accepted 2013 May 21. Received 2013 May 20; in original form 2013 February 15}

\maketitle

\clearpage
\label{firstpage}

\begin{abstract}

We investigate nuclear light profiles in 135 \atlas galaxies for which the Hubble Space Telescope (HST) imaging is available and compare them to the large scale kinematics obtained with the SAURON integral-field spectrograph. Specific angular momentum, $\lambda_R$, correlates with the shape of nuclear light profiles, where, as suggested by previous studies, cores are typically found in slow rotators and core-less galaxies are fast rotators. As also shown before, cores are found only in massive galaxies and only in systems with the stellar mass (measured via dynamical models) M $\gtrsim8\times10^{10}$ M$_\odot$. Based on our sample, we, however, see no evidence for a bimodal distribution of nuclear slopes. The best predictor for finding a core is based on the stellar velocity dispersion within an effective radius, $\sigma_e$, and specific angular momentum, where cores are found for $\lambda_R \lesssim 0.25$ and $\sigma_e \gtrsim 160$ km/s. We estimate that only about 10 per cent of nearby early-type galaxies contain cores. Furthermore, we show that there is a genuine population of fast rotators with cores. We also show that core fast rotators are morphologically, kinematically and dynamically different from core slow rotators. The cores of fast rotators, however, could harbour black holes of similar masses to those in core slow rotators, but typically more massive than those found in core-less fast rotators. Cores of both fast and slow rotators are made of old stars and found in galaxies typically lacking molecular or atomic gas (with a few exceptions). Core-less galaxies, and especially core-less fast rotators, are under-luminous in the diffuse X-ray emission, but the presence of a core does not imply high X-ray luminosities. Additionally, we postulate (as many of these galaxies lack HST imaging) a possible population of core-less galaxies among slow rotators, which can not be explained as face-on discs, but comprise a genuine sub-population of slow rotators. These galaxies are typically less massive and flatter than core slow rotators, and show evidence for dynamical cold structures and exponential photometric components. Based on our findings, major non-dissipative (gas poor) mergers together with black hole binary evolution may not be the only path for formation of cores in early-type galaxies. We discuss possible processes for formation of cores and their subsequent preservation.

\end{abstract}


%%%%%%%%%%%%%%%%%%%%%%%%%%%%%%%%%%%%%%%%%%%%%%%%%%%%%%%%%%%
%
% SECTION 1 SECTION 1 SECTION 1 SECTION 1 SECTION 1 SECTION 1 
%
%%%%%%%%%%%%%%%%%%%%%%%%%%%%%%%%%%%%%%%%%%%%%%%%%%%%%%%%%%%

\section{Introduction}

Dichotomies of {\it physical} parameters offer safe anchor points to which one can tie theoretical scenarios of galaxy evolution. It is, therefore, no wonder that a lot of effort was devoted to establish the existence of separate classes of early-type galaxies, in particular, ellipticals. Based purely on imaging, early-type galaxies can be divided into those that have and do not have discs, and this gave rise to the classical distinction between elliptical and S0 galaxies \citep{1920MNRAS..80..746R, 1922ApJ....56..162H, 1926ApJ....64..321H, 1928QB43.J4........} and the Hubble sequence of galaxies \citep[][ for a review]{1936RNeb..........H, 2005ARA&A..43..581S}. As the appearance of a galaxy is strongly dependant on its chance orientation in space, separating ellipticals and S0s as two separate classes is, however, not entirely founded on their physical properties \citep[e.g][]{1994A&A...288..401M, 1994ApJ...433..553J}. 

A promising path was established by using the disciness parameter \citep[e.g][]{1985MNRAS.216..429L,1988A&AS...74..385B}, which showed there are discs in galaxies classified as ellipticals. However, when the inclination is below $\sim60$\degr, the discy deformation of the isophotes ceases to be seen even in high signal-to-nose data (or models) of elliptical galaxies \citep[][]{1990ApJ...362...52R,1996MNRAS.279..993G}. While disciness enables one to connect these systems to disc dominated galaxies, S0s and spirals \citep{1996ApJ...464L.119K, 2012ApJS..198....2K}, its degeneracy with inclination does not provide a robust picture. 

A more promising path is to consider kinematics and, hence, the dynamical state of galaxies. Indeed, as soon as samples of early-type galaxies with resolved kinematics were available \citep[e.g.][]{1977ApJ...218L..43I,1982ApJ...256..460K,1983ApJ...266...41D}, galaxies were plotted in theoretically motivated $V/\sigma$ diagrams \citep{1978MNRAS.183..501B}, where $V$ is the maximum rotational speed, a measure of ordered motion, and $\sigma$ the central velocity dispersion, a measure of random motion. The combination of isophotal parameters and kinematic properties accumulated during the 1980s and 1990s clearly showed that there are two types of elliptical galaxies: luminous ellipticals with round or boxy isophotes that rotate slowly, and faint ellipticals with discy isophotes  that rotate fast \citep[e.g.][]{1989A&A...217...35B, 1989A&A...215..266N, 1989ARA&A..27..235K, 1991A&A...244L..37N, 1996ApJ...464L.119K}. 

Next to the issue of a strong degeneracy with inclination for isophotal shapes\footnote{The degeneracy applies really to the disciness parameters. Its counter-part,  the boxiness parameter \citep{1985MNRAS.216..429L,1989A&A...215..266N}, is less susceptible, but suffers from two setbacks. As galaxies showing boxy isophotes are often triaxial, there are projections at which the isophotes will look more round \citep{1991ApJ...383..112F}. Furthermore, bars viewed edge-on often have a shape of a peanut, and hence show strong boxy distortion to the isophotal shape. Galaxies harbouring such bars should, naturally, not be confused with triaxial ellipticals as they are dynamically different.}, the kinematic results of the 1980s and 1990s were based on information along long-slits, often only one (along the major axis) or, at best, two (along minor and major axes), but rarely more \citep[but see][]{1988ApJS...68..409D}. While the main kinematic properties of galaxies can be inferred in this way  \citep[e.g. the existence of kinematically distinct cores, hereafter KDCs;][]{1982MNRAS.201..975E,1988A&A...193L...7B,1988ApJ...330L..87J,1988ApJ...327L..55F}, the characterisation of the kinematic properties, such as angular momentum, is difficult and not robust \citep[e.g.][]{2005CQGra..22S.347C}. 

This was improved with the advent of integral-field spectrographs which can cover a significant part of the galaxy body. One of these instruments is SAURON \citep{2001MNRAS.326...23B}, initially used to survey a representative sample of nearby-early types galaxies \citep[SAURON Survey,][]{2002MNRAS.329..513D}. The SAURON survey showed it is possible to derive a measure of the specific angular momentum, $\lambda_R$ \citep{2007MNRAS.379..401E}, and use it to divide early-type galaxies in two classes, slow and fast rotators, in a way that is robust to inclination effects \citep{2007MNRAS.379..418C}. 

The \atlas Project \citep[][ hereafter Paper I]{2011MNRAS.413..813C} surveyed a complete and volume limited sample of nearby early-type galaxies, providing a statistical view of the relative numbers of ETGs belonging to the fast and slow rotator categories. In the nearby Universe the vast majority of early-type galaxies, including as much as 66 per cent of the galaxies classified as ellipticals, are fast rotators \citep[][ hereafter Paper III]{2011MNRAS.414..888E}, close to axisymmetric galaxies (modulo bars) with regular disc-like rotation \citep[][ hereafter Paper II]{2011MNRAS.414.2923K}.

The project showed that the fast rotator class is dominated by discs \citep[][ hereafter Paper XVII]{2012arXiv1210.8167K}. Furthermore, fast rotators form a smooth parallel sequence to spiral galaxies on the luminosity/mass--size plane (e.g. fig.~4 of Paper I). This was found to be due to a trend in the bulge fraction which appears the key driver for galaxy properties \citep[][ hereafter Paper XX]{2012arXiv1208.3523C}. This motivated a re-visitation of van den Bergh's proposed revision to Hubble's tuning fork, which emphasised a parallelism between S0 and spiral galaxies. We showed that the true parallelism is the one between fast rotator ETGs and spirals instead \citep[][ hereafter Paper VII]{2011MNRAS.416.1680C}.  Similar conclusions were also reached using photometry by \citet{2011AdAst2011E..18L} and  \citet{2012ApJS..198....2K}.

The improved imaging capabilities of the 1980s and 1990s also brought to light another distinction between early-type galaxies: those that contain steep and flat surface brightness profiles in the nuclei \citep[e.g.][]{1985ApJ...292..104L,1985ApJ...295...73K,1991A&A...244L..37N}. This field of research was revolutionised by Hubble Space Telescope (HST) imaging which confirmed the distinction between galaxies with {\it cores}, where a core is the region in which the surface brightness profiles flatten out, and those, often referred to as {\it power-laws}, that exhibit a rise in the surface brightness profile up to the resolution limit of 0.1\arcsec\, (or less), corresponding to $\sim8$ pc at the distance of Virgo \citep[e.g.][]{1994AJ....108.1598F, 1995AJ....110.2622L, 1997AJ....114.1771F, 2001AJ....121.2431R,2001AJ....122..653R}. 

An alternative way of looking at the distinction in the surface brightness profiles is to emphasise the difference between those galaxies that exhibit excess (related to power-laws) to those that show a deficit (related to cores) of light compared to a S\'ersic fit to large radial range \citep[e.g.][]{2003AJ....125.2951G, 2004AJ....127.1917T, 2006ApJS..164..334F,2009ApJS..182..216K}.

An obvious next step was to compare the three physical properties which suggest two populations of ellipticals: large-scale structure (i.e. isophotal shapes - disciness and boxiness), kinematics (dominance of ordered or random motions) and nuclear profiles (cores and power-laws). An initial comparison (not including kinematics) of \citet{1994AJ....108.1598F}, based on a small sample, indicated a link between power-law (Type II in their study) and discy ellipticals, as well as core galaxies (Type I in their study) and ``classical disk-free ellipticals". As a step further, \citet{1997AJ....114.1771F} compared the nuclear light profiles of a larger sample with both global structural and kinematic properties \citep[see also for a contemporary study of KDC galaxies][]{1997ApJ...481..710C}. Their conclusion was that cores were found in luminous, boxy\footnote{Note that in \citet{1997AJ....114.1771F} {\it boxy} are those galaxies with boxy, round and strongly varying isophotes.} and slowly rotating galaxies, while power-law galaxies were associated with less luminous discy and rapidly rotating ellipticals and S0s. A confirmation of this trend was given in \citet{2001AJ....121.2431R} who found only one of nine core galaxies with discy isophotes. 

The aim of this work is to connect the angular momentum based separation of early-type galaxies from the \atlas survey with properties of their nuclear surface brightness profiles. Paper III showed there is a clear trend that cores (or deficits of light) are found in slow rotators, while power-laws (or excesses of light) in fast rotators. Recently, this was expanded by \citet{2012ApJ...759...64L}, who, based on a larger sample, concluded that the division in slow and fast rotators essentially follows the division in core and power-law galaxies, and therefore reflects the \citet{1996ApJ...464L.119K} division of ellipticals based on isophote shape, rotation and central structure. We agree with the motivating statement of \citet{2012ApJ...759...64L} that the division of galaxies in power-laws and cores is significant and offers important clues about galaxy evolution,  but we prefer to retain a pure kinematic classification of early-type galaxies and not to move the boundary between fast and slow rotators such as to include all core galaxies. Here we look in more detail into those cases which seem to spoil the clean separation of early-type galaxies in two classes. 

Our work differs from that of \citet{2012ApJ...759...64L} in two major points. First, in this work we extend the data base of \atlas galaxies with core/power-law classification by almost a factor of 2. This allows us to show the existence of significant populations of power-law slow rotators and core fast rotators. Secondly, the separation of fast and slow rotators from Paper III is robust and core fast rotators are not kinematically misclassified galaxies. Additionally, we show that inclination effects can not be used to explain the existence of power-law galaxies among slow rotators. 

Based on our larger sample, and using additional \atlas data, we reach some conclusions that differ from \citet{2012ApJ...759...64L}. The main finding could be summed up in the following way: differences between early-type galaxies are unquestionable, and they lie along the lines summarised by \citet{1996ApJ...464L.119K}, Paper VII and \citet{2012ApJS..198....2K}. Nevertheless, the actuality of both core fast rotators and (likely a significant sub-population of) core-less slow-rotators has serious implications for the range of core formation mechanisms as well as the whole assembly history of early-type galaxies, which might be more varied than heretofore appreciated. 

A particular strength of our work is in relying on a clearly selected volume limited sample of nearby early-type galaxies using the largest available sample of HST observations. Regarding the later point, our results are only as robust as the match between \atlas and HST observed samples. Only about half of \atlas galaxies are actually observed with the HST, but the half that is in the archive clearly points out that with observations of an additional two dozen galaxies all remaining questions could be removed (as we argue in Section~\ref{ss:nohstSR}). 

Our study  is divided into six sections. In Section~\ref{s:obs} we briefly describe the \atlas sample and our search through the HST archive. In Section~\ref{s:ana} we discuss in detail our preferred choice of analysis of nuclear profiles and define the two classes of core and core-less galaxies. In Section~\ref{s:reza} we present the main results of this study, namely the relations between nuclear profiles and other global properties of galaxies, such as angular momentum, kinematics, mass, stellar populations, X-ray content and environment. This is followed by a discussion in Section~\ref{s:discusa} where we look into implications of our results with respect to the galaxy evolution scenarios. In Section~\ref{s:cona} we summarise our results. An interested reader can find a comparison of our classification with the literature and other possible parameterisations in Appendix A, images of galaxies for which we were not able to extract robust light profiles in Appendix B and a table with the results in Appendix C.

%%%%%%%%%%%%%%%%%%%%%%%%%%%%%%%%%%%%%%%%%%%%%%%%%%%%%%%%%%%
%
% SECTION 2 SECTION 2 SECTION 2 SECTION 2 SECTION 2 SECTION 2 
%
%%%%%%%%%%%%%%%%%%%%%%%%%%%%%%%%%%%%%%%%%%%%%%%%%%%%%%%%%%%

\section{Observations}
\label{s:obs}

The \atlas sample is defined in Paper I and consists of 260 early-type galaxies, visually selected from a magnitude (brighter than -21.5 mag in the {\it K}-band) and volume limited (within 42 Mpc) parent sample. The full sample was observed with the integral-field spectrograph (IFS) SAURON \citep{2001MNRAS.326...23B} mounted on William Herschel Telescope. The extraction of kinematics from the SAURON data is described in Paper I and \citet[][ for the subsample of 48 galaxies previously presented in the SAURON survey]{2004MNRAS.352..721E}. 

We searched the HST archive for observations of \atlas galaxies with three imaging instruments: {\it Wide-field Planetary Camera 2} \citep[WFPC2, ][]{1995PASP..107.1065H}, {\it Advanced Camera for Surveys} \citep[ACS,][]{1998SPIE.3356..234F} and {\it Wide-field Camera 3} (WFC3).  

We imposed the requirements that the nucleus is positioned on the central chip (PC1) of the WFPC2, and that it is close to its centre. The first requirement ensures the data are of high spatial resolution and sampling, and the second that one can derive large enough radial light profiles. While our choice of using only the PC1 chip ensures that the three instruments have similar and highest available spatial resolution (0.045\arcsec/pixel for WFPC2, 0.049\arcsec/pixel for ACS and 0.05\arcsec/pixel for WFC3), the radial extent of the data is different. ACS and WFC3 have a larger field-of-view of $\sim$100\arcsec compared to 34\arcsec\, for WFPC2 PC1. This distinction between the data sets will be used later as an argument for the analysis method (see Section~\ref{s:ana}). 

The search yielded 135 galaxies which satisfied these basic criteria, originating in a variety of observing programmes. A number of galaxies were observed with multiple instruments and we always favoured ACS data, unless the ACS imaging contained nuclear artefacts which could not be accounted for. While WFPC2 provides somewhat higher spatial resolution and a better behaved point-spread function (PSF), ACS images have a larger extent. As there were no galaxies observed only with WFC3, we again decided to favour ACS data to keep the analysis as uniform and similar as possible.

Some of the observing programmes were initiated with the analysis of the nuclear profiles in mind and there are 86 \atlas objects with already published light profiles. However, the characterisation of their light profiles was heterogeneous, and we decided to (re)analyse a total of 104 galaxies, often to exploit the more recent ACS imaging, while we used the published values for 31 profiles. In the following two subsections we outline our approach to previously published and unpublished analysis of the nuclear properties of \atlas galaxies. A table with results of our analysis can be found in Appendix~\ref{c:table}.

%%% Table 1.%%%%%%%%%%%%%%%%%%%%%%%%%%%%%%%%%%%%%%%%%%%%%%%%%%%%%%%%%%%%%
\begin{table}
\caption{Summary of observations analysed in this work from the HST Legacy Archive.}
\label{t:obs}%\\
\begin{tabular}{lcllll}
\hline
\hline
Observations& no. of objects & instrument & filter\\% & proposal PI \\
%(1)& (2)&(3) &(4)\\%&(5)\\
\hline
GO-9353   & 1  &  ACS/WFC1      & F555W  \\    % Phillips et al.
GO-9293   & 1  &  ACS/WFC2      & F814W  \\    % Ford et al.
GO-9399   & 1  &  ACS/WFC       & F606W  \\    % Carter et al.
GO-9788   & 4  &  ACS/WFC1      & F814W  \\    % Ho et al.
GO-10003  & 1  &  ACS/WFC       & F475W  \\    % Sarazin et al.
GO-10554  & 3  &  ACS/WFC       & F475W  \\    % Sharples et al.
GO-11679  & 2  &  ACS/WFCENTER  & F475W  \\    % Sarazin et al.
GO-10594  & 3  &  ACS/WFCENTER  & F475W  \\    % Goudfrooij et al.
GO-10705  & 1  &  ACS/WFC       & F555W  \\    % Noll et al.
GO-6554   & 2 &  WFPC2/PC1     & F555W  \\    % Brodie et al.
GO-5741   & 1 &  WFPC2/PC1     & F555W  \\    % Westphal et al.
GO-6357   & 8 &  WFPC2/PC1-FIX & F702W  \\    % Jaffe et al.
GO-6633   & 2 &  WFPC2/PC1-FIX & F555W  \\    % Carollo et al.
GO-8212   & 1 &  WFPC2/PC1     & F814W  \\    % Ajhar et al.
SNAP-5479 & 4 &  WFPC2/PC1     & F606W  \\    % Malkan et al.
GO-7403   & 1 &  WFPC2/PC1     & F702W  \\    % Filippenko et al.
SNAP-5446 & 1 &  WFPC2/PC1     & F555W  \\    % Illingworth et al.
SNAP-5446 & 7 &  WFPC2/PC1     & F606W  \\    % Illingworth et al.
SNAP-8597 & 2 &  WFPC2/PC1     & F606W  \\    % Regan et al.
SNAP-9042 & 1 &  WFPC2/WFALL   & F606W  \\    % Smartt et al.
GO-6785   & 1 &  WFPC2/PC1     & F702W  \\    % Malkan et al.
SNAP-5999 & 3 &  WFPC2/PC1     & F555W  \\    % Phillips et al.
GO-6107   & 1 &  WFPC2/PC1-FIX & F555W  \\    % Jaffe et al.
GO-5454   & 1 &  WFPC2/PC1     & F555W  \\    % Franx et al.
GO-7450   & 3 &  WFPC2/PC1     & F814W  \\    % Peletier et al.
GO-5920   & 1 &  WFPC2/PC1     & F555W  \\    % Brodie et al.
GO-8686   & 1 &  WFPC2/PC1     & F814W  \\    % Goudfrooij et al.
\hline
\end{tabular}
\\
\end{table}
%%%%%%%%%%%%%%%%%%%%%%%%%%%%%%%%%%%%%%%%%%%%%%%%%%%%%%%%%%%%%%%%%%%%%%%%%

\subsection{Published ACS and WFPC2 surface brightness profiles}
\label{ss:puba}
Forty-four \atlas galaxies are in the ACS Virgo Cluster Survey \citep[ACSVCS,][]{2004ApJS..153..223C} and their deconvolved surface brightness profiles were already published in \cite{2006ApJS..164..334F}. The data were kindly provided to us by the ACSVCS team (L. Ferrarese, private communication). A full description of their data reduction, PSF deconvolution and isophote fitting are described in \cite{2006ApJS..164..334F}. In Section~\ref{s:ana} we describe our choice of profile parameter fitting, which is different to the analysis presented in \cite{2006ApJS..164..334F}. Finally, we used ACS imaging for 43 galaxies, excluding one object due to a nuclear artefact on the image. 

A comprehensive collection of WFPC2 data was presented in \citet{2007ApJ...664..226L} and of their 219 objects collected from the literature there are 61 in common with the \atlas sample \citep[used by][]{2012ApJ...759...64L}, excluding two galaxies with only {\it  Wide Field/Planetary Camera 1} imaging. Of these, we used the information for 18 objects presented in \citet{2005AJ....129.2138L}, which include some data previously published by the same team in \citet{1995AJ....110.2622L} and \citet{1997AJ....114.1771F}. Furthermore,  we used 10 objects from \citet{2001AJ....121.2431R}, 4 from \citet{2001AJ....122..653R} and 1 from \citet{2000ApJS..128...85Q}. The latter five objects were actually observed with {\it NICMOS}, but a further search through the archive did not return galaxies only observed with {\it NICMOS}. The remaining objects have ACS imaging, either already published or in the archive. For NGC4660 we used the published WFPC2 profile of \citet{1995AJ....110.2622L,2005AJ....129.2138L} instead of the ACS image which had an artefact in the nuclear region. For further 14 galaxies we used archival WFPC2 images to extract profiles (see Section~\ref{ss:archa}) and performed our own fits. This was done based on disagreement between \citet{2007ApJ...664..226L} and subsequent studies (see Appendix~\ref{a:comp}) or to test for the influence of dust (see Appendix~\ref{b:nofit}).

\subsection{Archival HST/ACS and WFC2 data}
\label{ss:archa}
A search through the Hubble Legacy Archive (HLA) revealed additional 58 galaxies in common with the \atlas sample. In Table~\ref{t:obs} we list the observing programs, instruments and filters used. Of these 42 galaxies have WFPC2/PC1 imaging and the remaining 16 have ACS images. In Sections~\ref{ss:prof} and~\ref{ss:nuki} we outline the extraction of surface brightness profiles and the analysis.

%%%%%%%%%%%%%%%%%%%%%%%%%%%%%%%%%%%%%%%%%%%%%%%%%%%%%%%%%%%
%
% SECTION 3 SECTION 3 SECTION 3 SECTION 3 SECTION 3 SECTION 3 
%
%%%%%%%%%%%%%%%%%%%%%%%%%%%%%%%%%%%%%%%%%%%%%%%%%%%%%%%%%%%

\section{Analysis}
\label{s:ana}

\subsection{Surface-brightness profiles}
\label{ss:prof}

Surface-brightness profiles were constructed using the \textsc{iraf.stsdas} task \textsc{ellipse}. A full description of the task is given in \citet{1987MNRAS.226..747J}. Brießy, the intensity $I(\Phi)$ is azimuthally sampled along each ellipse, described by the semi-major axis length from the ellipse centre, position angle $\Phi$ and ellipticity $\epsilon$.  As the fitting proceeds the semi-major axis is increased logarithmically such that the proceeding ellipse has a semi-major axis which is 10 per cent larger. The best-fitting parameters for each ellipse (ellipse centre, $\Phi$ and $\epsilon$) are determined by minimising the sum of squares of the residuals between the data and the first two moments in the Fourier expansion. Since these parameters may be significantly influenced by dust and foreground stars, bad object masks were created prior to fitting. This method is similar to that used in \citet{2006ApJS..164..334F}. 

Deconvolved profiles were approximated in the following way. For each instrument and filter combination a single PSF image was created (at the aperture centre of the ACS/WFC and WFC2/PC1 fields), using TinyTIM \citep{2011SPIE.8127E..16K}, representative of a galaxy with a blackbody spectrum with temperature of 6500K. The original source HST image was convolved with the PSF image, using the \textsc{iraf.stsdas} task \textsc{fconvolve}, to create a smoothed image. We applied the same best-fitting ellipse parameters (ellipse centre, $\Phi$ and $\epsilon$) from the source image to the convolved image to extract smoothed surface brightness profiles. Using convolution theory a good approximation to the deconvolved profile is described as: $SB_{\rm deco} \sim 2 \times SB_{\rm orig}  - SB_{\rm conv}$, where $SB_{\rm deco}$, $SB_{\rm orig}$ and $SB_{\rm conv}$ are the deconvolved, original and convolved (smoothed) surface brightness profiles, respectively. The errors on the  $SB_{\rm deco}$ are estimated as a quadrature sum of the errors on $SB_{\rm orig} $ and $SB_{\rm conv}$ profiles. 

To make comparisons with previous work \citep{2001AJ....121.2431R,2005AJ....129.2138L,2007ApJ...664..226L}, the HST/ACS $g-$ (F475W) band surface brightness profiles (both ACSVCS profiles and archival observations) were converted to F555W (broadband V) photometry, following the photometric transformations in \citet{2005PASP..117.1049S} and using $g-$band and $r-$band colours (Petrosian magnitudes) from SDSS DR5.

\subsection{Nuker Profile}
\label{ss:nuki}

A number of analytic fitting functions can be used to probe the internal structure of galaxies. To explore the core region of galaxies we fit the ``Nuker law'' \citep{1995AJ....110.2622L}, to our surface brightness profiles: 
\begin{equation}
I (r)  = 2^{(\beta - \gamma)/\alpha}  I_b \left(\frac{r_b}{r}\right)^\gamma \left[ 1 + \left(\frac{r}{r_b}\right)^{\alpha}\right]^{(\gamma- \beta)/\alpha}\\
\end{equation}
where  $\gamma$ is the inner cusp slope as $r \rightarrow 0$ and is distinguished from $\gamma^\prime$, which is the purely local (logarithmic) gradient of the luminosity profile evaluated at the HST angular resolution limit, $r^\prime$ (by default we adopt $r^\prime$ = 0\farcs1, except for objects taken from the literature where we keep the published values), where \citep{2001AJ....121.2431R,2004AJ....127.1917T}:

\begin{equation}
\label{e:gamma}
\gamma^\prime \equiv - \frac{d~log~I}{d~log~r} \bigg|_{r = r^\prime}   = - \frac{\gamma + \beta(r^\prime/r_b)^\alpha}
{1 + (r^\prime/r_b)^\alpha} \\
\end{equation}

\noindent In this description, core galaxies exhibit a ``break" radius, $r_b$ which marks a rapid transition (moderated by $\alpha$) to a shallower cusp in surface brightness. We also follow \citet{1997ApJ...481..710C} and \citet{2007ApJ...662..808L} in defining a ``cusp radius":

\begin{equation}
\label{e:cusp}
r_\gamma \equiv r_b \left(  \frac{0.5 - \gamma}{\beta - 0.5}\right)^{1/\alpha}\\
\end{equation}

\noindent With this definition $r_\gamma$ is a radius at which the negative logarithmic slope of the galaxy surface brightness profile, $\gamma^\prime$, as defined by the parameters of the ``Nuker" fit, equals 0.5. For galaxies with cores, $r_\gamma$ can be used as a core scale parameter, as advocated by \citet{1997ApJ...481..710C} and demonstrated by \citet{2007ApJ...662..808L} on a large sample. As recently shown by \citet{2012ApJ...755..163D},  $r_\gamma$ can be used as an approximate break radius of the core-S\'ersic model, or a transition radius between the inner core and the outer S\'ersic profile\footnote{We note that for calculation of $r_\gamma$ we use the parameters of the ``Nuker" model, while \citet{2012ApJ...755..163D} use a non-parametric estimate. }. For galaxies with $\gamma\ge0.5$ at $r=0.5$\arcsec we adopt $R_\gamma<0.1$\arcsec, or the size that was used in the original publication, reflecting the spatial resolution of our study.

When necessary, light profiles were previously prepared as described in Section~\ref{ss:archa}. The fits were done using our own least squares minimization routine which is based on the {\sc IDL} routine \textsc{mpfit.pro}\footnote{http://purl.com/net/mpfit} \citep{2009ASPC..411..251M}, an \textsc{idl} implementation of the \textsc{minpack} algorithm \citep{1980ANL.....80.74M}. To minimise the effect of the PSF and when working with deconvolved profiles, we fit for radii beyond $2-3\sigma$ of the ACS/WFPC2 point-spread function (0.1--0.2\arcsec). We apply the same weighting to each data point in order to limit any bias from the outermost isophotes which may depend significantly on the sky determination. A similarly robust technique was adopted by \citet{2011MNRAS.414..445S} in their study of the profiles of bright cluster galaxies. Table \ref{tab:results} contains the Nuker best-fitting parameters and additional parameters used in the study. 

In order to classify the profiles depending on their local slope $\gamma^\prime$, we begin by adopting the same nomenclature from \citet{2005AJ....129.2138L}: {\it core} galaxies are defined to have  $\gamma^\prime \leq 0.3$ and {\it power-law} (``cuspy'') galaxies are defined to have a steeper inner slope with $\gamma^\prime\ge0.5$. Galaxies with $0.3 < \gamma^\prime<0.5$, first introduced by \citet{2001AJ....121.2431R}, we also call {\it intermediate} galaxies, but we stress that this does not represent a physical transition between power-law and core galaxies. Likewise, we consider power-law and intermediate galaxies to be galaxies that have no resolved cores. This does not exclude the possibility of nuclear cores on smaller spatial scales than probed by our data. Galaxies which are not fitted well with this functional form we call {\it uncertain} and we discuss them in more details below and show their images in Section~\ref{b:nofit}. 

The largest number of galaxies can be classified as power-law (78/135 or 58 per cent), second most numerous are cores (24/135 or 18 per cent), followed by the number of intermediate galaxies (20/135 or 15 per cent). There were also 13 (9 per cent) uncertain galaxies, namely: NGC2824, NGC3032, NGC3073, NGC3607, NGC4111, NGC4233, NGC4435, NGC4526, NGC4694, NGC4710, NGC5866, NGC7465 and UGC05408. All these galaxies have strong dust features in the nuclei, which we show in Fig~\ref{f:thumb} and discuss in Appendix~\ref{b:nofit}.  For these galaxies we simply do not offer a classification and we exclude them from analysis, although a reasonable assumption would be that they do not harbour cores. 

\subsection{The choice of parametrisation}
\label{ss:para}

While the ``Nuker law" might not be the best choice to describe the global light profiles of galaxies \citep{2003AJ....125.2951G,2004AJ....127.1917T}, a consensus as to what method one should use when describing the galaxy nuclei does not seem to be reached in the recent literature \citep[e.g.][]{2005AJ....129.2138L, 2006ApJS..164..334F, 2007ApJ...664..226L, 2007ApJ...671.1456C, 2009ApJS..182..216K,2012ApJ...755..163D, 2012ApJ...759...64L}. As an alternative to the ``Nuker law", \citet{2003AJ....125.2951G} proposed a hybrid function, combing a power-law with the \citet{1968adga.book.....S} function into a so-called ``core-S\'ersic" model\footnote{See also \citet{2012ApJ...744...74G} for a discussion on differences between core-S\'ersic and \citet{1962AJ.....67..471K} models.}. The main benefit of this model is that it can fit well a large radial range of the light profiles (using the S\'ersic profile), as well as the depleted cores of elliptical galaxies (using a power-law function). With this parametrisation, power-law and intermediate galaxies are typically fitted with a pure S\'ersic model, while core galaxies require the core-S\'ersic model. 

The underlying problem is that none of the proposed fitting formulas have a physical foundation, and it is natural to expect that different methods will give different parameters, even if these parameters characterise the same property, such as the break radius. The classification into core or core-S\'ersic profiles, as well as into power-law (and intermediate) and S\'ersic, could differ by up to 20 per cent \citep[][ who fitted Sersic models only to the inner $\sim10$\arcsec\,of light profiles]{2012ApJ...755..163D}, but this is argued against by \citet{2012ApJ...759...64L}, who suggest some 10 per cent discrepancies when outer regions of the light profiles are included in the fit, as advocated by \citet{2003AJ....125.2951G} and \citet{2009ApJS..182..216K}. We find a discrepancy between the two approaches in a few special cases. For these galaxies, it is not clear if this is actually driven by the way the fitting is done (i.e. radial extent used), by the data quality (e.g. PSF effects), or they can be similarly well described by both approaches. Furthermore, one should keep in mind that depleted cores from the S\'ersic formalism and the ``Nuker" cores are not exactly the same structures.  In Appendix~\ref{a:comp} we compare our classification with results in recent literature and show that, if one is only interested to separate cores from the rest of the profiles within the data collected in this work, there are no significant difference between the two approaches.  

%%%%%% Figure 1%%%%%%%%%%%%%%%%%%%%%%%%%%%%%%%%%%%%%%%%%%%%%%%
\begin{figure*}
\includegraphics[width=0.85\textwidth]{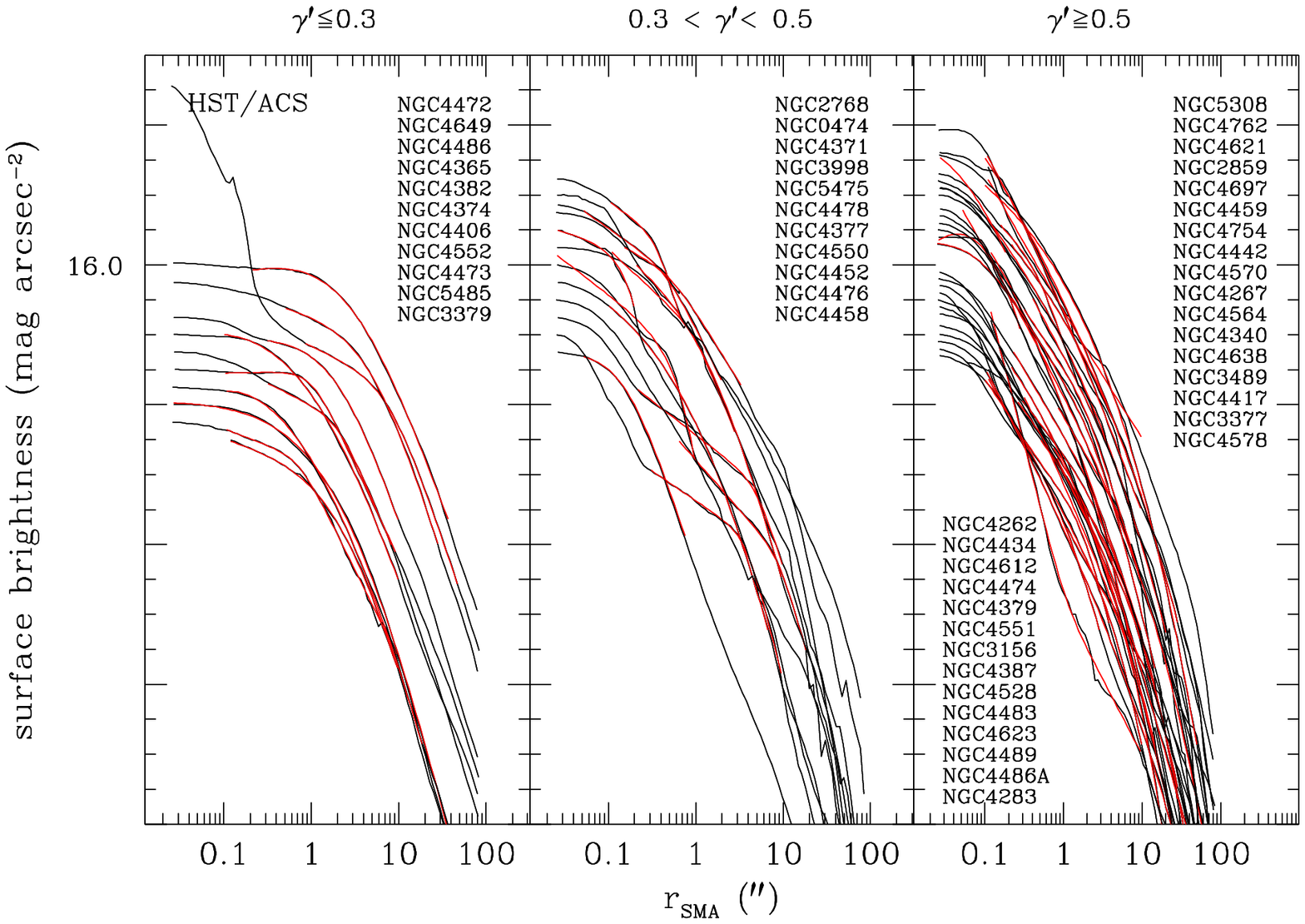}
\includegraphics[width=0.85\textwidth]{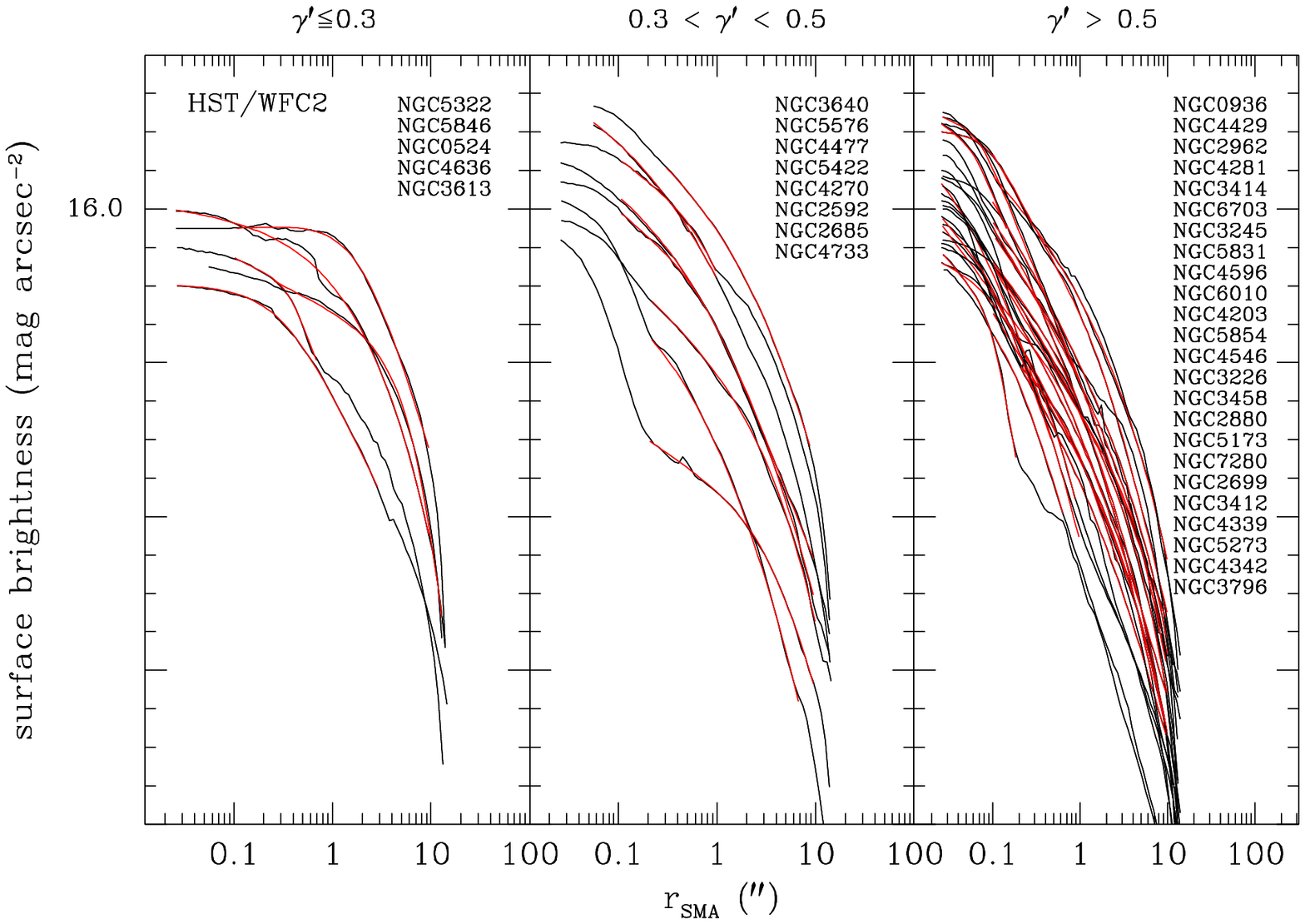}
\caption{Surface brightness profiles as a function of semi-major radius of all HST ACS ({\it top}) and WFPC2 ({\it bottom}) galaxies analysed in this work. On each panel are shown (from left to right): "core" $\gamma^\prime$ $\leq$0.3, "intermediate" 0.3 $< \gamma^\prime <$ 0.5,  "power-law"  $\gamma^\prime$ $\geq$0.5 galaxies. The region over which the Nuker fit is applied is shown in red, and corresponds to typically 0.1\arcsec -- 50\arcsec\, for ACS imaging and 0.1\arcsec -- 10\arcsec\, for WFPC2 PC1 imaging. In both top and bottom panels galaxies are ordered by their total K-band luminosity starting with the brightest at the top and offset downwards for 0.25 mag/\arcsec$^2$ (left panels) and 0.1 mag/\arcsec$^2$ (middle and right panels). Galaxy names follow the sequence of profiles. The information on the filters used is given in Table~\ref{tab:results}. The profile with the steep rise in the top left panel is NGC4486, where the inner profile is dominated by the contribution of the active nucleus. This component is ignored in the fit as well in the ordering of this galaxy in the figure. 
}
\label{f:hst_profiles}
\end{figure*}
%%%%%%%%%%%%%%%%%%%%%%%%%%%%%%%%%%%%%%%%%%%%%%%%%%%%%%%%%%

It is obvious that the ``Nuker law" can not fit the full radial range of a galaxy light profile, and it was intended to describe only its central regions. \citet{2001AJ....121.2431R} show that its double power law form is not adequate to fit those profiles that change smoothly (from steep to flat). Ultimately, the full light profiles of galaxies can not be parameterised in such a way \citep[e.g.][]{1953MNRAS.113..134D, 1968adga.book.....S,1993MNRAS.265.1013C}. To obtain robust physical parameters, such as the break radius, or the amount of light that is depleted, or in excess, one needs to parameterise robustly the full radial range. Therefore, S\'ersic and core-S\'ersic fitting functions are more suitable as they can reproduce most of the profile, although it is often necessary to add multiple components to \citep[e.g.][; Paper XVII]{2004MNRAS.355.1155D,2009ApJ...696..411W, 2011MNRAS.415.2158R, 2012ApJ...754...67F,2012ApJS..203....5T}. When using the S\'ersic function, however, it is imperative to have large radial extent of the light profiles as the overall results will depend on the actual radial range used during the fit \citep[e.g.][]{2003AJ....125.2951G,2004AJ....127.1917T,2009ApJS..182..216K}. Additionally, as shown by \citet{2006ApJS..164..334F} and \citet{2009ApJS..182..216K}, the local nature of ``Nuker" profiles can result in missing compact nuclear structures, such as large cores with extra light or extra-light profiles with small cores \citep[see fig.~111 in][ for an informative illustration of possible profiles]{2006ApJS..164..334F}.  

The light profiles we wish to analyse extend to different radii, due to observations with different HST cameras (e.g. limited extent of WFPC2/PC1 data), filters and exposure times. In Paper XVII, we showed that our sample consists of a large number of galaxies whose outer light profiles require multiple components (e.g. S\'ersic and exponential profiles and/or components describing bars, rings and ovals). As we are primarily interested in determining whether a galaxy has a core or not, in order to ensure a uniform and relatively simple analysis across the sample, we prefer to use the ``Nuker law" and fit only the nuclear regions. An alternative would be to assemble deep ground-based data (in various filters) and combine it with HST profiles, but this is beyond the scope of this paper.

\subsection{Core and core-less nuclei}
\label{ss:classa}

In Fig.~\ref{f:hst_profiles} we show light profiles of the 91 galaxies analysed in this work (13 "uncertain" galaxies are not shown here). They are grouped in those obtained with the ACS (top) and those with WFPC2 (bottom). Furthermore, we separate them according to standard classifications into core, intermediate and power-law profiles.  As shown by other studies \citep{2003AJ....125.2936G,2004AJ....127.1917T,2006ApJS..164..334F,2006ApJS..165...57C,2007ApJ...671.1456C}, this separation is not necessary related to a strong physical difference between these profiles, and by putting them all together emerges a continuous sequence of profiles. 
 
\begin{figure}
\includegraphics[width=\columnwidth]{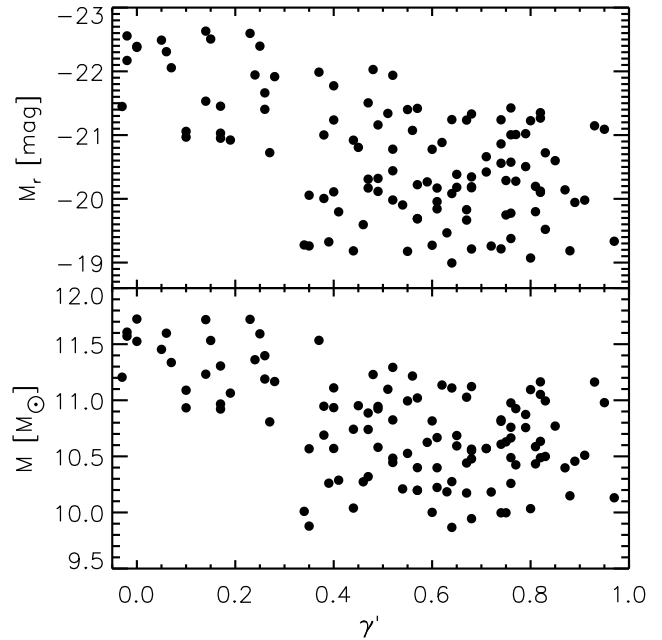}
\caption{Distribution of total r-band magnitude ({\it top}) and dynamical mass ({\it bottom}) with respect to the $\gamma^{\prime}$ slope of the ``Nuker" profiles at the angular resolution of 0.1\arcsec. }
\label{f:gamma}
\end{figure}

In the top panel of Fig.~\ref{f:gamma} we plot the $\gamma^{\prime}$ slope of the Nuker profiles against the total absolute luminosity in the {\it r}-band from Table~1 of \citet[][ hereafter Paper XV]{2012arXiv1208.3522C}, which was derived from the MGE models \citep{1994A&A...285..739E} of the \atlas sample by \citep[][ hereafter Paper XXI]{2012arXiv1211.4615S}. As previously noted \citep[][]{1994AJ....108.1579V,1997AJ....114.1771F,2001AJ....121.2431R,2001AJ....122..653R,2005AJ....129.2138L}, there is an overlap in luminosity between of galaxies with different nuclear slopes, but there is also a clear trend that brighter galaxies have lower $\gamma^{\prime}$ slopes, with no galaxies fainter than $-20.6$ and  $\gamma^{\prime}<0.3$ and there are no galaxies brighter than $-21.5$ and $\gamma^{\prime}>0.5$, using our {\it r}-band parametrisation. There are a few galaxies with $-22 < M_r < 21.5$ and intermediate values of 
 $\gamma^{\prime}$. 

In the bottom panel, we use mass estimates from Paper XV to investigate the $\gamma^\prime$ dependence further. The mass is obtained from Jeans anisotropic models \citep{2008MNRAS.390...71C} and comprises both the stellar and dark matter, although the dark matter typically does not contribute with more than 12 per cent (see Paper XV for details). The mass\footnote{The mass we use is the one called $M_{\rm JAM}$ in Paper XV.} is defined as $M = L \times (M/L)_e \approx 2 \times M_{1/2}$, where $M_{1/2}$ is the mass within a sphere enclosing half of the galaxy light and $(M/L)_e$ is the mass to light ratio within the same region. As is the case in the magnitude plot, there is an overlap zone in mass between $\sim10.8$ and $\sim11.2$ (in $\log$M$_\odot$), with no core galaxies below and no power-law or intermediate (except one) galaxies above this zone, respectively. 

%%%%%% Figure 2%%%%%%%%%%%%%%%%%%%%%%%%%%%%%%%%%%%%%%%%%%%%%%%
%%%%%%%%%%%%%%%%%%%%%%%%%%%%%%%%%%%%%%%%%%%%%%%%%%%%%%%%%%

\begin{figure}
\includegraphics[width=\columnwidth]{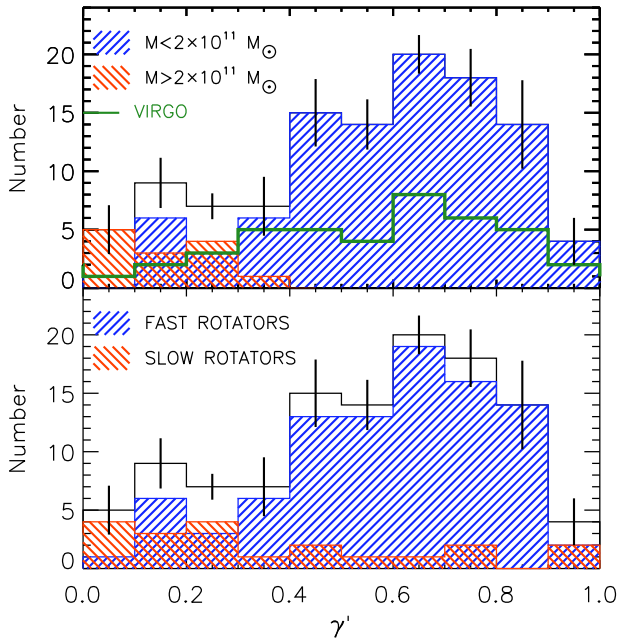}
\caption{Histograms of the $\gamma^{\prime}$ distribution. The top histogram  divides galaxies into those with mass greater (red - hashed to the left) or smaller (blue - hashed to the right) than $2\times10^{11}$ M$_\odot$. The distribution for the galaxies belonging to the Virgo cluster are shown with a green line. The bottom histogram divides galaxies into slow (red - hashed to the left) or fast (blue - hashed to the right) rotators. Vertical bars on both histograms show an estimate of the systematic uncertainty when different radii are used for estimating $\gamma^\prime$.   }
\label{f:gamma_hist}
\end{figure}

While there seems to be a continuous sequence in mass and magnitude, similar to what was found before \citep{2007ApJ...671.1456C,2011ApJ...726...31G}, the continuity of $\gamma^\prime$ parameters seems to be interrupted at just above $\gamma^\prime=0.3$. To investigate this further we plot in Fig.~\ref{f:gamma_hist} two histograms of $\gamma^\prime$, highlighting the dependence on mass and angular momentum in the top and bottom panels, respectively. Significantly more than reported in previous studies \citep{2001AJ....121.2431R, 2001AJ....122..653R,2007ApJ...664..226L}, we find in our sample a number of galaxies populating the intermediate range of $0.3<\gamma^\prime<0.5$, but we also see a mild excess of galaxies with $\gamma^\prime \sim 0.15-0.2$. This excess is, however, statistically not significant, but it is strongly dependant on the definition of $\gamma^\prime$. 

$\gamma^\prime$ is not a physically well defined parameter as it is a measure of the profile curvature at a radius fixed by the resolution of the HST.  Studies applying the ``Nuker law" used a range of radii to calculate $\gamma^\prime$. \citet{2005AJ....129.2138L}  and \citet{2007ApJ...664..226L} used 0\farcs02 or 0\farcs04 depending on the quality of the imaging, attempting to use the smallest angular radius at which a profile's slope could be estimated. \citet{2003AJ....126.2717L} used 0\farcs05, while \citet{2001AJ....121.2431R} used 0\farcs1. As noted above, we also use 0\farcs1 for our galaxies, unless their ``Nuker" parameters were taken directly from other studies. Crucially, however, the radius is not chosen with respect to the galaxy distance or size \citep[see][]{2007ApJ...671.1456C}. We investigated the dependence of our $\gamma^\prime$ determinations on the distance, but we did not find a significant correlation, in spite the fact that about half of \atlas galaxies with the HST imaging are found at distance less or similar to the Virgo Cluster, while the other half between 20 and 40 Mpc (for distance determinations of the \atlas sample see Paper I). However, we selected \atlas galaxies which belong to the Virgo cluster (and are all at comparable distances) and over-plot their distribution of $\gamma^\prime$, which is mostly continuous rising towards the higher values of $\gamma^\prime$.

%%%%%% Figure 3%%%%%%%%%%%%%%%%z%%%%%%%%%%%%%%%%%%%%%%%%%%%%%%%
%%%%%%%%%%%%%%%%%%%%%%%%%%%%%%%%%%%%%%%%%%%%%%%%%%%%%%%%%%

We also investigated the robustness of the $\gamma^\prime$ with respect to the radius at which it is measured, as a simple test for the influence of the galaxy size. This was done by estimating $\gamma^\prime$ at $r=0.05, 0.1, 0.15$ and two fixed physical scales of $r=R_e/100$, when $R_e>10$\arcsec and $r=R_e/200$, when $R_e>20$\arcsec \footnote{Note that the formal median half-light radius of \atlas galaxies with HST imaging is 23.6\arcsec, using the size estimates of Paper I.} (so that the minimum radius is comparable to the resolution of our data 0.1\arcsec). A standard deviation of these estimates is a measure for the systematic variation of $\gamma^\prime$, and it is shown with vertical bars in histograms of Fig.~\ref{f:gamma_hist}. These systematic uncertainties confirm the non-significance of the double peaked structure in the distribution of $\gamma^\prime$. As it is not clear at what physical size one should actually measure $\gamma^\prime$, we keep the practice as in previous studies, but would like the highlight the importance of this particular choice. 

We found, however, that dividing galaxies by mass or by angular momentum is not influenced by the details of $\gamma^\prime$ estimation, and two clear trends can be seen. We use a characteristic mass of $2\times10^{11}$ M$_\odot$, which was highlighted in Paper XX as separating early-type galaxies between typically disc-dominated fast rotators from more round and massive slow rotators. Considering only lower mass galaxies, there is a tail monotonically falling off from the peak in $\gamma^\prime \sim 0.65$ all the way to $\gamma^\prime = 0$. The most massive galaxies, however, have only $\gamma^\prime$ values associated with cores. This indicates that the sample selection is crucial when considering the distribution of the nuclear slopes, as stressed by \citet[e.g.][]{2007ApJ...671.1456C} and \citet{2011ApJ...726...31G}.

As expected, the division of galaxies into fast and slow rotators reflects, to the first order, the division by mass (Paper III). There are, however, two notable differences: some slow rotators have large $\gamma^\prime$ values and are therefore core-less galaxies, while also a number of fast rotators have cores. The latter issue was pointed out by \citet{2012ApJ...759...64L}, who argued that this warrants a change in the definition of line that separates slow and fast rotators, such that the line could be raised to include also core fast rotators present above the line defined in Paper III. In the next section we investigate this further and offer an alternative view. 

As $\gamma^\prime$ is not a physically well defined parameter, and as we are primarily interested in the presence of cores, we proceed by separating our sample into {\it core} ($\gamma^{\prime} \leq 0.3$) and {\it core-less} ($\gamma^{\prime} > 0.3$), but we keep the power-law and intermediate classifications in Table~\ref{tab:results} for clarity and comparison with previous studies based on the ``Nuker" law.

%%%%%%%%%%%%%%%%%%%%%%%%%%%%%%%%%%%%%%%%%%%%%%%%%%%%%%%%%%%
%
% SECTION 4 SECTION 4 SECTION 4 SECTION 4 SECTION 4 SECTION 4 
%
%%%%%%%%%%%%%%%%%%%%%%%%%%%%%%%%%%%%%%%%%%%%%%%%%%%%%%%%%%%

\section{Nuclear light profiles of \atlas galaxies}
\label{s:reza}

In order to put the following discussion in context, we note that $\lambda_R$ tries to capture in a single parameter the large amount of information provided in the kinematic maps. Also, the distinction between fast and slow rotators in the $\lambda_R - \epsilon$ diagram is an empirical one (Paper III). It is based on the fact that within one effective radius the velocity maps of early-type galaxies can be divided on the basis of their similarity with those of inclined discs \citep{2008MNRAS.390...93K}. In this respect the slow--fast separation was done by analysing the \atlas velocity maps (Paper II) with \textsc{kinemetry}\footnote{http://www.davor.krajnovic.org/idl} \citep{2006MNRAS.366..787K}, which is based on a Fourier expansion of the velocity profiles along the best fitting ellipses. Those maps that have larger Fourier coefficients, also show disturbed velocity maps or non-regular rotation, and typically have lower specific angular momentum. Galaxies with small Fourier terms (typically 4 per cent of the first Fourier term describing the rotational velocity) are, on the other hand, characterised by regular (disc-like) rotation, while their angular momentum is also dependent on the inclination. The separatrix line between fast and slow rotators, drawn in Paper III, minimises the overlap between galaxies with regular and non-regular rotation. Hence, slow rotators have low specific angular momentum, but their velocity maps are also irregular, whether being disturbed, containing kinematically distinct cores, counter-rotating structures or showing little or no rotation \citep[for velocity maps see][{ and Paper II}]{2004MNRAS.352..721E} 

%%%%%% Figure 4%%%%%%%%%%%%%%%%%%%%%%%%%%%%%%%%%%%%%%%%%%%%%%%
\begin{figure*}
\includegraphics[width=0.497\textwidth]{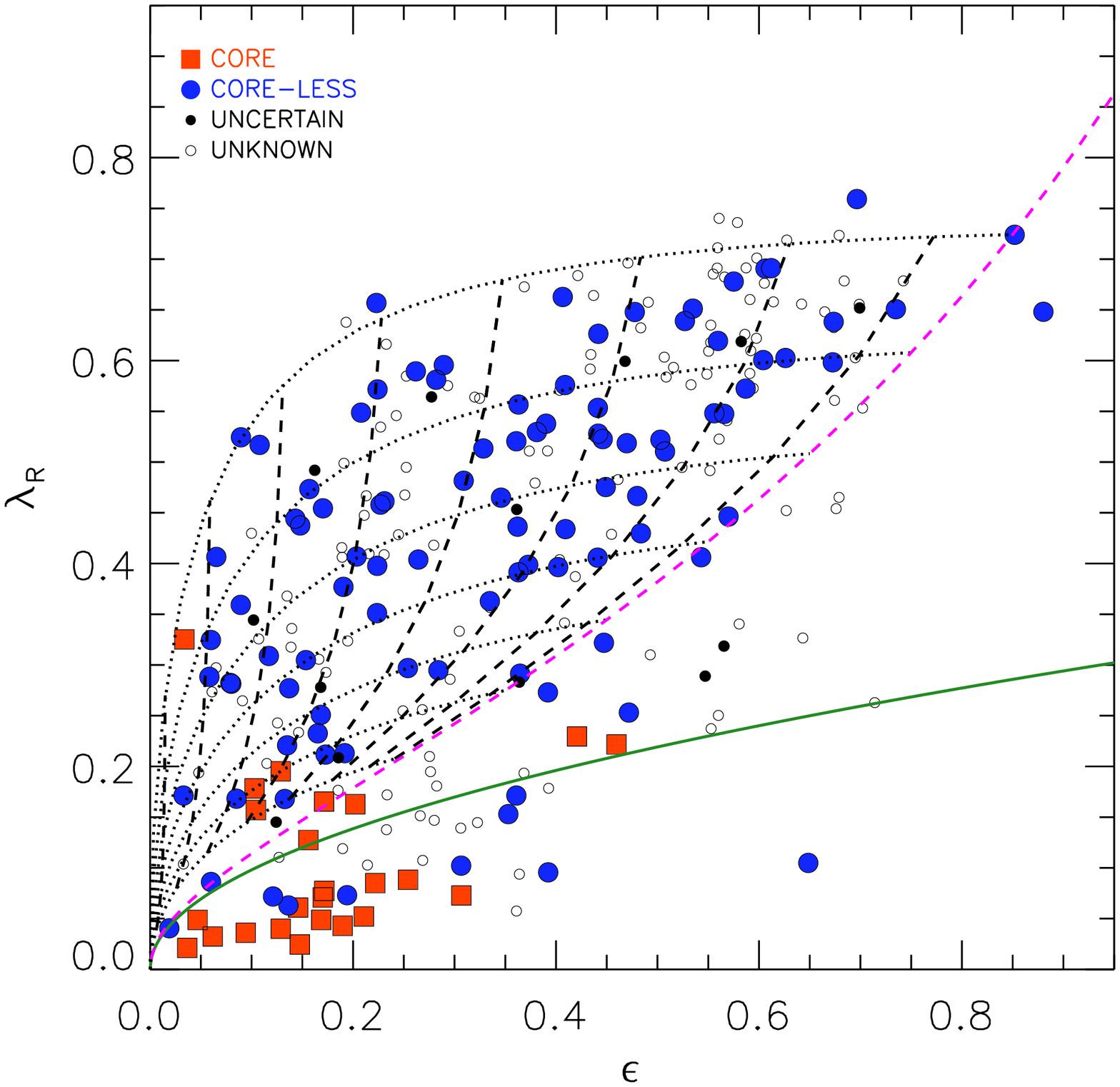}
\includegraphics[width=0.497\textwidth]{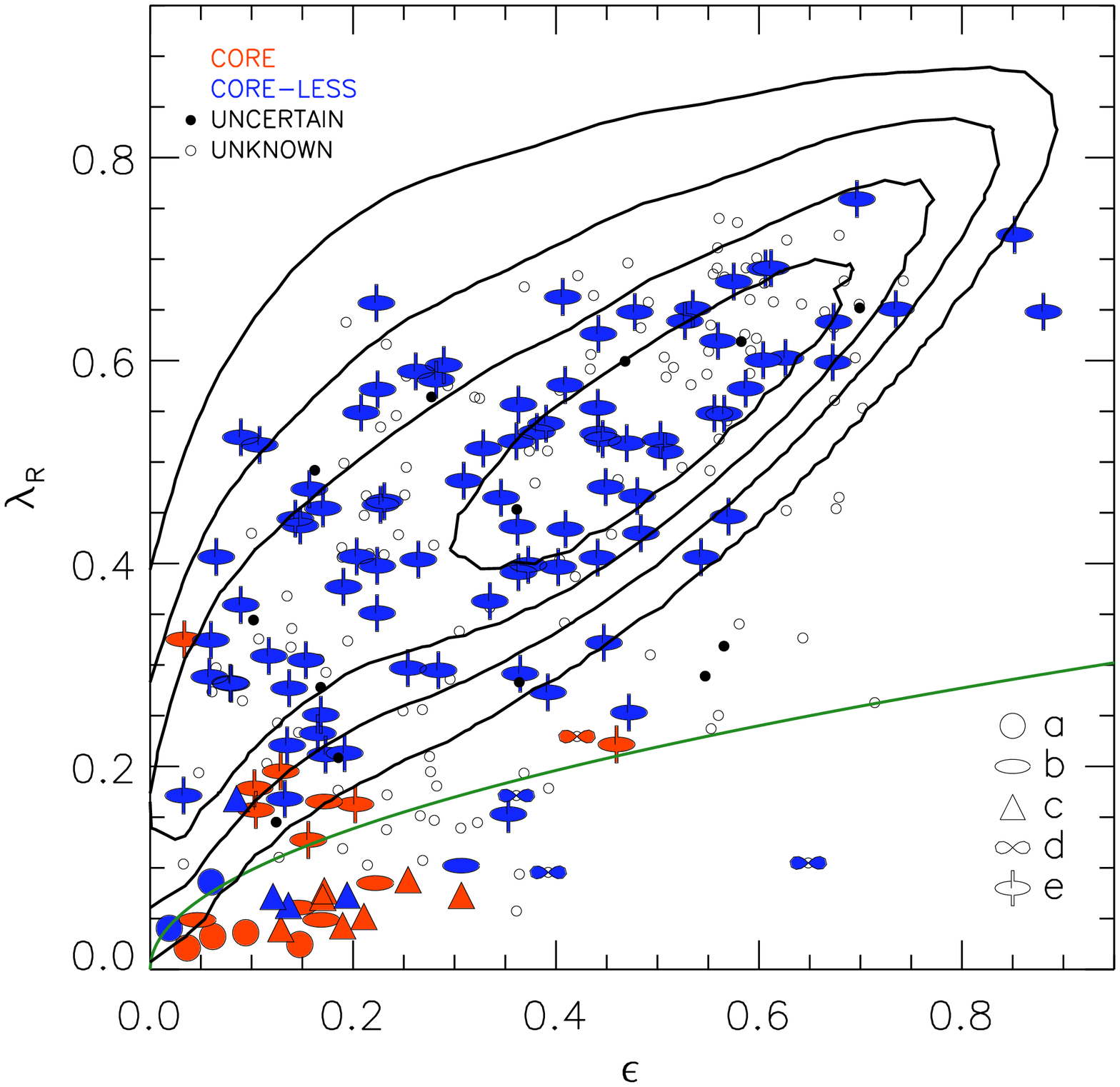}
\caption{$\lambda_R$ versus the ellipticity $\epsilon$ for 260 \atlas galaxies. On both panels, open small symbols are galaxies with no available HST observations, and filled small symbols are galaxies for which the classification was not possible (uncertain). Also on both panels, colours of symbols indicate the class of the nuclear profiles: red -- core ($\gamma^\prime\le 0.3$), blue -- core-less ($\gamma^\prime>0.3$). The green solid line separates fast from slow rotators (Paper III). {\it Left:}  Core galaxies are shown with squares and core-less galaxies with circles. The dashed magenta line shows the edge-on view for ellipsoidal galaxies integrated up to infinity with $\beta =0.7 \times \varepsilon_{\rm intr}$, as in \citet{2007MNRAS.379..418C}. Other dashed lines show the same relation at inclinations of 80\degr, 70\degr, 60\degr, 50\degr, 40\degr, 30\degr, 20\degr and 10\degr (from right to left). The dotted curves show the change of location for galaxies of intrinsic $\epsilon=$ 0.85, 0.75, 0.65, 0.55, 0.45, 0.35, 0.25 (from top to bottom). {\it Right:} Shapes of symbols indicate the kinematic group (Paper II): group a -- non-rotating galaxies, group b -- featureless non-regularly rotating galaxies, group c -- kinematically distinct cores, group d -- $2\sigma$ galaxies made of two counter-rotating discs, and group e - regularly rotating (disc-like) galaxies. Kinematic classification is not provided for galaxies with no HST data.  The contours show the distribution of a family of oblate objects with an intrinsic shape of $\epsilon_{\rm intr} = 0.7 \pm 0.2$ (as in fig. 15 of Paper III). }
\label{f:nuk}
\end{figure*}
%%%%%%%%%%%%%%%%%%%%%%%%%%%%%%%%%%%%%%%%%%%%%%%%%%%%%%%%%%

This is important for the consideration of inclination effects in the $\lambda_R - \epsilon$ diagram.  These we address in Section~\ref{ss:projection} \citep[for previous discussions see][]{2007MNRAS.379..401E, 2007MNRAS.379..418C,2009MNRAS.397.1202J}. The volume limited \atlas sample allowed us to define the curve between the slow and fast rotators taking into account the apparent shape of objects (Paper III). The point to keep in mind is that, while it is still possible that a few face-on fast rotators are misclassified as slow rotators (likely one or two galaxies in \atlas sample), the vast majority of slow rotators are objects with intrinsically different shape, kinematics and dynamics (hence, also likely different formation scenarios) with respect to fast rotators, which make up the majority of early-type galaxies. Other separations of early-type galaxies are possible, for example one involving the galaxies masses or the steepness of the nuclear light profiles, but they are complex. Our approach here is to keep the classification as simple as possible (as in Paper III), expecting that a comparison with the nuclear structure can highlight possible formation paths for ETGs. Therefore, in this section we investigate a range of properties of early-type galaxies and look for correlations with observed nuclear light profiles.

\subsection{$\lambda_R - \epsilon$ diagram}
\label{ss:lamR}

In Fig.~\ref{f:nuk} we show two $\lambda_R - \epsilon$ diagrams separating core (red) and core-less (blue) galaxies according to their measured $\gamma^\prime$ values, as defined in Section~\ref{ss:classa}.  As anticipated by Fig.~\ref{f:gamma_hist}, there are nine core galaxies above the green line which separates fast (above) from slow (below) rotators. As also pointed out by \citep{2012ApJ...759...64L}, they are preferentially found at both low $\lambda_R$ and $\epsilon$ values. More specifically, core fast rotators are clustered around $\epsilon \sim0.15$ and $\lambda_R \sim 0.15$ (six of nine currently known fast rotators with cores). Outside of this group, there are two flat objects and one galaxy with high angular momentum. 

Among slow rotators core galaxies are the dominant population, but there are also core-less galaxies. They notably occur for $\epsilon>0.35$, but a few are found around $\epsilon \sim0.15$. They seem to be present at all, but the very lowest, values of $\lambda_R$. The distribution of galaxies with HST imaging is mostly uniform in the  $\lambda_R - \epsilon$ diagram, except in the region approximately centred on $\lambda_R~\sim0.15$ and $\epsilon\sim0.28$. Most of the galaxies lacking HST imaging in this regions are slow rotators, close to the slow-fast separatrix. We do not know their nuclear profiles, but just the distribution of core and core-less galaxies around them suggest a possible mixed population, and an additional number of core-less slow rotators. We highlight these galaxies here as they will feature prominently in the rest of the paper. 

In the right panel we add the information about the types of velocity maps of galaxies with HST imaging. These are divided in five groups (Paper II): group a -- non-rotating galaxies, group b -- featureless non-regularly rotating galaxies, group c -- kinematically distinct cores, group d -- $2\sigma$ galaxies made of two counter-rotating discs, and group e - regularly rotating galaxies. This shows that cores can be present also in galaxies which have regularly rotating, disc-like velocity maps. Similarly, core-less galaxies are also possible in KDCs, but this is not the case for non-rotating galaxies, which are always cores, and found at the lowest values of $\lambda_R$. 

 \citet{2007MNRAS.379..418C} used \citet{1979ApJ...232..236S} orbit-superposition axisymmetric dynamical models of a subsample of 24 galaxies with SAURON data, to study the $(V/\sigma,\varepsilon)$ diagram of the full SAURON sample. They found fast rotators to be consistent with systems characterised by an approximately oblate ($\sigma_\phi\approx\sigma_R\ga\sigma_z$) average velocity ellipsoid, satisfying an upper limit in anisotropy approximated by $\beta_z\equiv1-\sigma_z^2/\sigma_R^2\approx0.7\times\varepsilon_{\rm intr}$. This trend was also independently found using similar models by \citet{2009MNRAS.393..641T}. Constructing simple axisymmetric models based on the \citet{1922MNRAS..82..132J} equations, \citet{2008MNRAS.390...71C} and \citet{2009MNRAS.398.1835S} explicitly showed that the SAURON kinematics (both $V$ and $\sigma$) of fast rotators can indeed be predicted in quite some detail under the oblate velocity ellipsoid assumption. A much more extensive comparison between the predictions of these dynamical models and the ATLAS$^{\rm 3D}$ kinematics for the full sample was presented in Paper~XV. It confirms that the kinematics, within about 1$R_{\rm e}$, of real fast rotators is well captured by the simple models with oblate velocity ellipsoid.

The projection for different inclinations of these galaxy models with oblate velocity ellipsoid and following the $\beta_z=0.7\times\varepsilon_{\rm intr}$ relation is shown using an analytic formalism in the left panel of Fig.~\ref{f:nuk} (grid of dotted and dashed lines) and using Monte Carlo simulations (from fig.~15 of Paper III) in the right panel (contours). These plots show that fast rotators need to have very low inclination to have sufficiently low $\lambda_R$ to be classified as slow rotators. Similar tests on the robustness of the fast -- slow rotator classification were performed using realistic galaxy $N$-body simulations by \citet{2009MNRAS.397.1202J}, confirming that only a handful of fast rotators may be misclassified in a sample of the size of the present one.

As already shown in Paper III, the grid on the left panel of Fig.~\ref{f:nuk} encloses the majority of fast rotators (note that for clarity we did not plot the cases with intrinsic ellipticities of less than 0.25). Notably the lines avoid (are above) the boundary between fast and slow rotators, although the green fast-slow separatrix was constructed only with regard to how regular (or irregular) the velocity maps are (see beginning of Section~\ref{s:reza} and Papers II and III). Cores are typically found only in galaxies which lie below the dotted line for the intrinsic $\epsilon=0.3$ \citep[given the anisotropy trend found by][]{2007MNRAS.379..418C}. 

In the left panel of Fig~\ref{f:nuk} we show the contours of Monte-Carlo simulation from fig.15 of Paper III, made by assumption that all fast rotators have a similar intrinsic shape, $\epsilon = 0.7 \pm 0.2$ (Gaussian distribution), and are randomly projected into $\lambda_R - \epsilon$ diagram. The contours, which do not overlap (significantly) with the region of slow rotators, enclose majority of fast rotators. However, in the region where core fast rotators are found, we see a change in the shape of the contours, indicating that objects of that particular intrinsic shape are not often found there. Galaxies in this region could have different formation scenarios from the majority of fast rotators \citep[][ hereafter Paper VI]{2011MNRAS.416.1654B}, and we will return to this point in Section~\ref{ss:diff}.

\subsection{Kinematic and morphological properties of core-less slow rotators and fast rotators with cores}
\label{ss:kina}

There are nine fast rotators with cores and there are nine core-less slow rotators among \atlas galaxies with HST imaging. In the next three sub-sections we analyse their respective global morphologies and kinematics.

\subsubsection{Core-less slow rotators}
\label{sss:corelessSR}

Core-less slow rotators are (in order of increasing ellipticity): NGC6703, NGC4458, NGC5831, NGC3414, NGC5576, NGC4476, NGC3796, NGC4528 and NGC4550. These galaxies could be separated in two main groups comprising flat and relatively round objects. We start from the group of flat slow rotators ($\epsilon\gtrsim0.35$), characterised by very special kinematic structure.  A number of these galaxies are counter-rotating discs and classified as $2\sigma$ galaxies (Paper II). The most obvious example is NGC4550 \citep{1992ApJ...394L...9R,1992ApJ...400L...5R,2007MNRAS.379..418C}. The fact that they do not have cores is consistent with their generally disc-like appearance while the low angular momentum is the consequence of the opposite spins of two rotating components. The only non $2\sigma$ galaxy in this group is NGC4476, actually classified as a regularly-rotating object.

 At $\epsilon=0.3$, and just outside the group of flat slow rotators, is NGC5576, one of the galaxies for which our classification disagrees with that of \citet[][ see Appendix~\ref{a:comp} for a discussion of our fit]{2005AJ....129.2138L}, {\bf who detect a core}. This is a galaxy with non-regular kinematics, but with global rotation, as well as a significant kinematic misalignment (Paper II). Its outer (beyond 2.5\arcsec) light profile is best fitted with a single S\'ersic component of high S\'ersic index (e.g. Paper XVII and \citealt{2012ApJ...759...64L}), while \citet{2013ApJ...768...36D} show that the profile can be fitted with two S\'ersic functions.
 
The second group of core-less slow rotators is found at low ellipticities ($0.1<\epsilon<0.2$). In this group there is also one galaxy (NGC4458) for which our classification disagrees with that of \citet[][ who classify it as core]{2005AJ....129.2138L} and we discuss our fit in Appendix~\ref{a:comp}. All galaxies harbour KDCs, while NGC3414 and NGC4528 have also significant exponential components (about 65 and 28 per cent of light, respectively, Paper XVII), and NGC3414 shows signatures of a recent interaction. 

Finally, the galaxy with the lowest ellipticity in the \atlas sample, NGC6703, is also core-less, but this one could be a rare case of a face-on disc. This is consistent with the dynamical models of Paper XV, which can only fit the kinematics if this galaxy is nearly face-on ($i\approx18$). However the disc-bulge decomposition in Paper XVII does not recover an exponential component (see also Paper III for a specific discussion on this galaxy). 
 
Therefore among the investigated galaxies, core-less slow rotators are those that are flat (not taking into account the misclassified face-on disc) or possibly a few special cases. However, a significant number of slow rotators was not observed with the HST and remains unclassified into core and core-less galaxies. We will address them in Section~\ref{ss:nohstSR}.

%%%%%% Figure 5%%%%%%%%%%%%%%%%%%%%%%%%%%%%%%%%%%%%%%%%%%%%%%%
\begin{figure*}
\includegraphics[width=\textwidth]{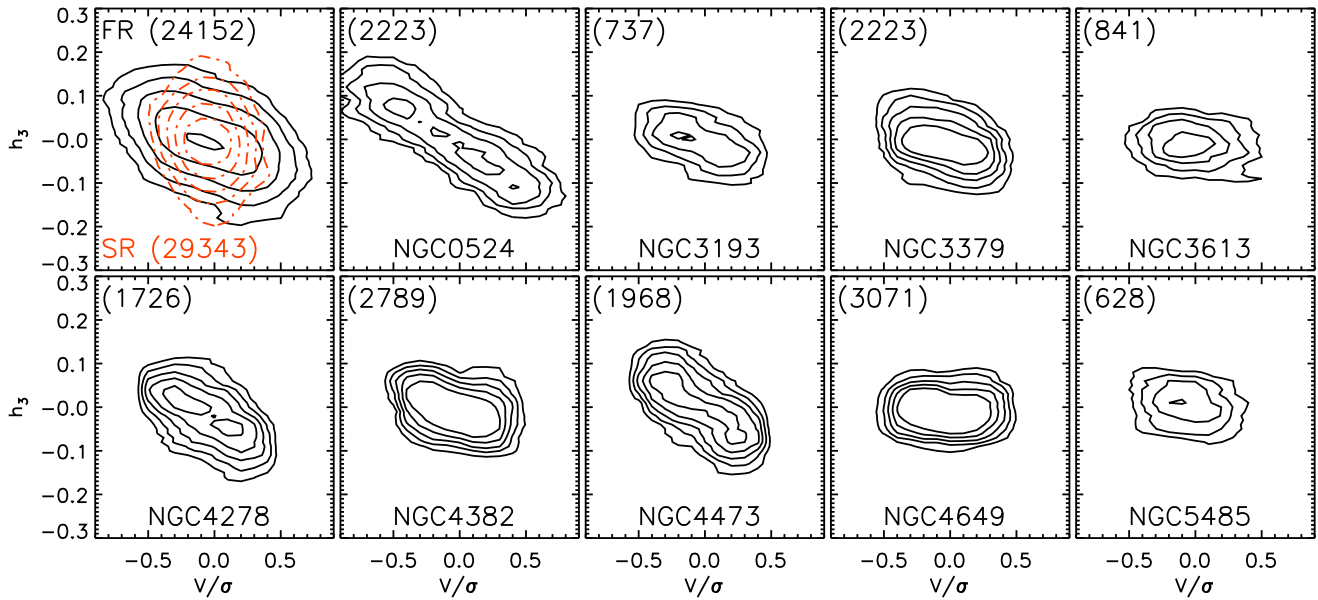}
\caption{Local $V/\sigma - h_3$ relation for non-barred \atlas galaxies with $\lambda_R < 0.3$ (upper left panel) and for individual core fast rotators. Only spatial bins with $\sigma > 120$ km/s and an error on $h_3<0.05$ are used in all panels (number of bins used is indicated by numbers in parentheses). On the upper left panel fast rotators are shown with solid black contours while slow rotators with dot-dashed red contours. The contours show distribution of values in bins of 0.1 in $V/\sigma$ and 0.01 in $h_3$, smoothed with a boxcar filter of a window of 2 pixels in both dimensions. The lowest level is 0.25 and the step is 0.25 in logarithmic units. Note that NGC0524 has $\lambda_R > 0.3$.}
\label{f:vs_h3}
\end{figure*}
%%%%%%%%%%%%%%%%%%%%%%%%%%%%%%%%%%%%%%%%%%%%%%%%%%%%%%%%%%

\subsubsection{Cored fast rotators}
\label{sss:coreFR}

There are nine fast rotators with cores (in order of increasing ellipticity): NGC0524, NGC4278, NGC3379, NGC3193, NGC4649, NGC5485, NGC4382, NGC4473 and NGC3613 (please see discussion on light profiles of NGC3193 and NGC4473 in Appendix~\ref{a:comp}). The velocity maps for all but two are classified as regular (Fig.~\ref{f:nuk} and Paper II); they typically have outer exponential components (Paper XVII), but do now show (obvious) signatures of bars (Paper II). This is significant as at least 30 per cent of \atlas galaxies are barred, and all except two are found among the fast rotators\footnote{NGC4528, a barred, $2\sigma$ slow rotator has notably a core-less nuclear profile.}. A few galaxies deserve a special mention.  NGC0524 has the highest $\lambda_R$ and a significant amount of gas and dust distributed in a spiral configuration. NGC4473 does not warrant a core-S\'ersic fit \citep{2006ApJS..164..334F,2012ApJ...755..163D}. \citet{2009ApJS..182..216K} detect an excess above a S\'ersic fit to specific and disconnected regions of the profile, while \citet{2013ApJ...768...36D} fitted it with an inner exponential and outer S\'ersic models. Therefore, this galaxy is one the few galaxies for which our classification, as well as the previous classification by \citet{2005AJ....129.2138L}, do not agree with those based on the (core-)S\'ersic fits. NGC4382 shows signs of a recent major merger, while NGC4649 is a massive galaxy, but with ordered rotation within the observed effective radius (with an indication that this might not be the case outside the SAURON FoV), and it sits close to the fast -- slow separatrix. Its classification is, therefore, uncertain. Finally, NGC5485 is a prolate rotator with an inner dust disc in a "polar ring" configuration. 

Typically low angular momentum and low ellipticity, together with cores and some peculiarities outlined above, suggest that these galaxies are very similar to slow rotators. Their dynamics is obviously somewhat different as it follows the difference in velocity maps on which the fast slow separation is made. However, the crucial distinction between slow and fast rotators is in the signature of embedded discs. In the next section we try to answer if discs could be present in core fast rotators. 

\subsubsection{$V/\sigma - h_3$ diagram}
\label{ss:vs_h3}

The kinematic information on the embedded high angular momentum components can be extracted from the line-of-sight velocity distribution (LOSVD). Specifically, they are found in the steep leading wings of the LOSVD. When the LOSVD is parametrized with a Gauss-Hermite series \citep{1993MNRAS.265..213G,1993ApJ...407..525V}, the third coefficient, $h_3$, measures the anti-symmetric deviations of the LOSVD from a Gaussian. \citet{1994MNRAS.269..785B} showed that galaxies with (embedded) discs and showing ordered rotation typically show an anti-correlation between $V/\sigma$ and  $h_3$, where $V$, $\sigma$ and $h_3$ are used to describe the local LOSVDs measured at different locations in galaxies. Therefore the so-called ``local" $V/\sigma - h_3$ diagram, constructed from spatially resolved spectra, can be used to indicate those galaxies that are likely to have embedded discs. In Paper II, we showed that \atlas galaxies with regular rotation show a strong anti-correlation pattern\footnote{Barred galaxies show a less pronounced anti-correlation, as bars are often characterised by correlation between  $V/\sigma$ and  $h_3$ within the bar \citep[e.g.][]{2004AJ....127.3192C}}, while galaxies with irregular kinematics, KDCs, or with no net rotation, do not show it. Furthermore, in Paper XVII we showed that $V/\sigma - h_3$ is strongly anti-correlated for those galaxies with structural components that can be fitted with an exponential profile (as opposed to a general S\'ersic profile of a large index $n$).

In the upper left panel of Fig.~\ref{f:vs_h3} we present the $V/\sigma - h_3$ diagram for all galaxies with $\lambda_R < 0.3$ without bars, dividing them into slow and fast rotators to illustrate the general difference in the kinematics between the objects in these two classes. Fast and slow rotators, in spite of having similar extent of $\vert h_3 \vert$, have different distributions of combined $V/\sigma - h_3$: slow rotators do not show the anti-correlation. Note that $\vert V/\sigma \vert$ is different between fast and slow rotators almost by definition as it enters the equations for calculating the specific angular momentum, $\lambda_R$, \citep{2007MNRAS.379..401E}. Therefore, we are primarily interested in the shape of the contours on Fig.~\ref{f:vs_h3} and not their extent along the $x-$axes. In particular, we want to see to what extent fast rotators with cores follow the two trends observed in the first panel, as this would imply to which class they are dynamically similar. 

In other panels of Fig.~\ref{f:vs_h3} we show $V/\sigma - h_3$ diagrams for individual core fast rotators. Most objects show an anti-correlation indicative of disc components. The exceptions are NGC3613, NGC4649 and NGC5485. The last one is a prolate rotator, NGC4649 is one of the most massive \atlas galaxies and on the border with slow rotators, while NCG3613 is also unusual as it is the flattest core galaxy. These three galaxies are indeed kinematically and structurally (no embedded disks) more similar to slow rotators. 

Among other core fast rotators, the anti-correlation trend is the strongest in NGC0524, which also has the largest $\lambda_R$ (actually $\lambda_R \sim0.33$, hence more than any fast rotator used in making of the upper left panel). 

In summary, fast rotators with cores typically have regular kinematics, while a few have some peculiar features. Most show a significant $V/\sigma - h_3$ anti-correlation and exhibit regular velocity maps. This indicates they contain embedded disc-like structures, which makes them morphologically, kinematically and dynamically different from (core) slow rotators, typically harbouring KDCs (or are not rotating at all) and showing no $V/\sigma - h_3$ anti-correlation. The existence of cores in rapidly rotating galaxies was reported before \citep{1997AJ....114.1771F}, and discs were seen in core galaxies \citep{2005AJ....129.2138L}. Recently it was also emphasised by \citet{2013ApJ...768...36D} that there are S0 galaxies (also most likely fast rotators) with cores. This suggests that a core and a disc can be found in the same object although their formation scenarios  (e.g. non-dissipative and dissipative processes, respectively) differ dramatically, if not mutually exclude each other, and as such present a puzzle.

\subsection{Mass dependence}
\label{ss:mass}

\begin{figure}
\includegraphics[width=\columnwidth]{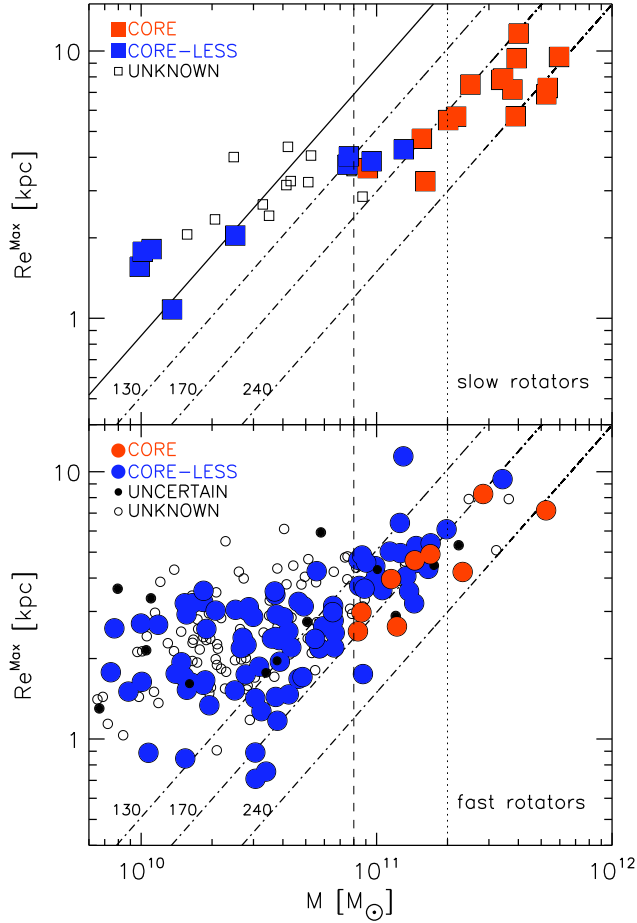}
\caption{Mass -- size relation for \atlas galaxies (Paper XX). The top and bottom panels show slow and fast rotators, respectively. Small open symbols are galaxies for which HST imaging is not available. Colour of symbols indicate the class of the HST nuclear profiles: red -- core, blue -- core-less, black -- uncertain (only on bottom panel).  Vertical lines are drawn at characteristic masses of 0.8 and $2\times10^{11}$ M$_\odot$, and dashed-dotted lines are for constant $\sigma_e=130$, 170 and 240 km/s. }
\label{f:mass}
\end{figure}

We investigate further the properties of \atlas galaxies with HST imaging by plotting them in the mass -- size diagram in Fig.~\ref{f:mass}. Both dynamical masses and sizes were previously reported in Paper XV (and we already used masses for Fig.~\ref{f:gamma}), while the \atlas mass--size plot was presented previously in Paper XX as a non edge-on projection of the Mass plane (Paper XV), which is, in essence, the fundamental plane where luminosity is substituted with mass.  For a general discussion on the mass -- size relation and its demographics with respect to Hubble types we refer the reader to Paper XX and its figs. 9 and 14.

%%%%%% Figure 6%%%%%%%%%%%%%%%%%%%%%%%%%%%%%%%%%%%%%%%%%%%%%%%
%%%%%%%%%%%%%%%%%%%%%%%%%%%%%%%%%%%%%%%%%%%%%%%%%%%%%%%%%%

Slow and fast rotators, coloured by the type of nuclear profiles, are separated into the top and bottom panels, respectively. As expected \citep{1997AJ....114.1771F}, core galaxies are found in massive and large galaxies, populating the narrow tail in the mass -- size diagram, to which early-types extend from the general population of galaxies.

This dependence on high mass for the existence of cores is, particularly, the property of slow rotators, while there is, at present, some room for uncertainty for fast rotators. Namely, out of eight fast rotators more massive than $2\times10^{11}$ M$_\odot$, three do not have HST data, one has strong dust features which prevent its classification (NGC3607), and one is classified as an intermediate case (NGC2768). The most massive fast rotator (cored) is NGC4649 which sits very close to the fast--slow separatrix in the $\lambda_R - \epsilon$ diagram. The other two core fast rotators are (in order of decreasing mass): NGC4382 and NGC0524. The three galaxies with no HST data have $\lambda_R > 0.4$, and are also flatter compared to all other galaxies in this mass range, which would imply, according to the trends in diagrams of Fig.~\ref{f:nuk}, they are most likely core-less galaxies. This suggests that the occurrence of cores in massive galaxies is strictly true only for slow rotators. Massive core-less fast rotators can exist. 

Continuing down in mass, in the regime between $8\times10^{10}$ and $2\times10^{11}$ M$_\odot$ we find a mixture of nuclear profiles, both for slow and fast rotators. For fast rotators this could just be the continuation of the trend seen at the high mass, but for slow rotators this is the region where both core and core-less profiles occur. For masses lower than $8\times10^{10}$ M$_\odot$ only core-less galaxies seem to exist. The lack of HST data for slow rotators in the same region prohibits a strong statement; the four observed slow rotators, mostly of the lower masses, do not have cores, but they also mostly belong to the category of the flattest slow rotators and counter-rotating discs. 

A property which is found in both slow and fast rotators with cores (above M $=  8\times10^{10}$ M$_\odot$) is the alignment of the host galaxies on the lines of constant velocity dispersion. As Paper XX demonstrated \citep[see also][]{2012ApJ...751L..44W}, various properties of early-type galaxies remain constant along lines of constant $\sigma$.  This was shown to be due to the fact that $\sigma$ traces the bugle fraction, as a large bulge is needed to quench star formation. Therefore, a contrast in the appearance between the narrow tail at high masses (and large sizes) and the region with the bulk of the galaxy population, is related to the difference in the evolutionary processes. Paper XX suggested that the distribution of galaxies on this diagram is due to two main processes: (i) bulge growth, which increases the galaxy concentrations and $\sigma$ while decreasing $R_e$ and increasing the likelihood for the star formation to be quenched and for the galaxy to appear as an early-type galaxy; (ii) dissipationless mergers, which move galaxies along lines of nearly constant $\sigma$, while increasing $R_e$ and mass.

The dominance of core early-type galaxies above M $=  2\times10^{11}$ M$_\odot$ and the lack of low-mass core slow rotators, seems consistent with this picture in which fast rotators need a sufficient number of dissipationless mergers to scour their cores and also transform into slow rotators. It indicates that not all slow rotators are the same and it emphasises the physical importance of the characteristic mass  M$ =  2\times10^{11}$ M$_\odot$ (as in fig.14 of Paper XX).

%%%%%% Figure 7%%%%%%%%%%%%%%%%%%%%%%%%%%%%%%%%%%%%%%%%%%%%%%%
\begin{figure}
\includegraphics[width=\columnwidth]{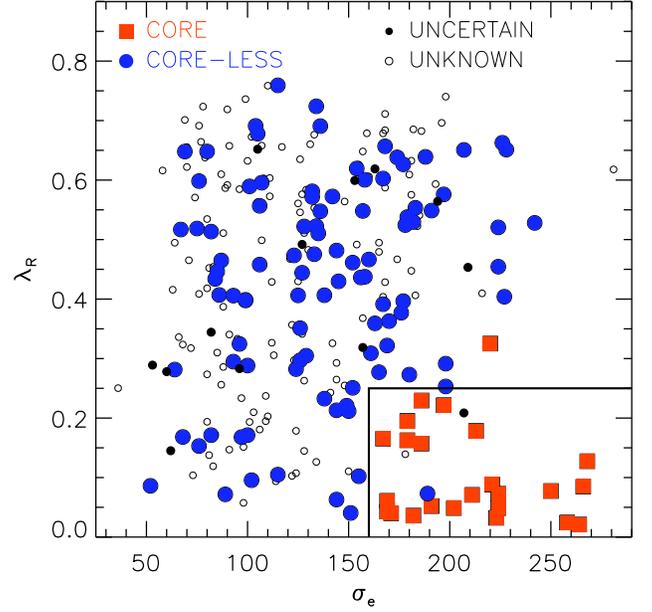}
\caption{A correlation of $\lambda_R$ with the velocity dispersion measured within the effective radius, $\sigma_e$. Galaxies with cores are shown with red squares, core-less galaxies with blue circles, galaxies with an uncertain nuclear profiles with small black circles, while galaxies with no HST data with small open circles. The lines delineate the region where mostly cores occur.  The core galaxy outside the box and the core-less galaxy within the box are NGC0524 and NGC3414, respectively. The uncertain galaxy within the box is NGC3607.}
\label{f:sigma}
\end{figure}
%%%%%%%%%%%%%%%%%%%%%%%%%%%%%%%%%%%%%%%%%%%%%%%%%%%%%%%%%%

Furthermore, there is a notable alignment of core fast rotators with lines of constant $\sigma$, similar to that seen for massive (and core) slow rotators. Semi-analytic models \citep[][ hereafter Paper VIII]{2011MNRAS.417..845K} suggest that there are two types of fast rotators, with the main difference in the availability of the gas for star formation. It is tempting to interpret the existence of cores in fast rotators above certain mass, as well as their alignment with constant $\sigma$ lines as an indication of this separation between the two sub-classes, where the core galaxies represent clear cases for no additional star formation which would refill the cores. This is also interesting if one takes into account that the distribution of fast rotators in the mass -- size plot forms a smooth parallel sequence and lower boundary to the distribution of spirals (Paper XX). 

As it can be seen from the mass -- size diagram there are no core galaxies for M $<8\times10^{10}$ M$_\odot$, but there also seems to be a well defined lower limit in the global velocity dispersion. In Fig.~\ref{f:sigma} we correlate $\lambda_R$ with the observed velocity dispersion within an effective radius, $\sigma_e$ (from Paper XV). Indeed, core galaxies are clustered in the lower right corner, of high velocity dispersion and low angular momentum values. Specifically, cores dominate for $\lambda_R \lesssim 0.25$ and $\sigma_e \gtrsim160$ km/s.  Fast rotators with cores are found in the upper part of the boxed region with $\lambda_R > 0.15$ (and $\sigma_e < 220$ km/s). Taking into account that top three core galaxies in the boxed region are (in order of decreasing $\lambda_R$): NGC4473, NGC3613 and NGC3193, of which both NGC4473 and NGC3913 are marginal cases in terms of their core classification (see Appendix~\ref{a:comp}), a conservative upper limit for separating cores from core-less galaxies is around $\lambda_R \lesssim 0.2$. 

The relatively clean separation of core and core-less galaxies in Fig.~\ref{f:sigma} reflects the main finding of this paper: cores are typically found in galaxies with low specific angular momentum and high mass (high $\sigma$). Most of the information in Fig.~\ref{f:sigma} is already visible from Fig.~\ref{f:mass}, but we highlight it here as particularly interesting as it has a predictive power to separate core from core-less galaxies based on two direct observables of any survey with integral-field spectrographs: $\lambda_R$ and $\sigma_e$.

\subsection{Correlation with other parameters}
\label{ss:razno}

We now investigate potential correlations between the presence of cores in fast and slow rotators with other properties such as stellar populations, presence of the atomic and molecular gas,  X-rays and the influence of the environment. 

\subsubsection{Stellar populations}
\label{sss:ssp}

We investigated the residuals from the best-fitting linear relations between age, metallicity and abundances and the velocity dispersion (McDermid et al. in prep) with respect to the presence of cores, and found no significant trends. Cored and core-less galaxies in the same $\sigma_e$ (or mass) range have consistent distributions in age, metallicity and abundances. The same is true if one only selects core fast and core slow rotators, or core-less fast and slow rotators. We looked for the differences in single stellar population parameters measured within apertures of one effective radius and one eight of the effective radius. This test gave consistent results. 

Stellar populations indicate that the cores and core-less nuclei in massive galaxies, both in fast and slow rotators, are typically made of old stars. This indicates that  cores were either created early ($z> 1.5$, with a few exceptions) and survived until present, or they were scoured in dissipationless mergers which did not involve creation of new stars.

\subsubsection{Molecular and atomic gas}
\label{sss:co}

A comparison with \citet[][ hereafter Paper IV]{2011MNRAS.414..940Y} shows that carbon monoxide (CO) is detected in only one core galaxy (NGC0524). This galaxy has spiral dust structure and it also has the highest $\lambda_R$ among core objects. The lack of molecular gas in core galaxies is not surprising, as this gas is typically associated with star-formation, which would fill in the cores. In the \atlas sample it is only detected in fast rotators, although not all of them experienced a strong star forming period recently (McDermid et al. in prep). Notably, all galaxies for which we were not able to extract reliable profiles and classify their nuclear structures ({\it uncertain}), as they were very dusty, are also strong CO detections. Of particular interest here is NGC3607, which is one of the most massive fast rotators and it is found close to the region in $\lambda_R - \epsilon$ diagram populated by other core fast rotators. It is also found in the regions where mostly core galaxies occur in both Fig.~\ref{f:mass} and Fig.~\ref{f:sigma}. There are a number of similarities between NGC0524 and NGC3607, such as the existence of prominent dust spiral structure, CO detection \citep{2003ApJ...584..260W}, the galaxy mass and similar position in the above mentioned diagrams. NGC3607 has some what higher inferred mass of $H_2$ \citep[Paper IV,][]{2003ApJ...584..260W}, and this might have made a difference for the shape of their nuclear profile. While NGC0524 has a core, the dust content of NGC3607, unfortunately, impedes this analysis at present, but we note that \citet{2005AJ....129.2138L} detected a core in this galaxy.

HI is found in similar quantities in both slow and fast rotators \citep[][ hereafter Paper XIII]{2006MNRAS.371..157M,2010MNRAS.409..500O,2012MNRAS.422.1835S}, but it is not often found among core galaxies. A comparison with Table B1 of Paper XIII shows there are three core slow rotator with HI detections (NGC4406, NGC5198 and NGC5557) all of which are in unsettled configurations. Additionally, NGC3608 has some HI clouds in the vicinity. Noteworthy is also that a core-less slow rotator and three further slow rotators with no HST imaging (see Section~\ref{ss:nohstSR}) have ordered HI structures (discs or rings). Core fast rotators show a similar fraction of HI detections: two galaxies, NGC3193 and NGC4278, as unsettled clouds and a large scale settled HI disc, respectively.

\subsubsection{X -- rays}
\label{sss:xrays}

%%%%%% Figure 8%%%%%%%%%%%%%%%%%%%%%%%%%%%%%%%%%%%%%%%%%%%%%%%
\begin{figure}
\includegraphics[width=0.9\columnwidth]{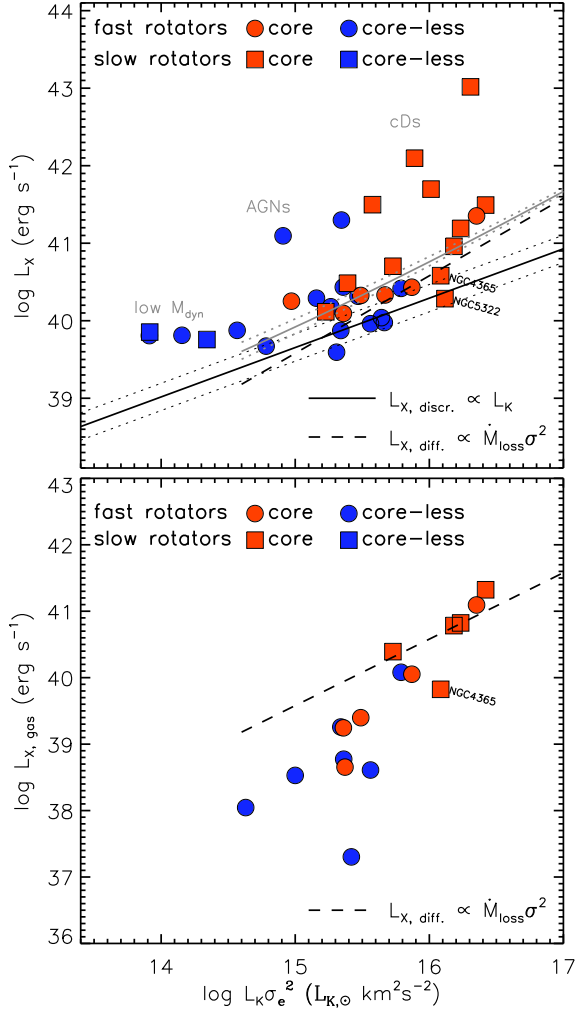}
\caption{ {\it Top:} $L_K\sigma_e^2$ vs. $L_X$ diagrams for low ({\it top panel}) and high ({\it bottom panel}) X-ray resolution \atlas sub-samples galaxies with HST imaging (based on figs.~3 and~6 in Paper XIX, respectively). On both panels, core and core-less galaxies are represented by red and blue colours, respectively, while symbols corresponds to fast and slow rotators. On both panels the dashed line shows the contribution to the X-ray luminosity from hot-gas emission sustained by the thermalisation of the kinetic energy that stellar-mass loss material inherit from their parent stars, which follow a simple $L_{X,diff}=3/2\dot{M}\sigma^2$ law. On the top panel, the solid line shows also the contribution of unresolved X-ray binaries \citep[$L_{X,discr}$, as in][ dotted lines shows uncertainties of such a model]{2011ApJ...729...12B}, while the grey solid line traces the sum of the both diffuse and discrete components that ought to be compared with the low-resolution X-ray measurements. The grey labels indicate low-mass galaxies or objects with the X-ray measurements significantly contaminated by an AGN or the X-ray emission from the cluster medium.}
\label{f:xray}
\end{figure}
%%%%%%%%%%%%%%%%%%%%%%%%%%%%%%%%%%%%%%%%%%%%%%%%%%%%%%%%%%

Within the earlier context of a division of early-type galaxies into massive boxy galaxies with nuclear cores and less massive discy objects with nuclear cusps (core-less), a number of papers have also looked into the hot-gas content of early-type galaxies \citep[e.g.][]{1989A&A...217...35B,2005MNRAS.364..169P,2009ApJS..182..216K} with \citet{2005MNRAS.364..169P}, in particular, being the first to recognise how galaxies with central cores tend to display higher X-ray luminosity values, L$_X$, than core-less galaxies.

In \citet[][ hereafter Paper XIX]{2013arXiv1301.2589S} we have looked into the hot-gas content of early-type galaxies in the \atlas sample by using L$_X$ measurements from X-ray observations of both low and high angular resolution, as measured with {\it ROSAT} or {\it Einstein} \citep{2001MNRAS.328..461O} and {\it Chandra} \citep{2011ApJ...729...12B}, respectively. Based on these X-ray data and our integral-field measurements we found that slow-rotators display L$_X$ values that are consistent with hot-gas emission that is sustained by the thermalisation of the kinetic energy carried by stellar mass-loss material, whereas fast rotators generally fall short from such a simple prediction.

Considering that fast rotators are intrinsically flatter than slow-rotators \citep[see fig. 2 of Paper VII and][ hereafter Paper XXIV]{Weijmans2013}, in Paper XIX we concluded that the intrinsic shape of an early-type galaxy is the most likely driver for the X-ray luminosity of its hot-gas halo, consistent with the suggestion of \citet{1996MNRAS.279..240C}, whereby flatter systems are less capable of holding on to their hot gas.

Following the work of Paper XIX and the earlier suggestions of a connection between nuclear properties and X-ray luminosity, here we have also looked into the hot-gas content of the core and core-less galaxies in the \atlas sample. The overlap with the HST subset of our sample with the subsets with either low or high X-ray angular resolution from Paper XIX is small, but it enables us to recognise some general trends.

Including the information regarding the presence of cores into plots similar to those presented in Paper XIX (Fig.~\ref{f:xray}) shows that core-less galaxies are indeed X-ray deficient, and that cores are found in the most X-ray luminous galaxies. Yet, the presence of a core does not imply high X-ray luminosities, as in the case of the relatively flat slow-rotators NGC4365 and NGC5322 and in fast-rotators such as NGC3379 or NGC4278 which may all be X-ray deficient by virtue of their intrinsic flat shape.

The presence of a core in the X-ray luminous galaxies of the \atlas sample is consistent with the point made by \citet{2009ApJS..182..216K}, where the presence of halo of hot gas, or even more the fact of being deeply embedded in the hot medium of a galaxy cluster, would prevent the accretion of cold gas and the reforming of a central stellar cusp. In fact, as noticed in Paper XIX, the presence of a hot medium would also prevent the cooling of stellar-mass loss material and its recycling into new stars. Conversely, the finding of cores in X-ray deficient galaxies does not necessarily pose a problem since the accretion of fresh gas would depend also on environmental effects and would thus not be guaranteed. In contrast, the processes leading to the formation of a core do not have to significantly alter the overall shape of a galaxy, for instance making it rounder and more capable to retain a halo of hot gas.

\subsubsection{Environment dependence}
\label{sss:envir}

We also investigated the influence of galaxy environment on the type of nuclear profiles. Using the density estimators of Paper VII, which probe cluster and group environments, we did not see clear correlations, which are not related to the fact that slow rotators are found in large numbers only in the densest environments. A possible exception are less massive slow rotators, for which HST imaging is not available and we do not know their nuclear light profiles. These galaxies are typically found in low density environments. In Section~\ref{ss:nohstSR} we discuss if these objects contain cores or are a core-less subpopulation of slow rotators.

\subsection{A caveat: are there more cores?}
\label{sss:more}

As the definition of core/core-less galaxies depends on the choice of radius at which $\gamma^\prime$ is evaluated it is likely that we do not recognise smaller cores in more distant galaxies. The sizes of our cores, using the ``cusp radius"  $r_\gamma$ as a measure, range from tens to hundreds of parsecs, and for some galaxies we probe the light profiles down to a few parsec. Still, as mentioned earlier, a galaxy with a ``Nuker" power-law at the HST resolution could still harbour a core at smaller scales. Therefore we wish to compare how our conservative radius limit for estimating $\gamma^\prime$ biases our ability to detect cores. 

\begin{figure}
\includegraphics[width=\columnwidth]{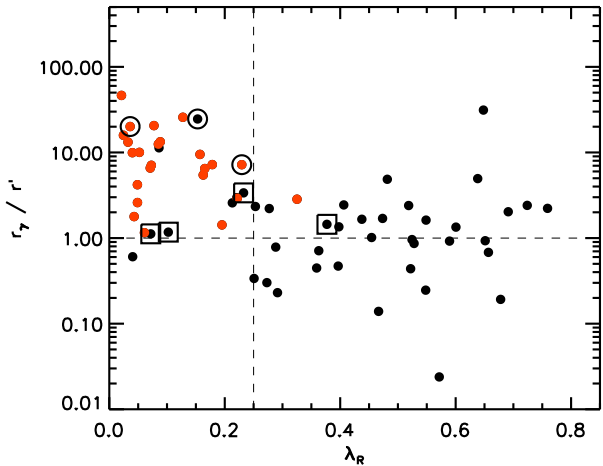}
\caption{Ratio of the ``cusp radius", $r_\gamma$, and $r^\prime = 0.1$\arcsec, the radius at which $\gamma^\prime$ was evaluated (see eq.~\ref{e:gamma}) as a function of the global angular momentum, $\lambda_R$. The dashed horizontal line highlights the ratio of 1, and the vertical line is an indication of the angular momentum below which cores occur in our sample (with the exception of NGC0524). Large open squares and open circles show galaxies for which our classification differs from those of the ``Nuker" fits and core-S\'ersic/S\'ersic parametrisation, respectively, as found in the literature (see Appendix~\ref{a:comp} for details).}
\label{f:cusp}
\end{figure}

In Fig.~\ref{f:cusp} we show the ratio of $r_\gamma$ and the adopted radius for $\gamma^\prime$\footnote{Note that for galaxies for which we use \citet{2005AJ....129.2138L} values, $r^\prime$ is not the radius at which $\gamma^\prime$ was estimated, but 0\farcs02 or 0\farcs04 were used instead. See that paper for details.}, $r^\prime=0.1$\arcsec\, as a function of the global angular momentum, $\lambda_R$, for all galaxies that have profiles less steep than 0.5 at 0.5\arcsec. When this ratio is below 1, our probe of the nuclear region is larger than a possible core, given the ``Nuker" fit, hence, we cannot detect a core. As the figure shows, this is indeed the case (no cores for  $r_\gamma/r^\prime<1$). In Appendix~\ref{a:comp} we discuss galaxies for which our classification differs from the literature and in Fig.~\ref{f:cusp} we highlight these galaxies as open squares (difference with ``Nuker" fits) and open circles (difference with core-S\'ersic fits). It is evident that at least some galaxies which were previously classified as core using the ``Nuker" fit have the ratio close to 1, and, therefore, the scale at which they are characterised is important. We postpone the further discussion about these objects to Appendix~\ref{a:comp}. 

%%%%%% Figure 9%%%%%%%%%%%%%%%%%%%%%%%%%%%%%%%%%%%%%%%%%%%%%%%
%%%%%%%%%%%%%%%%%%%%%%%%%%%%%%%%%%%%%%%%%%%%%%%%%%%%%%%%%%

The cores we detect typically occur for $\lambda_R \lesssim 0.25$ (except NGC0524). It is noteworthy that below this value for $\lambda_R$ there is only one galaxy with $r_\gamma/r^\prime<1$: formally a slow rotator and face-on candidate NGC6703, which we classify as core-less. Other galaxies with $r_\gamma/r^\prime<1$ occur for $\lambda_R >0.25$, the first three being NGC0821, NGC4434 and NGC4621. According to their respective $\sigma$, $\lambda_R$ and their position in Figs.~\ref{f:mass} and~\ref{f:sigma}, NGC4434 is unlikely to have a core (too small $\sigma$), while NGC0821 and NGC4621 are close to the space occupied by cores in these figures. For NGC0821 we use the data from \citet{2005AJ....129.2138L} and their derivation of $\gamma^\prime$ at radius 0\farcs04, and this galaxy remains core-less. For NGC4434 we used an ACS image (and 0\farcs1 radius for $\gamma^\prime$), but analysis in \citet{1996AJ....111.1889B} and \citet{2007ApJ...664..226L} suggests that our conservative resolution does not change the classification of this galaxy. Furthermore, as galaxies with higher $\lambda_R$ values are typically fainter and less massive, they are also less likely to have cores \citep[e.g.][ and results in previous sections]{1997AJ....114.1771F}. Therefore, we conclude that our estimates of core/core-less nuclear profiles does not suffer greatly due to our conservative approach in estimating $\gamma^\prime$, and that this is not the prime reason why there are no cores for  $\lambda_R >0.25$.

%%%%%%%%%%%%%%%%%%%%%%%%%%%%%%%%%%%%%%%%%%%%%%%%%%%%%%%%%%%
%
% SECTION 5 SECTION 5 SECTION 5 SECTION 5 SECTION 5 SECTION 5
%
%%%%%%%%%%%%%%%%%%%%%%%%%%%%%%%%%%%%%%%%%%%%%%%%%%%%%%%%%%%

\section{Discussion}
\label{s:discusa}

\subsection{The influence of projection effects}
\label{ss:projection}

A simple question difficult to answer is: are projections effects responsible for the observed mismatch between structural and kinematic properties? \citet{2012ApJ...759...64L} suggested that core-less (power-law) galaxies that fall among galaxies with cores in the $\lambda_R - \epsilon$ diagram (basically all power-laws with $\lambda_R<0.25$) are indeed there due to projection effects. \citet{2007MNRAS.379..401E} and \citet{2007MNRAS.379..418C} argued that fast and slow rotators are two separate galaxy populations and that inclination effects can not move galaxies from one to the other group, except in a not very common case of {\it perfectly} face-on discs, where no radial velocity gradients could be detected. This is supported by the over-plotted lines in the right panel of Fig~\ref{f:nuk} (see also Section~\ref{ss:lamR}), which encompass fast rotors and suggest that the majority of these objects can be explained as a single population of objects, with specific dynamics, seen at different inclinations. Moreover, the contours in the right panel of Fig~\ref{f:nuk}, indicate the distribution of a family of objects with an intrinsic shape, $\epsilon = 0.7 \pm 0.2$ (Gaussian distribution). These contours avoid the region of slow rotators and suggest that these two classes of galaxies have different intrinsic shapes (as explicitly shown in Paper XXIV), structural and dynamical properties. 

\citet{2009MNRAS.397.1202J} showed that the intrinsic shape of galaxies has a limited influence on the actual classification of galaxies as fast or slow rotators. Investigating the $\lambda_R$ parameter of simulated galaxies they showed that intrinsically oblate objects are almost always classified as fast rotators independently of projection, which is in agreement with the results of \citet{2007MNRAS.379..418C}. Similarly, intrinsically triaxial objects are always classified as slow rotators. Only prolate objects can change from fast to slow rotators, depending on the viewing angles. \citet{2009MNRAS.397.1202J} argue that this is a consequence of their specific orbital structure. Based on the investigation of kinematic misalignment angle (Paper II) ,there are only two prolate galaxies in the \atlas sample of 260 galaxies. Indeed, one is classified as a fast and the other as a slow rotator, and both of them harbour cores. 

The two plots in Fig.~\ref{f:nuk} and the findings of \citet{2009MNRAS.397.1202J} suggest that projection effects can not explain the existence of slow rotators with core-less profiles or fast rotators with cores. The evidence for embedded discs (Fig.~\ref{f:vs_h3}, Section~\ref{ss:vs_h3}) in cored fast rotators also challenges the grouping of all cored galaxies under the same class (both morphologically and dynamically). Raising the fast-slow separatrix at low ellipticities to a value of $\lambda_R\sim 0.2$ (or similar) as suggested by \citet{2012ApJ...759...64L}, and similarly by \citet{2012ApJS..198....2K}, in order to include all core galaxies among slow rotators, is, however, not advisable as in that case one would include a large number of fast rotators which are indeed at low inclinations\footnote{This was noted by \citet{2012ApJ...759...64L}.}. Misclassification due to projection is expected only in a few rare objects: face on discs and prolate rotators. The expected frequency of these objects  (one to two cases of both types in the \atlas sample, Papers II and III) does not explain the observed number of, specifically, core fast rotators. The answer to the question at the beginning of this section is: inclination effects do not (typically) change a fast into a slow rotator. The mismatch between global angular momentum and nuclear profiles, more likely, indicates variations in assembly history within the class.

\subsection{Two types of fast and slow rotators}
\label{ss:diff}

Evidence presented in papers of the SAURON survey and of the \atlas project show that fast and slow rotators are two separate populations of galaxies. \citet{2007MNRAS.379..401E} and \cite{2007MNRAS.379..418C} showed that the distinction between fast and slow rotators is not sensitive to the projection effects. Paper III put the separation of fast and slow rotators on a more statistical basis in the nearby Universe. In Paper II we showed that all fast rotators are nearly axisymmetric (modulo bars), while slow rotators are not. \citet{2008MNRAS.390...93K} suggested that fast rotators contain discs, while in Paper XVII we showed that fast rotators contain exponential components in their large-scale light profiles (or are best fitted with a S\'ersic model of low $n$). Additionally, the different intrinsic shapes of fast and slow rotators were shown in Paper XX (via dynamical models) and Paper XXIV (via statistical deprojection). Fast and slow rotators also differ in the presence of molecular gas \citep[Paper IV,][]{2012arXiv1210.5524A}, while this is not the case  in terms of ionised \citep{2006MNRAS.366.1151S} and atomic gas at large scales \citep[][ Paper XIII]{2006MNRAS.371..157M,2010MNRAS.409..500O}. Fast and slow rotators, however, differ in their X-ray emission originating in the hot gas component (Paper XIX). 

Nevertheless, these pieces of evidence do not exclude that there could be sub-populations among fast and slow rotators as result of somewhat different formation paths, with a continuous range of parameters. The existence or the lack of cores can be used as an indication for the differences between the sub-populations among fast and slow rotators, respectively. In particular, one should consider the evolution of galaxies within the two phase formation scenario \citep{2010ApJ...725.2312O}, where the early phase is dominated by gas inflows and formation of stars within galaxies \citep[e.g][]{2005MNRAS.363....2K, 2009ApJ...703..785D}, and the further evolution is seen in the second phase of assembly of starts created in other galaxies dominated by frequent, and often non-dissipative, merging \citep[e.g.][]{2012ApJ...754..115J, 2012MNRAS.425..641L,2012ApJ...744...63O}. 

The large range of angular momentum values, coupled with the full range of ellipticities, indicates that properties of fast rotators could be explained as a combination of projection effects and different formation processes. This is supported by the contours in the right panel of Fig.~\ref{f:nuk}. The relatively regular and elsewhere ellipsoidal shape of the second contour is twisted such that it does not include values around (0.2, 0.15) in ($\lambda_R, \epsilon$). Galaxies in that region could come from a population with a different intrinsic shape or have different formation histories from the majority of fast rotators. The latter can be the case as this is exactly the location of a number of fast rotators exhibiting cores. For example, re-mergers of major disc mergers of Paper VI (see their fig.~11) fall in this region, suggesting that indeed galaxies which suffered more significant major merging (either dry and wet) could populate it. Additionally, the semi-analytic models of Paper VIII predict that the class of fast rotators comprises two sub-populations with different histories in terms of availability of cold gas. These two different classes are not easy to recognise, neither using morphology (e.g. disc-bulge decomposition of Paper XVII) due to inclination effects, nor by considering kinematics (Paper II), which might not be sensitive enough to the subtle differences in the star formations histories. The existence of cores in fast rotators, however, could point to, at least, a subset of formation scenarios still consistent with producing a fast rotator. 

As outlined in Papers III and VII, there is a clear case for different types among slow rotators. Slow rotators with $\epsilon \gtrsim 0.35$ are often made of counter-rotating discs and classified as S0s, while rounder galaxies are characterised by no net rotation or harbour large KDCs (Paper II), which are not necessarily separate components \citep{2008MNRAS.385..647V}, but could originate in complex orbital structure, and projections of triaxial bodies \citep{1991AJ....102..882S}. Here we also show that flat slow rotators typically have core-less nuclear profiles. This supports the notion that their assembly histories are different from those of more round slow rotators with cores, with implication to their ability to retain hot gas and inhibit further star formation. The lack of cores in some slow rotators points to those extreme formation scenarios consistent with producing a slow rotator, but incapable of producing or maintaing the core.

\subsection{Are there more core-less slow rotators?}
\label{ss:nohstSR}

Open symbols plotted in Figs.~\ref{f:nuk}, ~\ref{f:mass} and~\ref{f:sigma} represent galaxies for which HST data suitable for this analysis is unavailable. They are present over the full extent of the $\lambda_R - \epsilon$ diagram, but as highlighted in Section~\ref{ss:lamR}, there is a specific region where they seem to dominate: for $0.1 \lesssim \lambda_R \lesssim  0.2$ and $0.2 \lesssim \epsilon \lesssim  0.35$. This region is of particular interest as it is a transitional region between slow and fast rotators. It also comprises a population of flatter slow rotators of similar angular momentum as the majority of core fast rotators. They are particularly interesting as they are less massive than other slow rotators ($10^{10.5} - 10^{11}$ M$_\odot$), typically have $\sigma_e<160$ km/s, and a number of them are morphologically classified as S0. They are bounded by core fast rotators on the left and core slow rotators below. 

%%%%%% Figure 10%%%%%%%%%%%%%%%%%%%%%%%%%%%%%%%%%%%%%%%%%%%%%%%
\begin{figure}
\includegraphics[width=\columnwidth]{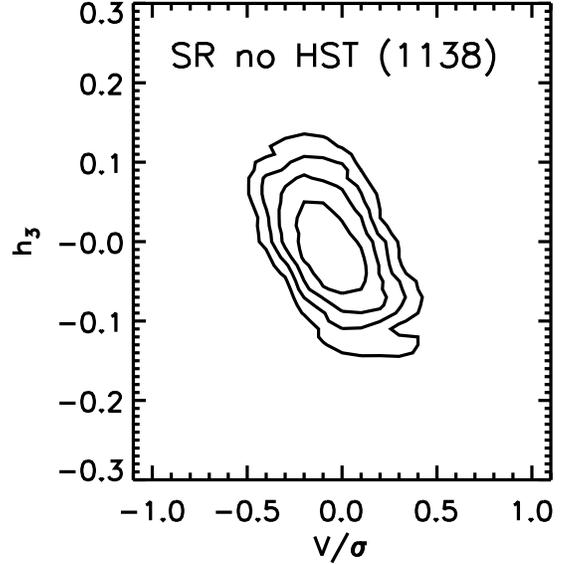}
\caption{ Local $V/\sigma - h_3$ diagram for all slow rotators with no HST imaging. The same selection criteria as in Fig.~\ref{f:vs_h3} were used to select the spatial bins suitable for plotting, but typically low $\sigma$ ($<$120 km/s) values remove a significant number of bins.}
\label{f:nohst}
\end{figure}
%%%%%%%%%%%%%%%%%%%%%%%%%%%%%%%%%%%%%%%%%%%%%%%%%%%%%%%%%%

Following trends on Fig.~\ref{f:nuk}, one could expect these galaxies to have flat nuclear profiles. According to Fig.~\ref{f:nohst} their $V/\sigma$ and $h_3$ are weakly anti-correlated; not as much as for a typical fast rotator, but significantly more than for other slow rotators. Their global structures suggest they contain exponential components or are well fitted with a single S\'ersic function of low $n$ (Paper XVII). These trends present them as different from typical slow rotators, but more similar to fast rotators, and perhaps even to core fast rotators (at least in the sense that having a core and an embedded disc seems to be possible). However, according to the trends in the mass -- size diagram (Fig.~\ref{f:mass} and discussion in Section~\ref{ss:mass}, low mass, low $sigma$), and the fact they are found in low density environments, these galaxies are most likely core-less, and hence very special cases for understanding the formation of slow rotators.  

If they are indeed core-less galaxies, then the conjecture of \citet{2012ApJ...759...64L} that below $\lambda_R<0.25$ only slow rotators and face-on fast rotators exist cannot be true; these galaxies are too flat to be considered face on discs. There is no doubt that these galaxies are different from other fast rotators, including those with similar $\lambda_R$ values. Their velocity maps are disturbed, although not as irregular as of other slow rotators. Obtaining high resolution imaging of these galaxies would settle the issue, and robustly calibrate the separation between core and core-less galaxies in the $\lambda_R - \sigma$ diagram.

\subsection{Are cores in fast and slow rotators different?}
\label{ss:mbh}

We showed above that core fast rotators cover a similar range in masses as core slow rotators. On the other hand, the significant kinematic difference between fast and slow rotators suggests that properties of cores could also be different in these two classes of galaxies. \citet{2012ApJ...759...64L} compared the sizes of cores in fast and slow rotators using the ``cusp radius", $r_\gamma$, and found that there is no difference between core sizes in fast and slow rotators. 

Here we look at the relation between the mass of the central black hole, M$_{BH}$, and a global property of the host galaxy, namely, its global velocity dispersion: the M$_{BH} - \sigma$ relation \citep{2000ApJ...539L...9F,2000ApJ...539L..13G}. We use the recent compilation of black hole masses by \citet{2013ApJ...764..151G}. The sample consists of 72 measurements, but the overlap with the \atlas sample is rather small: only 32 galaxies. As before, we divide galaxies according to their angular momentum and nuclear structure. We do not attempt to assign other galaxies with measured M$_{BH}$ into fast and slow rotator classes as they do not have the necessary integral-field data. 

%%%%%% Figure 11%%%%%%%%%%%%%%%%%%%%%%%%%%%%%%%%%%%%%%%%%%%%%%%
\begin{figure}
\includegraphics[width=\columnwidth]{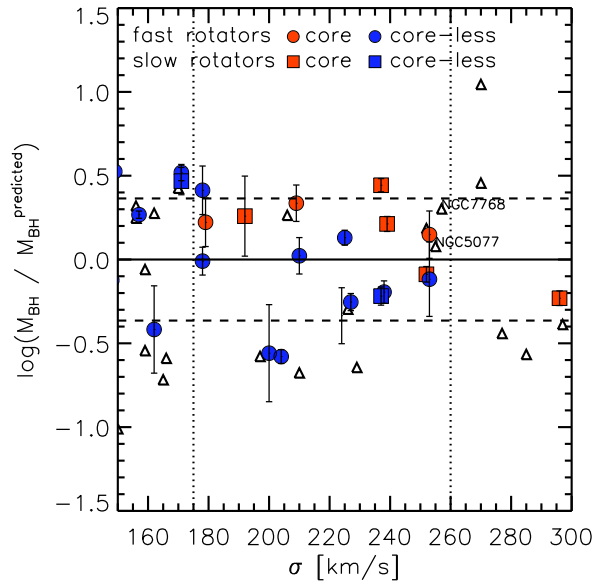}
\caption{Residuals from the best fitting M$_{BH} - \sigma$ scaling relation of \citet{2013ApJ...764..151G} for all galaxies in their sample. Horizontal dashed lines show the scatter of the M$_{BH} - \sigma$. Core fast rotators are found only (see footnote~\ref{fn:ngc4649}) in the limited $\sigma$ range bounded by the vertical dotted lines and used for statistical analysis (see the text). Open triangles represent galaxies not in the \atlas sample. Solid symbols represent galaxies in the \atlas sample. Core-less galaxies are plotted with blue symbols and core galaxies with red symbols, while fast and slow rotators are shown with circles and squares, respectively. Therefore, core fast rotators are shown as red circles, core-less fast rotators as blue circles, while blue squares represent core-less slow rotators and red squares represent core slow rotators. Of four open triangles above the best fitting relation, two objects have cores (their names are shown), while two do not. }
\label{f:mbh}
\end{figure}
%%%%%%%%%%%%%%%%%%%%%%%%%%%%%%%%%%%%%%%%%%%%%%%%%%%%%%%%%%

Recent compilations of data show a possible trend that at the high $\sigma$ end of the relations, mostly populated by core galaxies, M$_{BH}$ increases faster than $\sigma$ \citep[e.g.][]{2011Natur.480..215M}, as well as a possible different best fitting relation for core and core-less galaxies \citep[especially when plotted against spheroid mass or luminosity,][]{2013ApJ...764..184M, 2013ApJ...764..151G,2013arXiv1303.5490S}. This could be a consequence of selection effects \citep{2007ApJ...660..267B} or an indication of a different growth process for both black holes and host galaxies \citep[e.g. Paper XX; ][]{2007ApJ...662..808L, 2013ApJ...764..151G,2013arXiv1303.5490S}. 

In Fig.~\ref{f:mbh} we show residuals obtained by subtracting the best fit relation of \citet{2013ApJ...764..151G} from the measured values. We focus here on the regime within the range $170 < \sigma < 260$ km/s, where the core fast rotators occur. There are M$_{BH}$ estimates for three\footnote{\label{fn:ngc4649} The only other core fast rotator of the \atlas sample with a M$_{BH}$ is NGC4649 harbouring one of the most massive known black holes and with $\sigma\sim340$ km/s. We discussed this galaxy in Section~\ref{sss:coreFR}.} such galaxies (in order of increasing $\sigma$): NGC4473, NGC3379\footnote{Even when using the lower (axisymmetric) M$_{BH}$ estimate for NGC3379 of \citet{2006MNRAS.370..559S} instead of the more massive triaxial estimate of \citet{2010MNRAS.401.1770V}, this galaxy remains above the mean relations.}, and NGC524. All have M$_{BH}$ more massive than the best-fitting relations. This is irrespective of which relation is used (we investigated also those from \citealt{2005SSRv..116..523F}  and \citealt{2013ApJ...764..184M}).  Noteworthy is also that there are estimates for two core-less slow rotator (NGC3414 and NGC5576), and their M$_{BH}$ fall below and above\footnote{We remind the reader that NGC5576 is classified as core in \citet{2005AJ....129.2138L}.} the best fitting relations, respectively. 

Within the selected $\sigma$ range, the mean value of residuals for core fast rotators is $0.23 \pm 0.07$, while for core slow rotators is $0.22 \pm 0.02$. In contrast the mean value of residuals for core-less fast rotators is $-0.19 \pm 0.02$. This suggests that M$_{BH}$ of core fast rotators are similar to M$_{BH}$ of slow rotators in this limited $\sigma$ range, at which both types of galaxies occur. Core-less galaxies seem to typically have lower M$_{BH}$. This suggests that black holes in all core galaxies (of the same velocity dispersion or mass) are similar, regardless whether they live in a fast or a slow rotator, implying that core formation proceeded in a similar process.

There are significant systematics associated with determination of M$_{BH}$, such as using triaxial instead of axisymmetric models \citep{2010MNRAS.401.1770V}, or inclusion of a dark matter halo \citep[e.g.][]{2009ApJ...700.1690G,2011ApJ...729...21S}. The systematics influence the determination of M$_{BH}$ by at least a factor of 2. Additionally our statistic is based on a limited number of galaxies. Finally, in the investigated $\sigma$ range there are eight galaxies which we cannot classify as fast or slow rotators, four above and four below the best fitting relations. Only two of these galaxies have cores, NGC5077 \citep{2001AJ....121.2431R} and NGC7768 (\citealt{2012ApJ...756..179M}, but see \citealt{2003AJ....126.2717L}) and these are found above the best fitting relations. The other two galaxies above the best fitting relation, NGC3115 and NGC3585, are core-less \citep{2005AJ....129.2138L}, but including those to the core-less fast rotators would not change significantly the statistics (the mean of the core-less fast rotator residuals moves to $-0.17 \pm 0.02$). Therefore we conclude that there is a tentative result that black holes in core fast rotators are similar in mass to those in core slow rotators, but different from those in core-less fast rotators (of the same galaxy mass or $\sigma$). 

\subsection{Multiple scenarios for formation of cores}
\label{ss:cores}

The currently favoured scenario for formation of cores is based on evidence from numerical models of binary black hole interactions \citep{1991Natur.354..212E, 1996ApJ...465..527M, 1996NewA....1...35Q,1997NewA....2..533Q, 2001ApJ...563...34M, 2002MNRAS.331L..51M, 2007ApJ...671...53M, 2012MNRAS.422.1306K,2012ApJ...744...74G}. The idea \citep{1980Natur.287..307B} is that as black holes spiral down, the binary loses its angular momentum by ejecting stars found in the vicinity. The binary black hole might not fully merge, but the stars are quickly depleted form the nucleus  \citep[e.g.][]{2002MNRAS.331..935Y, 2003ApJ...596..860M, 2004ApJ...602...93M, 2005LRR.....8....8M, 2006ApJ...648..976M}. The deficit of the stellar mass can be estimated by counting the number of mergers and assuming a certain efficiency for depletion, $f$, \citep[e.g.][]{1997AJ....114.1771F}, which will depend on the initial nuclear density, types of stellar orbits and the peculiarities of the binary interaction \citep[e.g.][]{2012ApJ...744...74G,2012ApJ...749..147K}. For a review of dynamics and evolution of  black hole binaries see \citet{2006RPPh...69.2513M}. 

According to this scenario, the major ingredient for removing stars in galactic nuclei is a secondary black hole, which will interact gravitationally with the one that is already present \citep[but see][]{2012MNRAS.422.1306K}. This black hole has to be relatively massive (comparable with the host black hole) in order to sink rapidly to the nucleus. This implies that core formation happens in comparable-mass merger events. Finally, if the merger involves significant quantities of gas, it is likely that the core will be filled in by new stars created in a nuclear starburst associated with the gas-rich merger \citep[e.g.][]{1994ApJ...437L..47M,1996ApJ...471..115B,2004AJ....128.2098R, 2008ApJ...679..156H,2009ApJS..181..135H}. Therefore, the equal mass mergers should also be mostly non-dissipative (but see \citealt{2009ApJS..181..486H}, who show that it is possible to have some star-formation resulting in a younger component of the galaxy, but not sufficiently large to prevent the formation of a core.) 

On the other hand, the existence of massive black holes in core-less galaxies, as evident in the M$_{BH} - \sigma$ relation, suggests that the actual time-scales of the coalescence of black holes (and core formation) could be shorter than that of nuclear star-bursts \citep{1997AJ....114.1771F,2009ApJS..182..216K}. Additionally, the presence of gas could contribute in taking away the energy of the shrinking binary (instead of sling shot stars) and prevent the creation of a core \citep{2001ApJ...563...34M}. Furthermore, a core created early \citep[e.g.][]{2012MNRAS.422.1306K} could be erased by a later (even minor) gas rich accretion or merger events. There are other scenarios for the re-growth of steep core-less profiles. For example, through adiabatic growth of black holes \citep{1999AJ....117..744V} or via an energy exchange between stars moving in the gravitational field of the single black hole \citep{2006ApJ...648..890M}.

The hardening binary scenario for core formation is interesting also as it takes place on the right spatial scales of 1-100 pc, which closely corresponds to the observed sizes of cores \citep{1997AJ....114.1771F, 2006ApJS..164..334F, 2007ApJ...671.1456C}. It is, however, not the only possible scenario for formation of cores \citep[e.g. see also][]{2008ApJ...678..780G}. An alternative scenario is based on a rapid mass-loss  in the nucleus of a stellar system (e.g. globular cluster) which influences its dynamical evolution and, in particular, increase of size \citep{1980ApJ...235..986H}. A side effect of such an adiabatic expansion is that as the stars are redistributed, the light profile changes and becomes flatter in the centre \citep[e.g.][]{2010MNRAS.401.1099H}. 

Supernova feedback and AGNs are invoked as possible initiators of the mass-loss \citep{1996MNRAS.283L..72N, 2002MNRAS.333..299G, 2005MNRAS.356..107R, 2010Natur.463..203G, 2012arXiv1206.4895T}, specifically in the context of creation of cores in dark matter profiles \citep{2012MNRAS.421.3464P}. As stars and dark matter share collision-less dynamics, these processes could be responsible for production of both extended sizes of galaxies at low redshift, as well as stellar cores that are of interest here \citep[e.g.][]{2012arXiv1211.2648M}.

The simulations of \citet{2012arXiv1211.2648M,2012MNRAS.422.3081M}, which investigate the influence of repeatedly occurring AGN in a galaxy equivalent to a brightest cluster galaxy, show that resulting gravitational fluctuations can be very effective in creating a stellar core, in addition to a core in the dark matter density. These simulations, however, can not resolve the nuclei of galaxies (their stellar cores are $\sim10$ kpc in size). An additional significant complication is that, unlike in the case of dark mater, each of the cooling episodes, which follow the cessation of nuclear activity, could result in formation of new stars. Those would not have the memory of the previous perturbations, and thus prohibit the formation of stellar cores. More simulations of higher resolution are, however, needed to give weight to this process of stellar core formation.

\subsection{Using nuclear profiles to differentiate between formation histories}
\label{ss:formation}

Generally, cores are found in slow rotators and core-less galaxies are in fast rotators, adding to the separation of early-type galaxies based on their physical properties as outlined in \citet{1996ApJ...464L.119K}, Paper VII and \citet{2012ApJS..198....2K}. The most dominant processes that shape early-type galaxies are those that create (or maintain) core-less fast rotators and core slow-rotators. As a first approximation, they can be divided into dissipational and dissipationless, respectively, where the general properties of early-type galaxies can be explained by a sequence of the relative mass fractions and the quantity of gas involved. In addition to this global division, the existence of core fast rotators and core-less slow rotators, even if they are rare, points to possible additional scenarios, or specific combinations thereof. 

As seen in simulations of binary mergers \citep[Paper VI; ][]{2010ApJ...723..818H}, even dry mergers can spin up the remnant, and, depending on the orbital configuration of the merger, produce fast rotators. Furthermore, slow rotators can be produced in mergers with gas, where the gas fraction plays a less important role compared with the orbital arrangement of the progenitors \citep[e.g. Paper VI, ][]{2009ApJ...705..920H}. Following the core scouring scenario, all those galaxies suffering from major mergers, that end up either as fast or as slow rotators, should have cores. A crucial distinction is that dissipationless major mergers do not result in an anti-correlated $V/\sigma - h_3$, contrary to dissipative  major mergers  \citep[e.g.][]{2007ApJ...658..710N, 2009ApJ...705..920H}.

Given that fast rotators, including those with cores, show an anti-correlated $V/\sigma - h_3$, explaining the cores in fast rotators requires an additional constraint on the formation process. Cores in fast rotators point to either a gas-rich merger with a nuclear starburst that is shorter than the time needed for the hardening of the black hole binary \citep[][]{1997AJ....114.1771F,2009ApJS..182..216K}, or to scenarios in which the core was either created afterwards (e.g. AGN induced as outlined above), or the galaxy was spun up later in a process that did not destroy the core. The main point here is that there has to be a process that regulates the star formation in the nucleus, such that the core is not (completely) filled, as the spinning up of the main body, such that $V/\sigma$ and $h_3$ get anti-correlated, is likely possible only with acquisition of gas \citep[but for the effect  of multiple minor dissipationless mergers on the $V/\sigma$ only see][]{2005A&A...437...69B}. The current literature suggests two such processes: feedback from the black hole \citep[e.g.][]{2005Natur.433..604D} or morphological quenching \citep{2009ApJ...707..250M}. 

NGC0524 is a possible example of the latter case. The core of this galaxy is relatively small, but the galaxy contains a dusty disc as well as a significant amount of molecular gas in the nucleus \citep[Paper IV,][]{2012arXiv1210.5524A}. It is, however, not forming stars significantly \citep{2011MNRAS.410.1197C}, while \citet{2012arXiv1212.2288M} argue that this is due to the large bulge mass of this galaxy, which quenches star formation by stabilising the local turbulences. A small core in one of the most massive fast rotators with a mid-range $\lambda_R$, could be a relic of a larger core which was only partially refilled by a quenched starburst. Generally, it is most likely that formation scenarios differ from object to object, which could explain (the second order) peculiarities in the observed kinematics of core fast rotators (see Section~\ref{ss:kina}).

%%%%%% Figure 12%%%%%%%%%%%%%%%%%%%%%%%%%%%%%%%%%%%%%%%%%%%%%%%
\begin{figure}
\includegraphics[width=\columnwidth]{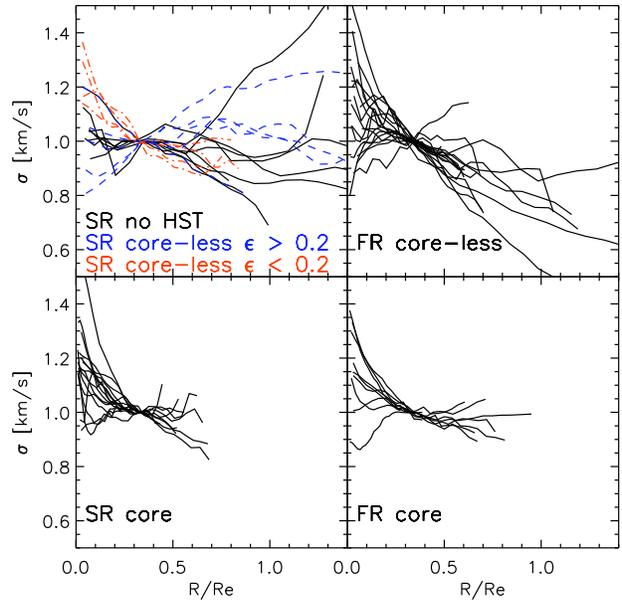}
\caption{Velocity dispersion radial profiles for galaxies with $\lambda_R < 0.3$ obtained with \textsc{kinemetry} from SAURON maps. Each panel shows profiles for a different population of objects: ({\it top left}) slow-rotators with no HST imaging together with core-less slow rotators separated in flat ($\epsilon > 0.2$, blue dashed lines) and round ($\epsilon < 0.2$, red, dashed-dotted lines), ({\it top right}) fast rotators with core-less profiles, ({\it bottom left}) slow rotators with cores and ({\it bottom right}) fast rotators with cores. All profiles are normalised to a third of the effective radius for presentation purpose.}
\label{f:disp}
\end{figure}
%%%%%%%%%%%%%%%%%%%%%%%%%%%%%%%%%%%%%%%%%%%%%%%%%%%%%%%%%%

In Section~\ref{ss:nohstSR} we suggested there is a sub-class of slow rotators with likely core-less light profiles (although no HST imaging is available). As these galaxies show evidence for anti-correlation in $V/\sigma - h_3$, their evolution has to be linked to those dissipative  (gas-rich) mergers \citep[e.g.][]{2006MNRAS.372..839N,2007ApJ...658..710N,2007MNRAS.376..997J}, which are, however, also able to decrease the overall angular momentum. The mechanism of making slow rotators in major mergers with gas is related to orbital configuration of galaxies (Paper VI, Naab et al. in prep.), but it is not trivial to detect it in simulations as it strongly depends on resolution \citep{2010MNRAS.406.2405B}. 

If these galaxies contained cores before the last major merger, cores could be re-filled with a central starburst fuelled by the gas, provided the above outlined quenching processes are not effective (indeed $\sigma_e$ of these galaxies are smaller), and there exist a fine-tuning between the duration of the nuclear starburst and the evolution of the binary black hole, such as that the coalescence of the binary is shorter than the starburst (a contrary case to one mentioned above for the creation of cores). The existence of gas could indeed speed up the hardening of the black hole binary \citep{2002ApJ...567L...9A}, although feedback effects need to be better understood

Simulations show that galaxies, which experienced a major dissipative merger that decreased the global angular momentum, should show evidence for drops in the central velocity dispersion associated with embedded disc-like structures created in the starburst (Naab et al. in prep). We investigate this by plotting in Fig.~\ref{f:disp} radial velocity dispersion profiles for core fast and slow rotators, core-less fast and slow rotators  (with $\lambda_R<0.3$) and slow rotators with no HST data (which we assume are also core-less). The profiles were obtained by running \textsc{kinemetry} (in even mode) on fixed ellipses corresponding to the measured global ellipticity and position angle (Paper II). As expected, the majority of velocity dispersion profiles are rising in the centre. This is in particular the case for core galaxies (with a few exceptions). Core-less fast rotators often have significant $\sigma$ drops. Assumed core-less slow rotators (those with no HST data) are perhaps best described as having flat central velocity dispersion profiles, making them marginally consistent with expectations. Similar $\sigma$ profiles are found in all flat ($\epsilon>0.2$) core-less slow rotators, but not in those that are more round. This suggests that all flat slow rotators in the \atlas sample experienced a gas rich merger event, which did not create a bona fide fast rotator, but it left cuspy nuclear light profiles (including in those galaxies with no available HST data), and created a sub-population of slow rotators. 

From the stellar populations point of view, cores follow the trends of host galaxies: as they are found in massive galaxies, cores are made of old and metal rich stars. This implies that at least some cores were made early, during the more violent, first phase of mass assembly, as outlined by \citet{2010ApJ...725.2312O}. They are subsequently kept frozen, while the main body of the galaxy grew during the second phase of galaxy evolution.  If this evolution induces regular rotation, then it could be a possible path for formation of core fast rotators. On the other hand, some cores were carved out in non-dissipative processes via black hole binaries with no new star formation, providing an applicable path for formation of cores at later times.

\begin{figure}
%Fig made by plot_results_kinemetry_sigma.pro
\includegraphics[width=\columnwidth]{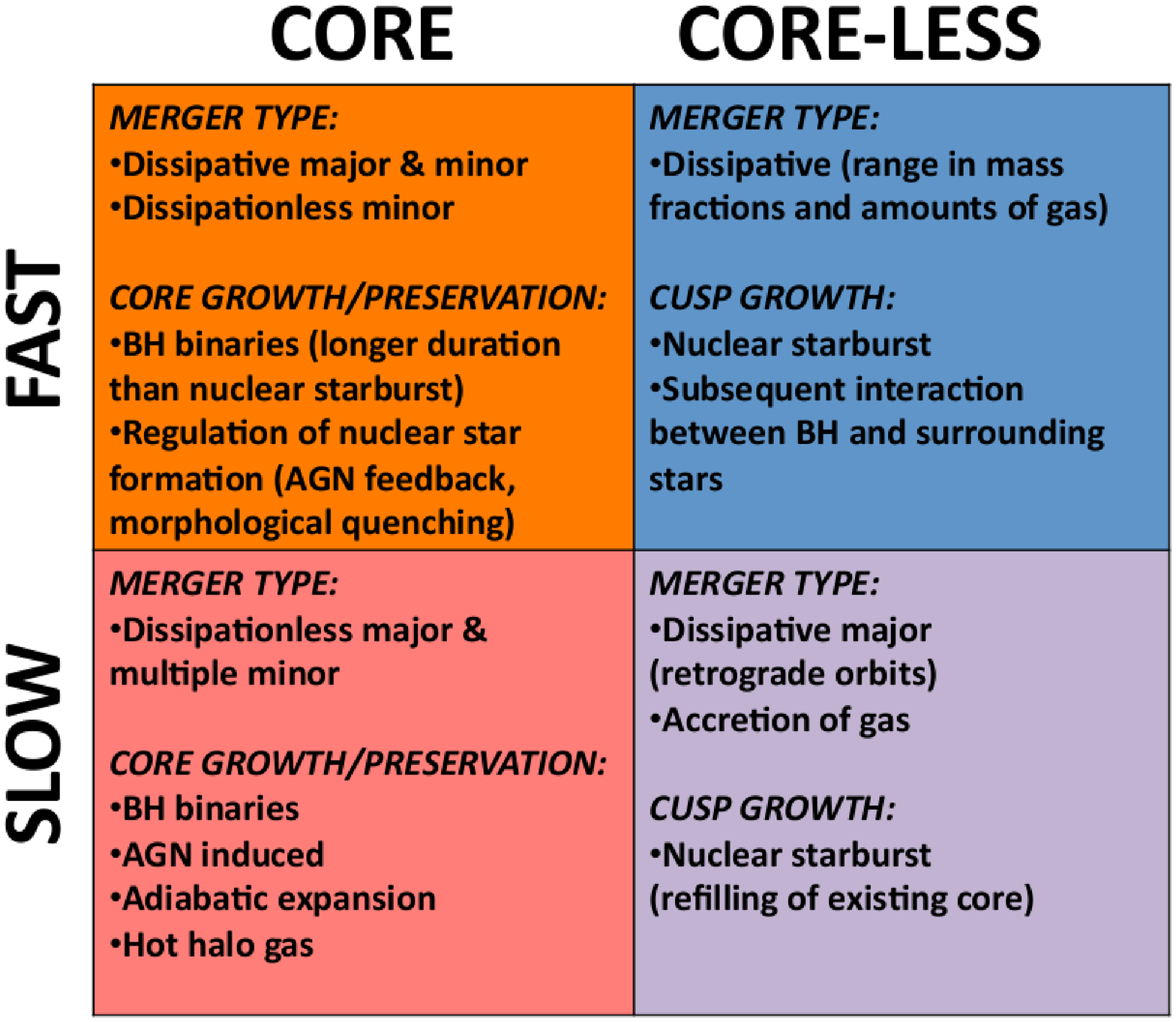}
\caption{A summary of the dominant processes that shape the angular momentum of a host galaxy and its nuclear profile.}
\label{f:prop}
\end{figure}

In Fig.~\ref{f:prop} we summarise processes that dominate in the creation of core and core-less early-type galaxies. The majority of galaxies are found in the upper right box (core-less fast rotators). Their mass assembly is dominated by dissipative  processes with varying gas fractions, while the nuclear cusps are created in a nuclear starburst or, perhaps, in interaction between the single black hole and surrounding stars (as mentioned in Section~\ref{ss:cores}). Similar processes act to produce core-less slow rotators (lower right box), comprising perhaps up to a quarter of the total population of slow rotators. Likely, there is a continuity of properties in the formation scenarios which create core-less fast and slow rotators, with a notable difference in the merging configuration, such that those processes forming slow rotators can significantly lower the angular momentum of the remnant.  

The second most numerous group in the \atlas sample is represented by the lower left box. These are the most massive, weakly triaxial galaxies living mostly in centres of clusters harbouring classical cores. As previous studies evoked \citep[e.g.][]{2007MNRAS.375....2D, 2007ApJ...658..710N,2009MNRAS.395..160K,2009MNRAS.397..506K,2009ApJ...703..785D,2010ApJ...709..218F}, their formation is dominated by non-dissipative major and multiple minor mergers. \citet{2011MNRAS.417..863D} showed, however, that major dissipative mergers are also possible among at least some of these objects. The cores are grown by scouring via black hole binaries, but they could also be induced by AGNs that remove significant amounts of mass in a short time-scale and hence change the potential. This process could adiabatically grow both cores and the host galaxies \citep[e.g.][]{1980ApJ...235..986H,2010MNRAS.401.1099H,2011MNRAS.414.3690R}. If the cores have been created early, the ability to maintain them is likely linked to the existence of hot halo gas \citep[][ Paper XIX]{2009ApJS..182..216K}, especially for those galaxies with an excess of diffuse X-ray emission, such as those living in cluster cores (Fig.~\ref{f:xray}).

Finally, the kinematic structure of the rare core fast rotators (i.e. velocity maps, $V/\sigma-h_3$ anti-correlation) indicates that they are also formed in gas rich assembly processes, but their cores could be explained as results of a fine tuning between the duration of the binary black hole coalescence and duration of the starburst. Alternatively, existing cores could be maintained through mechanisms that regulate (perhaps fully stop) star formation in the nucleus, but allow rebuilding of disc-like structures at large radii, such as the AGN feedback, or morphological quenching. Cored fast rotators seem to have similar diffuse X-ray emission like core-less galaxies. A simple possibility might be that their gas reservoirs are depleted (e.g. in dense environments), or are kept at large radii where the densities are not sufficient for star formation (Paper XIII). \citet{2013ApJ...768...36D} link these systems with compact galaxies at $z\sim1.5$, which were able to grow a disk. Indeed, multiple non-dissipative minor mergers \citep[e.g. mass ratios larger than 1:4,][]{2005A&A...437...69B} could provide the required large-scale kinematics and photometric structure \citep[e.g][]{2003MNRAS.338..880S,2012A&A...547A..48E}, but it is not clear to what extent they would preserve the existing core.

%%%%%% Figure 13%%%%%%%%%%%%%%%%%%%%%%%%%%%%%%%%%%%%%%%%%%%%%%%
%%%%%%%%%%%%%%%%%%%%%%%%%%%%%%%%%%%%%%%%%%%%%%%%%%%%%%%%%%

%%%%%%%%%%%%%%%%%%%%%%%%%%%%%%%%%%%%%%%%%%%%%%%%%%%%%%%%%%%
%
% SECTION 6 SECTION 6 SECTION 6 SECTION 6 SECTION 6 SECTION 6
%
%%%%%%%%%%%%%%%%%%%%%%%%%%%%%%%%%%%%%%%%%%%%%%%%%%%%%%%%%%%

\section{Summary}
\label{s:cona}

In this paper we connect data from the HST archive with \atlas results to investigate the links between the nuclear structure and global kinematic properties of early-type galaxies. The observations show: 
\begin{enumerate}

\item[(i)] There is a general correspondence between the classifications into fast -- slow rotators and core -- core-less profiles, as one would expect if both slow rotators and cores were related to dry merging. However, there are exceptions, which indicate that the detailed process that form a core galaxy do not always produce a slow rotator and vice versa.

\item[(ii)] Galaxies without cores dominate the population of early-type galaxies: of 135 \atlas galaxies with HST imaging there are 98 core-less galaxies (78 power-law, 20 intermediate cases), 24 cores and 13 galaxies for which we were unable to characterise the nuclear profiles. We do not find evidence for a dichotomy between core and core-less galaxies. 

\item[(iii)] The 135 galaxies we analysed are divided in 112 fast rotators and 23 slow rotators. We were able to investigate 50 per cent of the fast rotators and about 68 per cent of the slow rotators in the \atlas sample. A consequence is that the observed 72 per cent (98 of 135) of core-less galaxies is almost certainly a lower limit. Based on trends in Figs.~\ref{f:nuk}, ~\ref{f:mass}, ~\ref{f:sigma} and~\ref{f:cusp}, it is likely that only a few more slow and fast rotators contain cores. We estimate that they occur in about 10 per cent of nearby early-type galaxies.

\item[(iv)] Cores are found in the most massive and most luminous bodies. Fast rotators with cores are on average less massive than slow rotators with cores, but the lower mass limit for existence of cores is the same in both types:  there seem to be no cores in galaxies less massive than $8\times10^{10}$ M$_\odot$. All slow rotators above $2\times10^{11}$ M$_\odot$ have cores. The same might not be true for fast rotators. 

\item[(v)] A a good predictor for determining if a galaxy has a core is $\lambda_R - \sigma_e$ diagram, readily obtained with integral-filed spectrographs. More specifically, based on our data any galaxy with observed $\lambda_R \lesssim 0.25$ and $\sigma_e \gtrsim 160$ km/s will, most likely, have a core. As a more conservative estimate we suggest using $\lambda_R \lesssim 0.2$.  Future IFS surveys might be able to both distinguish between core and core-less profiles and differentiate fast and slow rotators, creating a more complete picture of possible processes forming early-type galaxies.

\item[(vi)] Slow and fast rotators with cores share similar projected shape ($\epsilon < 0.2$)\footnote{Note that among these there could be galaxies that are likely intrinsically flat}, if NGC4473 and NGC3613 are excluded. The majority of slow rotators and some fast rotators with cores populate the region in mass -- size diagram that is expected to be dominated by evolutionary processes, which do not change $\sigma$ of the galaxy, but change its mass and size. Lower mass  (below $2\times10^{11}$ M$_\odot$) core galaxies (both slow and fast rotators) are in the regime where dissipative  (gas rich) processes also influence the evolution. 

\item[(vii)] Compared to core slow rotators, X-ray luminosities of core fast rotators are typically smaller, and similar to those found in core-less galaxies. While the presence of a hot medium prevents cooling of external and internal gas, and, therefore inhibits star formation, the lack of this medium does not imply a core-less structure. Similarly, the creation of a core does not require a formation of a rounder galaxy, which would be more capable of retaining the hot gas halo.

\item[(viii)] Slow and fast rotators with cores share some additional characteristics: similar stellar populations, and a general lack of atomic and molecular gas. Additionally, we conjecture, based on only a few cases, that masses of central black holes in core fast rotators are similar to those found in core slow rotators. The implication of this results may be that cores in both fast and slow rotators were made through the same process (i.e. black hole binary interaction), but the subsequent evolution of the host galaxies was different.

\item[(ix)] Slow and fast rotators with cores do not have the same dynamics. The majority of core fast rotators shows evidence of embedded disc components seen in regularly rotating velocity maps, exponential S\'ersic components and anti-correlation in  $V/\sigma-h_3$ distribution.

\item[(x)] Based on the present data, core-less slow rotators are rare and often found in slow rotators with large ellipticities  ($\epsilon > 0.35$). There is, however, another potential sub-population of slow rotators which could also contain core-less nuclei, even though they share similar ellipticity range as core slow rotators ($0.2< \epsilon<0.35$) and similar angular momentum as core fast rotators ($0.1 < \lambda_R < 0.2$). These systems are dynamically different from fast rotators because their velocity maps are irregular and they are too flat to be considered discs at low inclinations, but they show a minor anti-correlation in $V/\sigma-h_3$ distribution and have central drops in the velocity dispersion maps. Furthermore, they are less massive ($<8\times10^{10}$ M$_\odot$) and have low $\sigma$ ($<$160 km/s).

\item[(xi)] The lack of cores in some slow rotators suggest the existence of a sub-population of slow rotators. Similarly, fast rotators with cores could be part of a sub-populations of fast rotators with different formation scenario from the majority of objects in this class, as predicted by semi-analytic models. 

\item[(xii)] Cores do not occur only in non-rotating or triaxial objects with KDCs, but can also be found in quite regularly rotating galaxies, with embedded disc-like structures, provided they are massive. Therefore, core formation should not be linked only to dissipationless dry mergers and binary black holes. Additional possible processes include: dissipative  major mergers where the evolution of the black hole binary is longer than the duration of the nuclear starburst and an AGN induced growth of cores linked with the size evolution of the host. Furthermore, cores could be maintained by regulation of the nuclear star formation via AGN feedback or morphological quenching. The existence of core-less slow rotators also suggest different formation scenarios from those creating the majority of (round and cored) slow rotators. 

\end{enumerate}

\section*{Acknowledgements}

We thank Laura Ferrarese, Patrick C\^ot\'e and the ACSVCS team for providing the 44 light profiles of Virgo Cluster galaxies. MC acknowledges support a Royal Society University Research Fellowship. MS acknowledges support from an STFC Advanced Fellowship ST/F009186/1. RMcD is supported by the Gemini Observatory, which is operated by the Association of Universities for Research in Astronomy, Inc., on behalf of the international Gemini partnership of Argentina, Australia, Brazil, Canada, Chile, the United Kingdom, and the United States of America. SK acknowledges support from the Royal Society Joint Projects Grant JP0869822. TN and MBois acknowledge support from the DFG Cluster of Excellence `Origin and Structure of the Universe'. PS is an NWO/Veni fellow. LY acknowledges support from NSF AST-1109803. The research leading to these results has received funding from the European Community's Seventh Framework Programme (/FP7/2007-2013/) under grant agreement No 229517. This work was supported by the rolling grants `Astrophysics at Oxford' PP/E001114/1 and ST/H002456/1 and visitors grants PPA/V/S/2002/00553, PP/E001564/1 and ST/H504862/1 from the UK Research Councils. Based on observations made with the NASA/ESA Hubble Space Telescope, and obtained from the Hubble Legacy Archive, which is a collaboration between the Space Telescope Science Institute (STScI/NASA), the Space Telescope European Coordinating Facility (ST-ECF/ESA) and the Canadian Astronomy Data Centre (CADC/NRC/CSA).

%\bibliographystyle{mn2e}
%\bibliography{../../refs.bib}

\appendix

%%%%%%%%%%%%%%%%%%%%%%%%%%%%%%%%%%%%%%%%%%%%%%%%%%%%%%%%%%%
%
% APPENDIX A   APPENDIX A    APPENDIX A    APPENDIX A     APPENDIX A     APPENDIX A    APPENDIX A 
%
%%%%%%%%%%%%%%%%%%%%%%%%%%%%%%%%%%%%%%%%%%%%%%%%%%%%%%%%%%%

\section{Core classification comparisons}
\label{a:comp}

Here we compare our classification of galaxies into core and core-less (power-law and intermediate) galaxies with the classification of \citet{2005AJ....129.2138L,2007ApJ...664..226L} and then move on to compare with four recent works which use core-S\'ersic and S\'ersic methodology: \citet{2006ApJS..165...57C}, \citet{2011MNRAS.415.2158R}, \citet{2012ApJ...755..163D} and \citet{2013ApJ...768...36D}. We note that the cores obtained by fitting a ``Nuker" model are not necessarily the same as the partially depleted cores obtained using core-S\'ersic fits. Therefore it is natural to expect differences in classification (and parameters) when using these two approaches. Galaxies for which classifications differ (typically due to the difference in our and previous approaches) are interesting cases which deserve special consideration, and we also highlight those in Table~\ref{tab:results}.

\begin{figure}
\includegraphics[width=\columnwidth]{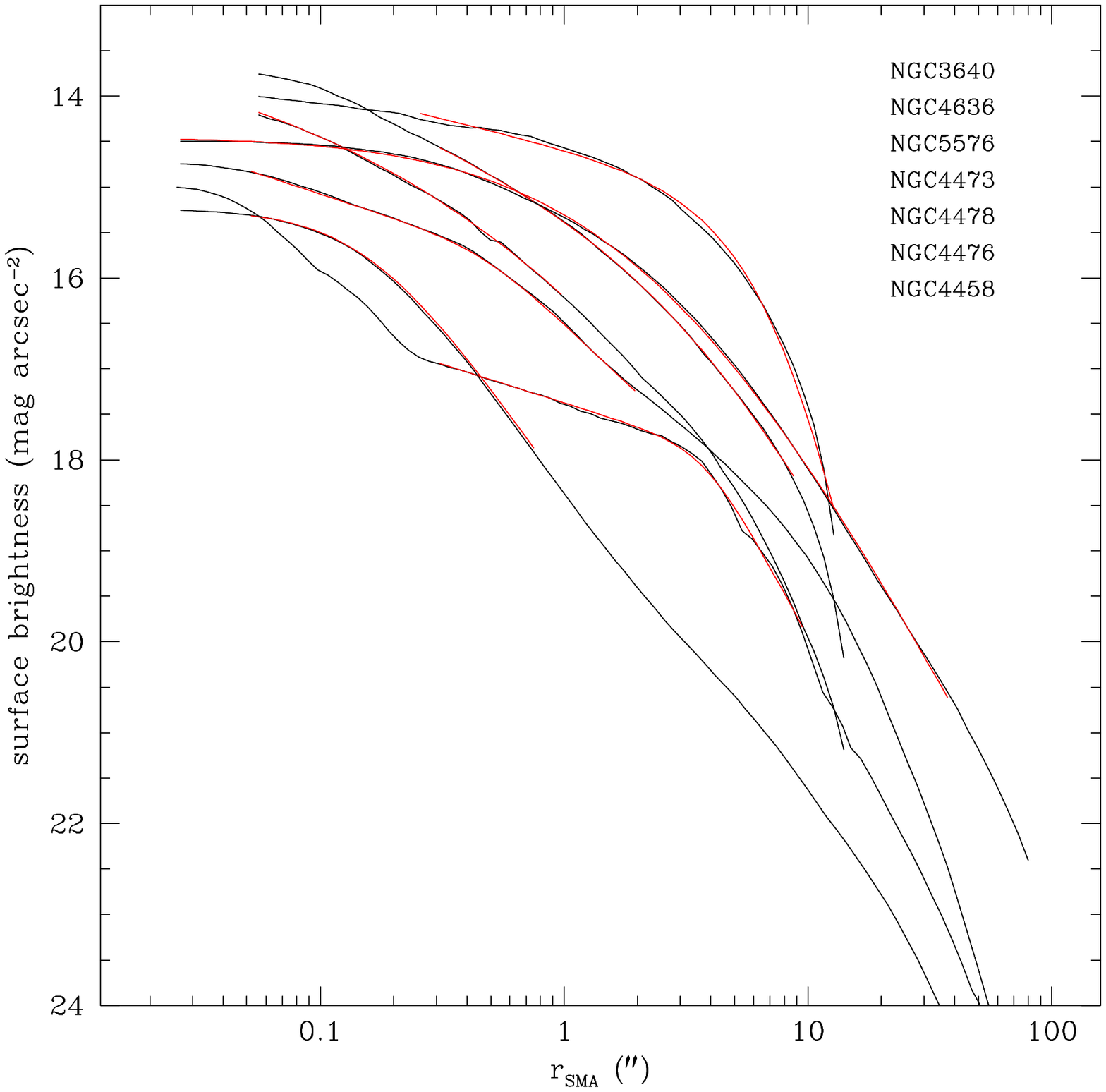}
\caption{Surface brightness profiles of galaxies for which our classification differs with those found in the literature. The region over which the Nuker fit is applied is shown in red. Galaxies are ordered by their total {\it K}-band luminosity starting with the brightest at the top and offset downwards for 0.25 mag/\arcsec$^2$. Galaxy names follow the sequence of profiles and are linked to the innermost points of the profile.}
\label{f:diff}
\end{figure}

There are 33 galaxies for which we can compare the classification based on our own fits using the ``Nuker law". These include 19 galaxies which overlap with the ACSVCS. The additional 14 galaxies we fitted to investigate the influence of the dust (see Appendix~\ref{b:nofit}) and if the classification by \citet{2005AJ....129.2138L, 2007ApJ...664..226L} differed from other works. Overall, the comparison is good, resulting in only four galaxies which we do not classify as cores: NGC3640, NGC4458, NGC4478 and NGC5576. We show our profiles and fits for these galaxies in Fig.~\ref{f:diff}. 
 
In the ``Nuker" model, the definition of core/core-less galaxies depends on the choice of radius at which $\gamma^\prime$ is evaluated. In Fig.~\ref{f:cusp} we show that the ratio of the ``cusp radius" and the radius at which we evaluate $\gamma^\prime$, $r_\gamma/r^\prime$, is typically large enough to detect cores in those galaxies where one could expect a large core (see Section~\ref{sss:more}). We remind the reader that our choice was to use $r^\prime = 0.1$\arcsec, as we used ACS data and otherwise our data set was very heterogeneous. The four galaxies for which our classifications differ have, however, the ratio close to (but larger than) 1 (all except NGC4478, which has a ratio of $\sim3$). This suggest that by using a smaller scale at which $\gamma^\prime$ is evaluated, as was done by \citet{2005AJ....129.2138L,2007ApJ...664..226L} using Nyquist-reconstructed images, could indeed classify them as cores. Furthermore, NGC3640 and NGC5576 occupy the region in mass-size diagram (Fig.~\ref{f:mass}) in which both core and core-less galaxies occur, while NGC4458 and NGC4478 are significantly below the lower mass limit of core galaxies, found in Section~\ref{ss:mass}. These galaxies also occupy a special region in fig.5 of \citet{2007ApJ...662..808L}, being fainter that typical cores, but typically having larger $r_\gamma$ than other galaxies at the same luminosities.

NGC3640 and NGC5576 also have different ``Nuker" fits reported by \citet{2001AJ....121.2431R}, who also estimate $\gamma^\prime$ at 0\farcs1, and we used our own analysis of archival WFCP2 data for these two galaxies. In both cases we derive a fit similar to \citet{2001AJ....121.2431R}, classifying them as intermediate. NGC5576 is controversial as recently \citet{2012ApJ...755..163D}, \citet{2012ApJ...759...64L} and \citet{2013ApJ...768...36D} argue on the best way to fit to this galaxy. It is obviously a special case worth keeping in mind when considering the results in the main text of this work.

%%%%%% Figure A1%%%%%%%%%%%%%%%%%%%%%%%%%%%%%%%%%%%%%%%%%%%%%%%
%%%%%%%%%%%%%%%%%%%%%%%%%%%%%%%%%%%%%%%%%%%%%%%%%%%%%%%%%%

For NGC4478, we use an ACS image and classify it as core-less (intermediate). In essence, this profile is hard to fit with the ``Nuker  law" in general, as it features two humps, the first peaking at  $\sim0.4$\arcsec and the second at $\sim7$\arcsec, where the inner component might be related to a flattened nuclear disc-like structure. Depending on the fitting range used, we can classify this profile either as a marginal intermediate ($\gamma^\prime=0.35$), or as a power-law ($\gamma^\prime=0.59$). A fit over a larger radial extent (encompassing both components) gives a poor fit. \citet{2006ApJS..165...57C} identified an extra nuclear component \citep[see also][]{2009ApJS..182..216K}, and \citet{2012ApJ...755..163D} fitted this galaxy with two S\'ersic functions. The ``Nuker-law" is not well suited to fit these kind of profiles (unless one limits the fit to a specific and often restricted region), as it was not designed for them. Therefore, we keep this galaxy as core-less. 

For NGC4458, we also use an ACS image. Its light profile is smooth within central $\sim5$\arcsec, but it is a clear composite of more components \citep[e.g. Paper XVII,][]{2006ApJS..164..334F, 2009ApJS..182..216K, 2012ApJ...755..163D}. Its $r_\gamma/r^\prime \sim 1$, and if there is a core, our approach is just marginally resolving it. Therefore, using smaller scale for evaluating the ``Nuker" fit \citep[as in][]{2005AJ....129.2138L} one could find the core. We wish to highlight this issue and urge the reader to keep this in mind, but we keep NGC4458 as a core-less galaxy according to our own analysis.

Finally, we would like to highlight NGC3193, which is also classified as a core by \citet{2007ApJ...664..226L} and as an intermediate by \citet{2001AJ....121.2431R}. Our fits suggest that it is a core, with $\gamma^\prime$ values between 0.28 and 0.3. It is clearly a difficult profile to fit and a marginal case, but we keep it as a core. 

Turning our attention to the comparison with S\'ersic based models, we note that the sample overlap with \citet{2006ApJS..165...57C} comprises 44 ACSVCS galaxies. Two of these we did not classify due to strong dust (see Section~\ref{b:nofit}) and one has a saturation artefact in the nucleus, but for the remaining 41 objects we used ACSVCS profiles to derive ``Nuker" fits parameters. Out of these galaxies there were two (NGC4473 and NGC4476) for which we provided different classifications. NGC4473 is in our case a core galaxy and a pure S\'ersc fit for \citet{2006ApJS..165...57C}. NGC4476 is an opposite case: an intermediate in our case and a core-S\'ersic in theirs. We will come back to these galaxies below. We also note that \citet{2006ApJS..165...57C} fitted NGC4365 with a core-S\'ersic model, but detect a small compact nucleus which they could not fit \citep[see also][]{2009ApJS..182..216K}. We also do not fit the nucleus and classify this galaxy as a core, in agreement with core-S\'ersic parametrisation.

The sample of \citet{2011MNRAS.415.2158R} contains 43 galaxies in common with ours (with a considerable overlap with the ASCVCS galaxies). They classify galaxies as core-S\'ersic, single S\'ersic and double S\'ersic, and indicate those galaxies for which the fits are uncertain (but still separate them in three classes). We assume that single S\'ersic and double S\'ersic objects can be grouped together as ``core-less" S\'ersic objects, in the same way as we combine power-law and intermediate cases as core-less objects with Nuker fits. In this case, there is one galaxy (NGC4636) for which our classification differs (core in our case) and there are further four which we are not able to fit (NGC3607, NGC4111, NGC4435 and NGC5866). NGC4111 and NGC5866 also are highlighted as uncertain by \citet{2011MNRAS.415.2158R}). We will discuss NGC4636 below. 

\citet{2012ApJ...755..163D} select their sample to contain only galaxies classified as core by \citet{2005AJ....129.2138L}. They perform the core-S\'ersic and S\'ersic fits to only the central 10\arcsec\,of HST light profiles and find a few galaxies classified differently compared to the previous work. If we compare 17 galaxies in common with our sample (again there is a substantial overlap with ASCVCS sample), we find a disagreement for one galaxy (NGC4473), which we classify as cores and \citet{2012ApJ...755..163D} as single S\'ersic. We note that for three other galaxies (NGC4458, NGC4478 and NGC5576), for which \citet{2012ApJ...755..163D} disagree with published values based on a ``Nuker" fit, we did our own fits (as mentioned above) and find a general agreement with the S\'ersic parametrisation. 

The smallest overlap of our sample is with the sample studied by \citet{2013ApJ...768...36D}. We have two galaxies in common (NGC3607 and NGC4382) and our classifications agree for both; the first one we cannot classify due to strong dust features (similar to that work), while the other one has a core.

Summarising, comparison of our core galaxies, based on the ``Nuker"  fit, with those classified as core-S\'ersic yields 3 out of 82 (the number of individual, non-repeating galaxies in the three works cited above) that do not agree. They are: NGC4473, NGC4476 and NGC4636. Their profiles and our fits are also shown in Fig.~\ref{f:diff}.
 
NGC4473 was already highlighted by \citet{2003ApJ...596..903P} as a curious case of a core galaxy with, generally, discy isophotes and significant rotation, and they suggested a merger origin. \citet{2004MNRAS.352..721E} show complex kinematics of this galaxy, which classifies it as a $2\sigma$ galaxy containing two counter-rotating discs (Paper II), while \citet{2007MNRAS.379..418C} constructed a dynamical model for the galaxy which requires two counter-rotating stellar components containing 30 and 70 per cent mass, respectively. There is disagreement between \citet{2012ApJ...755..163D}, \citet{2012ApJ...759...64L} and \citet{2013ApJ...768...36D} in how to classify this galaxy, where both authors show profiles fitted well with different S\'ersic models. \citet{2006ApJS..165...57C} show that its ACS profile can be well fitted with a single S\'ersic model (in agreement with \citealt{2012ApJ...755..163D}, but not \citealt{2012ApJ...759...64L} and \citealt{2013ApJ...768...36D}). Our fit to the same ACS profile indicates that a ``Nuker" fit is similarly well adapted for this galaxy, suggesting a core \citep[in agreement with][]{2005AJ....129.2138L}. The curvature of this profile, as well as of NGC4636, is such that both ``Nuker" and S\'ersic model fitted similarly  well. Nevertheless, NGC4473 is a very interesting case, as it is (according to the ``Nuker" fit) one of the two flattest core galaxies (and fast rotators), with consequences on the extent of properties that core galaxies have. 

NGC4476 is also clearly a special case.  \citet{2006ApJS..165...57C} fit it with a core-S\'ersic profile, but note that it is a clearly nucleated galaxy, as can be seen on Fig.~\ref{f:diff}. The core-S\'ersic fit is not ideal, as is the case with a single S\'ersic model. We fitted the ``Nuker" profiles excluding the very nucleus ($0.3$\arcsec $<r_{\rm fit}<100$\arcsec) and the resulting $\gamma^{\prime}$ puts this galaxy as an intermediate case, and clearly not a core. 

It is remarkable that there are only 3-4 per cent differences between the classifications using the ``Nuker" and core-S\'ersic models. Their $r_\gamma$ are all an order of magnitude larger than the radius at which we define cores, and the classification of their nuclear structure is not a question of resolution. Still, even though the ``Nuker'' and S\'ersic models address different characteristics of light profiles, the comparison of our classification with the S\'ersic based method gives us confidence in the robustness of our results.

%%%%%%%%%%%%%%%%%%%%%%%%%%%%%%%%%%%%%%%%%%%%%%%%%%%%%%%%%%%
%
% APPENDIX B   APPENDIX B    APPENDIX B   APPENDIX B     APPENDIX B     APPENDIX B    APPENDIX B
%
%%%%%%%%%%%%%%%%%%%%%%%%%%%%%%%%%%%%%%%%%%%%%%%%%%%%%%%%%%%

\section{Galaxies for which classification was not possible} 
\label{b:nofit}

%%%%% Figure 1%%%%%%%%%%%%%%%%%%%%%%%%%%%%%%%%%%%%%%%%%%%%%%%
\begin{figure*}
\includegraphics[width=\textwidth]{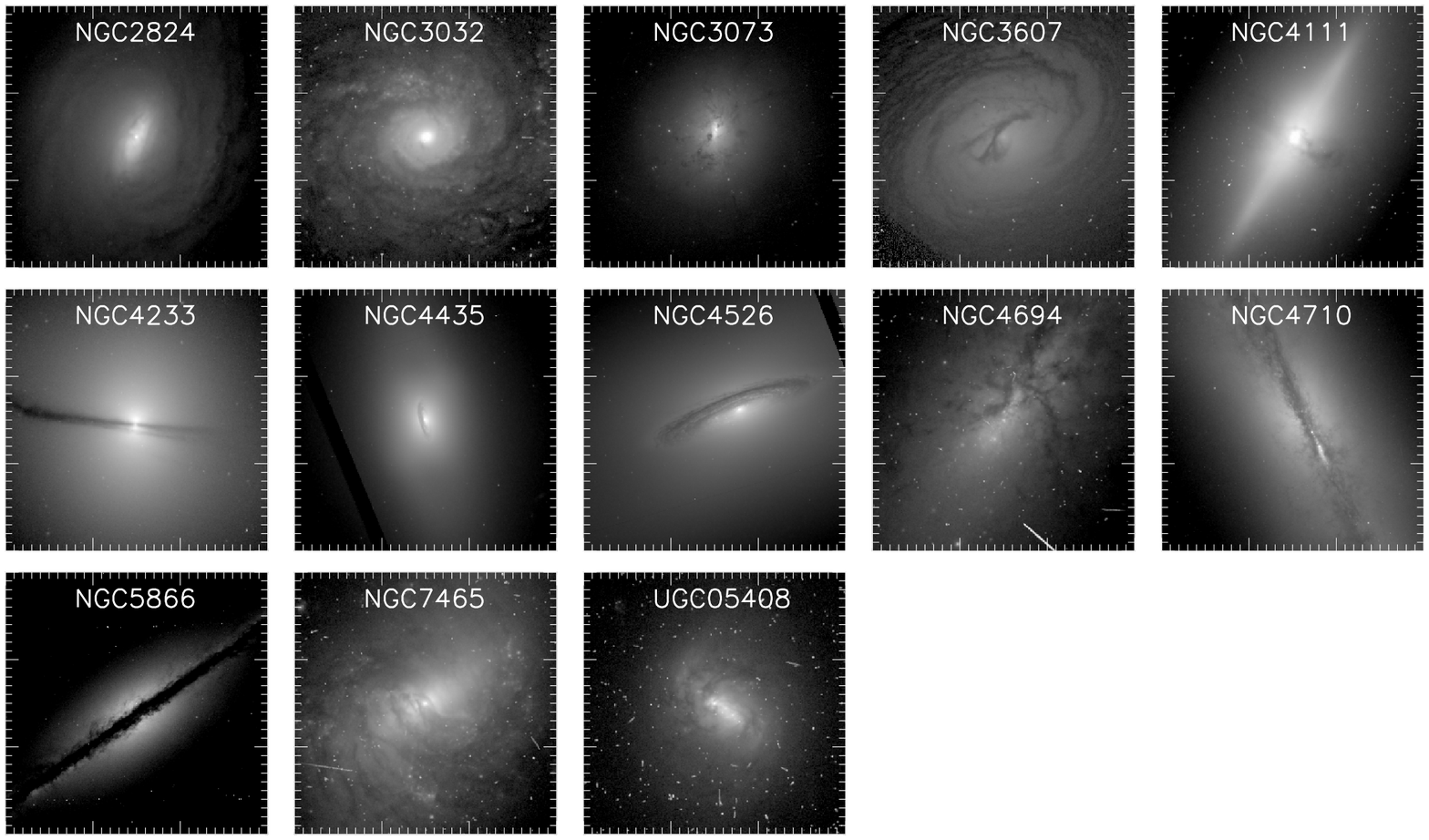}
\caption{Images of {\it uncertain} galaxies, showing strong dust features which make the characterisation of their profiles unreliable. We use WFPC/PC1 images for all galaxies, except NGC4435 and NGC4526, for which we used ACS images. The distance between two major tick marks corresponds to 100 pixels, corresponding to 4.55\arcsec for WFPC2 and 5.0\arcsec for ACS images. }
\label{f:thumb}
\end{figure*}
%%%%%%%%%%%%%%%%%%%%%%%%%%%%%%%%%%%%%%%%%%%%%%%%%%%%%%%%%%

We investigated the influence of dust on the robustness of the derived fits and found 13 galaxies for which we were not able to robustly derive light profiles, nor to fit them well. Images of these galaxies are provided in Fig.~\ref{f:thumb} and they clearly show a strong dust features, some in disc configurations seen at large inclinations (e.g NGC4710 and NGC5866), but mostly in irregular filamentary structures (e.g. NGC3073, NGC4694). Several galaxies have also visible spiral like patters (e.g. NGC3032, NGC3607), and a few also show dust in polar-ring configurations (e.g. NGC4111 and NGC4233). In a number of cases the double power-law can fit profiles extracted from these images, but the blind interpretation of the fit (typically $\gamma^\prime\leq0.3$) is not advisable. The underlying physical condition which makes the fit parameters unreliable is the presence of strong dust features which significantly change nuclear profiles within 1--2\arcsec, crucial for the determination of the $\gamma^{\prime}$ slope. Hence, we classify these galaxies as {\it uncertain}. All galaxies on this figure are fast rotators, and rich in CO and molecular gas. According to their general properties, such as $\lambda_r$, $V/\sigma - h_3$, mass, $\sigma_e$, they are most likely core-less galaxies.

%%%%%%%%%%%%%%%%%%%%%%%%%%%%%%%%%%%%%%%%%%%%%%%%%%%%%%%%%%%
%
% APPENDIX C   APPENDIX C   APPENDIX C   APPENDIX C     APPENDIX C     APPENDIX C    APPENDIX C 
%
%%%%%%%%%%%%%%%%%%%%%%%%%%%%%%%%%%%%%%%%%%%%%%%%%%%%%%%%%%%

\section{Result of  ``Nuker law" fits to HST imaging of 124 galaxies from the \atlas sample}
\label{c:table}

\clearpage
%%% Table 1.%%%%%%%%%%%%%%%%%%%%%%%%%%%%%%%%%%%%%%%%%%%%%%%%
\begin{deluxetable}{lccccccccccccccc}
\tablewidth{0pt}
\tabcolsep=5pt
\tabletypesize{\small}%scriptsize}
\tablecaption{Fitting parameters for ATLAS$^{\rm 3D}$ galaxies.}
\tablehead{ 
\colhead{Galaxy} & 
\colhead{Source} & 
\colhead{Class} & 
\colhead{Filter} & 
\colhead{$\Theta$} & 
\colhead{$\mu$} & 
\colhead{$\alpha$} & 
\colhead{$\beta$} & 
\colhead{$\gamma$} & 
\colhead{$\gamma^\prime$} & 
\colhead{RMS} &
\colhead{$R_\gamma$} &
\colhead{$R_\gamma$} &
\colhead{F/S} &
\colhead{Group}\\
\colhead{} & 
\colhead{} & 
\colhead{} & 
\colhead{} & 
\colhead{\arcsec} & 
\colhead{} & 
\colhead{} & 
\colhead{} & 
\colhead{} & 
\colhead{} & 
\colhead{} &
\colhead{\arcsec} &
\colhead{pc} &
\colhead{} &
\colhead{}\\
\colhead{(1)} &
\colhead{(2)} &
\colhead{(3)} &
\colhead{(4)} &
\colhead{(5)} &
\colhead{(6)} &
\colhead{(7)} &
\colhead{(8)} &
\colhead{(9)} &
\colhead{(10)} &
\colhead{(11)} &
\colhead{(12)} &
\colhead{(13)} &
\colhead{(14)} &
\colhead{(15)} &
}\startdata
 NGC0474  &   c  &    $\wedge$  &     F606W  &           1.83  &          14.46  &           1.20  &           1.97  &           0.36  &           0.40  &0.02  &           0.26  &          38.64  &   F  &   e \\
   NGC0524  &   d  &      $\cap$  &     F555W  &           0.37  &          17.00  &           5.00  &           1.47  &           0.24  &           0.24  &0.02  &           0.28  &          32.12  &   F  &   e \\
   NGC0821  &  b1  & $\setminus$  &         -  &           0.48  &          15.69  &           0.40  &           1.71  &           0.10  &           0.51  &   -  &           0.03  &           3.42  &   F  &   e \\
   NGC0936  &   d  & $\setminus$  &     F555W  &          33.64  &          24.00  &           0.70  &           5.88  &           0.43  &           0.52  &0.04  &           0.07  &           7.39  &   F  &   e \\
   NGC1023  &  b1  & $\setminus$  &         -  &           1.23  &          15.71  &           7.20  &           1.15  &           0.74  &           0.74  &   -  &           0.04  &           2.15  &   F  &   e \\
   NGC2549  &  b2  & $\setminus$  &         -  &           3.70  &          17.57  &           1.80  &           1.71  &           0.67  &           0.67  &   -  &$<$0.1  &$<$5.96  &   F  &   e \\
   NGC2592  &   d  &    $\wedge$  &     F702W  &          35.43  &          24.00  &           0.36  &           6.62  &          -0.38  &           0.38  &0.03  &           0.16  &          19.64  &   F  &   e \\
   NGC2685  &   d  &    $\wedge$  &     F555W  &          27.97  &          24.00  &           0.96  &           7.29  &           0.44  &           0.47  &0.04  &           0.20  &          16.43  &   F  &   e \\
   NGC2699  &   d  & $\setminus$  &     F702W  &           0.63  &          15.65  &           5.00  &           1.21  &           0.61  &           0.61  &0.06  &$<$0.1  &$<$12.7  &   F  &   e \\
   NGC2768  &   c  &    $\wedge$  &     F555W  &           0.37  &          15.82  &           5.00  &           1.79  &           0.37  &           0.37  &0.02  &           0.23  &          24.71  &   F  &   e \\
   NGC2778  &  b1  & $\setminus$  &         -  &           0.35  &          16.06  &           0.40  &           1.75  &           0.33  &           0.83  &   -  &           0.04  &           4.32  &   F  &   e \\
   NGC2859  &   c  & $\setminus$  &     F814W  &           0.11  &          14.00  &           0.20  &           4.14  &          -2.54  &           0.76  &0.02  &           0.04  &           5.85  &   F  &   e \\
   NGC2880  &   d  & $\setminus$  &     F555W  &           9.91  &          21.49  &           1.44  &           4.42  &           0.75  &           0.75  &0.02  &$<$0.1  &$<$10.3  &   F  &   e \\
   NGC2950  &  b2  & $\setminus$  &         -  &           2.43  &          16.77  &           2.40  &           1.81  &           0.82  &           0.82  &   -  &$<$0.1  &$<$7.02  &   F  &   e \\
   NGC2962  &   d  & $\setminus$  &     F814W  &           1.85  &          17.62  &           2.36  &           1.88  &           0.80  &           0.80  &0.03  &$<$0.1  &$<$16.4  &   F  &   e \\
   NGC2974  &  b1  & $\setminus$  &         -  &           0.31  &          15.17  &          25.10  &           1.05  &           0.62  &           0.62  &   -  &           0.02  &           2.03  &   F  &   e \\
   NGC3156  &   c  & $\setminus$  &     F475W  &           1.07  &          18.10  &           4.29  &           0.86  &           1.78  &           1.78  &0.08  &$<$0.1  &$<$10.5  &   F  &   e \\
   NGC3193  &  b2  &      $\cap$  &         -  &           0.81  &          16.07  &           0.60  &           1.89  &           0.01  &           0.28  &   -  &           0.14  &          22.87  &   F  &   e \\
   NGC3226  &   d  & $\setminus$  &     F814W  &           0.15  &          14.00  &           0.32  &           2.95  &          -1.02  &           0.83  &0.04  &           0.03  &           3.75  &   F  &   e \\
   NGC3245  &   d  & $\setminus$  &     F702W  &          35.18  &          24.00  &           0.92  &           7.09  &           0.71  &           0.74  &0.04  &$<$0.1  &$<$9.84  &   F  &   e \\
   NGC3377  &   c  & $\setminus$  &     F475W  &           0.17  &          14.00  &           0.92  &           1.37  &           0.25  &           0.68  &0.03  &           0.04  &           2.32  &   F  &   e \\
   NGC3379  &   c  &      $\cap$  &     F475W  &           2.41  &          16.84  &           1.34  &           1.72  &           0.15  &           0.17  &0.01  &           0.95  &          47.40  &   F  &   e \\
   NGC3384  &  b1  & $\setminus$  &         -  &           3.15  &          16.44  &          15.30  &           1.81  &           0.71  &           0.71  &   -  &           0.04  &           2.19  &   F  &   e \\
   NGC3412  &   d  & $\setminus$  &     F606W  &           3.39  &          18.18  &           1.24  &           2.76  &           0.64  &           0.67  &0.02  &$<$0.1  &$<$5.33  &   F  &   e \\
   NGC3414  &   d  & $\setminus$  &     F555W  &           0.23  &          15.79  &           5.00  &           1.15  &           0.64  &           0.64  &0.03  &$<$0.1  &$<$11.8  &   S  &   c \\
   NGC3458  &   d  & $\setminus$  &     F606W  &           1.73  &          18.06  &           1.40  &           2.40  &           0.56  &           0.59  &0.02  &$<$0.1  &$<$14.9  &   F  &   e \\
   NGC3489  &   c  & $\setminus$  &     F814W  &           0.41  &          14.00  &           0.64  &           4.84  &          -1.17  &           0.57  &0.05  &           0.09  &           5.23  &   F  &   e \\
   NGC3595  &  b2  & $\setminus$  &         -  &           2.30  &          18.04  &           2.30  &           1.52  &           0.75  &           0.76  &   -  &$<$0.1  &$<$16.8  &   F  &   e \\
   NGC3599  &  b1  & $\setminus$  &         -  &           1.25  &          17.45  &          50.00  &           1.45  &           0.75  &           0.75  &   -  &           0.04  &           3.84  &   F  &   e \\
   NGC3605  &  b1  & $\setminus$  &         -  &           0.97  &          17.21  &           2.40  &           1.27  &           0.59  &           0.60  &   -  &           0.04  &           3.90  &   F  &   e \\
   NGC3608  &  b1  &      $\cap$  &         -  &           0.48  &          15.73  &           0.90  &           1.50  &           0.09  &           0.17  &   -  &           0.18  &          19.27  &   S  &   c \\
   NGC3610  &  b1  & $\setminus$  &         -  &           2.84  &          16.39  &          48.50  &           1.86  &           0.76  &           0.76  &   -  &           0.02  &           2.02  &   F  &   e \\
   NGC3613  &   d  &      $\cap$  &     F702W  &           0.33  &          15.80  &           1.65  &           1.10  &           0.01  &           0.14  &0.01  &           0.29  &          40.05  &   F  &   e \\
   NGC3640$^\dag$  &   d  &    $\wedge$  &     F555W  &           4.05  &          18.00  &           1.12  &           2.17  &           0.46  &           0.48  &0.01  &           0.14  &          18.45  &   F  &   e \\
   NGC3796  &   d  & $\setminus$  &     F814W  &           0.20  &          14.00  &           3.24  &           1.48  &           0.66  &           0.74  &0.01  &$<$0.1  &$<$11.0  &   S  &   d \\
   NGC3945  &  b1  & $\setminus$  &         -  &           7.38  &          18.62  &           0.30  &           2.56  &          -0.06  &           0.57  &   -  &           0.10  &          10.80  &   F  &   e \\
   NGC3998  &   c  &    $\wedge$  &     F814W  &           0.28  &          14.00  &           2.77  &           6.32  &           0.15  &           0.49  &0.03  &           0.10  &           6.74  &   F  &   e \\
   NGC4026  &  b1  & $\setminus$  &         -  &           0.63  &          15.23  &           0.40  &           1.78  &           0.15  &           0.65  &   -  &           0.02  &           1.58  &   F  &   e \\
   NGC4143  &  b3  & $\setminus$  &         -  &           3.11  &          17.11  &           1.30  &           2.18  &           0.59  &           0.61  &   -  &$<$0.1  &$<$7.51  &   F  &   e \\
   NGC4150  &  b3  & $\setminus$  &         -  &           0.63  &          15.80  &           1.20  &           1.67  &           0.58  &           0.68  &   -  &$<$0.1  &$<$6.49  &   F  &   e \\
   NGC4168  &  b2  &      $\cap$  &         -  &           2.02  &          18.06  &           1.40  &           1.39  &           0.17  &           0.17  &   -  &           0.99  &         148.98  &   S  &   c \\
   NGC4203  &   d  & $\setminus$  &     F555W  &           4.52  &          19.13  &           1.61  &           2.56  &           0.73  &           0.74  &0.07  &$<$0.1  &$<$7.12  &   F  &   e \\
   NGC4261  &  b3  &      $\cap$  &         -  &           1.62  &          16.43  &           2.40  &           1.43  &           0.00  &           0.00  &   -  &           1.25  &         186.79  &   S  &   b \\
   NGC4262  &   a  & $\setminus$  &     F475W  &           2.59  &          17.07  &           5.00  &           1.96  &           0.76  &           0.76  &0.03  &$<$0.1  &$<$7.46  &   F  &   e \\
   NGC4267  &   a  & $\setminus$  &     F475W  &           2.97  &          17.52  &           2.29  &           1.90  &           0.71  &           0.71  &0.01  &$<$0.1  &$<$7.66  &   F  &   e \\
   NGC4270  &   d  &    $\wedge$  &     F606W  &           0.73  &          17.48  &           2.21  &           1.29  &           0.43  &           0.44  &0.02  &           0.24  &          41.61  &   F  &   e \\
   NGC4278  &  b1  &      $\cap$  &         -  &           1.26  &          16.20  &           1.40  &           1.46  &           0.06  &           0.10  &   -  &           0.72  &          54.58  &   F  &   e \\
   NGC4281  &   d  & $\setminus$  &     F606W  &           0.09  &          15.23  &           3.59  &           0.89  &           0.06  &           0.56  &0.01  &           0.09  &          11.01  &   F  &   e \\
   NGC4283  &   c  & $\setminus$  &     F475W  &           4.81  &          19.44  &           2.32  &           2.35  &           0.80  &           0.80  &0.04  &$<$0.1  &$<$7.41  &   F  &   e \\
   NGC4339  &   d  & $\setminus$  &     F606W  &          23.29  &          24.00  &           1.08  &           5.93  &           0.80  &           0.81  &0.03  &$<$0.1  &$<$7.75  &   F  &   e \\
   NGC4340  &   a  & $\setminus$  &     F475W  &           1.91  &          17.45  &           4.27  &           1.44  &           0.68  &           0.68  &0.03  &$<$0.1  &$<$8.92  &   F  &   e \\
   NGC4342  &   d  & $\setminus$  &     F555W  &           8.38  &          24.00  &           0.56  &           8.53  &          -0.12  &           0.55  &0.02  &           0.09  &           6.92  &   F  &   e \\
   NGC4350  &   a  &    $\wedge$  &     F475W  &           0.94  &          15.71  &           5.00  &           1.24  &           0.47  &           0.47  &0.06  &           0.50  &          36.97  &   F  &   e \\
   NGC4365  &   a  &      $\cap$  &     F475W  &           2.07  &          16.71  &           1.53  &           1.50  &          -0.01  &           0.00  &0.00  &           1.33  &         150.58  &   S  &   c \\
   NGC4371  &   a  &    $\wedge$  &     F475W  &           0.53  &          16.26  &           3.45  &           0.81  &           0.27  &           0.27  &0.02  &           0.49  &          40.06  &   F  &   e \\
   NGC4374  &   a  &      $\cap$  &     F475W  &           2.10  &          16.20  &           3.73  &           1.21  &           0.25  &           0.25  &0.01  &           1.59  &         142.37  &   S  &   a \\
   NGC4377  &   a  &    $\wedge$  &     F475W  &           1.49  &          16.80  &           0.83  &           2.20  &           0.22  &           0.41  &0.01  &           0.17  &          14.64  &   F  &   e \\
   NGC4379  &   a  & $\setminus$  &     F475W  &          59.50  &          23.24  &           0.39  &           4.15  &           0.16  &           0.46  &0.02  &           0.14  &          10.37  &   F  &   e \\
   NGC4382  &   a  &      $\cap$  &     F475W  &           0.90  &          14.00  &           0.80  &           1.56  &          -0.21  &           0.05  &0.01  &           0.55  &          47.33  &   F  &   e \\
   NGC4387  &   a  & $\setminus$  &     F475W  &          18.17  &          20.90  &           1.40  &           3.61  &           0.63  &           0.63  &0.01  &$<$0.1  &$<$8.67  &   F  &   e \\
   NGC4406  &   a  &      $\cap$  &     F475W  &           1.05  &          16.00  &           3.43  &           1.11  &          -0.02  &          -0.02  &0.01  &           1.00  &          81.63  &   S  &   c \\
   NGC4417  &   a  & $\setminus$  &     F475W  &           1.94  &          16.72  &           5.00  &           1.48  &           0.68  &           0.68  &0.01  &$<$0.1  &$<$7.75  &   F  &   e \\
   NGC4429  &   d  & $\setminus$  &     F606W  &           0.06  &          14.61  &           5.00  &           1.14  &           0.31  &           1.07  &0.01  &           0.05  &           3.76  &   F  &   e \\
   NGC4434  &   a  & $\setminus$  &     F475W  &           6.89  &          19.63  &           0.49  &           3.01  &           0.22  &           0.54  &0.03  &           0.08  &           8.51  &   F  &   e \\
   NGC4442  &   a  & $\setminus$  &     F475W  &          55.80  &          21.73  &           0.54  &           3.79  &           0.41  &           0.52  &0.02  &           0.07  &           5.28  &   F  &   e \\
   NGC4452  &   a  &    $\wedge$  &     F475W  &           4.91  &          18.87  &           5.00  &           1.55  &           0.39  &           0.39  &0.04  &           3.13  &         236.49  &   F  &   e \\
   NGC4458$^\dag$  &   a  &    $\wedge$  &     F475W  &           0.16  &          14.58  &           2.11  &           1.54  &           0.01  &           0.44  &0.00  &           0.11  &           8.91  &   S  &   c \\
   NGC4459  &   a  & $\setminus$  &     F475W  &           0.46  &          15.00  &           3.64  &           1.32  &           0.48  &           0.49  &0.01  &           0.17  &          12.94  &   F  &   e \\
   NGC4472  &   a  &      $\cap$  &     F475W  &           2.82  &          16.62  &           1.14  &           1.43  &          -0.15  &          -0.11  &0.02  &           2.06  &         170.75  &   S  &   c \\
   NGC4473$^\ddag$  &   a  &      $\cap$  &     F475W  &           2.66  &          16.47  &           0.86  &           2.10  &          -0.02  &           0.10  &0.02  &           0.72  &          53.40  &   F  &   d \\
   NGC4474  &   a  & $\setminus$  &     F475W  &           3.65  &          18.16  &           2.03  &           1.74  &           0.57  &           0.57  &0.01  &$<$0.1  &$<$7.56  &   F  &   e \\
   NGC4476$^\ddag$  &   a  &    $\wedge$  &     F475W  &           3.85  &          17.92  &           5.00  &           2.00  &           0.34  &           0.34  &0.03  &           2.46  &         209.97  &   S  &   e \\
   NGC4477  &   d  &    $\wedge$  &     F606W  &           4.89  &          22.31  &           1.48  &          10.00  &           0.35  &           0.38  &0.01  &           0.30  &          23.72  &   F  &   e \\
   NGC4478$^\dag$  &   a  & $\setminus$  &     F475W  &           0.47  &          15.86  &           3.69  &           1.01  &           0.35  &           0.35  &0.01  &           0.34  &          27.80  &   F  &   e \\
   NGC4483  &   a  & $\setminus$  &     F475W  &          83.07  &          24.00  &           0.61  &           3.25  &           0.84  &           0.88  &0.02  &$<$0.1  &$<$8.09  &   F  &   e \\
   NGC4486  &   a  &      $\cap$  &     F475W  &           8.19  &          17.90  &           2.32  &           1.51  &           0.23  &           0.23  &0.00  &           4.64  &         386.74  &   S  &   a \\
  NGC4486A  &   a  & $\setminus$  &     F475W  &          13.38  &          21.14  &           1.03  &           4.34  &           0.70  &           0.72  &0.60  &$<$0.1  &$<$8.87  &   F  &   e \\
   NGC4489  &   a  & $\setminus$  &     F475W  &           1.89  &          18.13  &           1.05  &           2.21  &           0.57  &           0.64  &0.01  &$<$0.1  &$<$7.46  &   F  &   c \\
   NGC4494  &  b1  & $\setminus$  &         -  &           2.82  &          17.19  &           0.70  &           1.88  &           0.52  &           0.55  &   -  &           0.04  &           3.22  &   F  &   e \\
   NGC4503  &  b2  & $\setminus$  &         -  &           1.65  &          17.14  &           1.80  &           1.30  &           0.64  &           0.65  &   -  &$<$0.1  &$<$7.99  &   F  &   e \\
   NGC4528  &   a  & $\setminus$  &     F475W  &          51.50  &          24.00  &           1.16  &           5.90  &           0.97  &           0.97  &0.10  &$<$0.1  &$<$7.66  &   S  &   d \\
   NGC4546  &   d  & $\setminus$  &     F606W  &           2.07  &          17.21  &           1.44  &           1.68  &           0.78  &           0.79  &0.01  &$<$0.1  &$<$6.64  &   F  &   e \\
   NGC4550  &   a  &    $\wedge$  &     F475W  &           9.32  &          18.52  &           5.00  &           1.83  &           0.57  &           0.57  &0.03  &$<$0.1  &$<$7.51  &   S  &   d \\
   NGC4551  &   a  & $\setminus$  &     F475W  &          54.71  &          24.00  &           1.09  &           5.91  &           0.75  &           0.76  &0.02  &$<$0.1  &$<$7.80  &   F  &   e \\
   NGC4552  &   a  &      $\cap$  &     F475W  &           0.10  &          14.69  &           0.26  &           2.91  &          -3.00  &          -0.03  &0.03  &           0.42  &          32.17  &   S  &   b \\
   NGC4564  &   a  & $\setminus$  &     F475W  &         100.00  &          23.97  &           0.79  &           4.92  &           0.79  &           0.81  &0.09  &$<$0.1  &$<$7.66  &   F  &   e \\
   NGC4570  &   a  & $\setminus$  &     F475W  &           4.53  &          17.51  &           3.22  &           1.56  &           0.85  &           0.85  &0.02  &$<$0.1  &$<$8.29  &   F  &   e \\
   NGC4578  &   a  & $\setminus$  &     F475W  &           3.95  &          18.39  &           4.75  &           1.55  &           0.89  &           0.89  &0.02  &$<$0.1  &$<$7.90  &   F  &   e \\
   NGC4596  &   d  & $\setminus$  &     F555W  &           5.30  &          19.40  &           2.99  &           2.70  &           0.77  &           0.77  &0.09  &$<$0.1  &$<$7.99  &   F  &   e \\
   NGC4612  &   a  & $\setminus$  &     F475W  &           1.77  &          17.14  &           4.46  &           1.71  &           0.64  &           0.64  &0.01  &$<$0.1  &$<$8.04  &   F  &   e \\
   NGC4621  &   a  & $\setminus$  &     F475W  &           7.95  &          18.09  &           0.06  &           5.47  &          -3.00  &           0.68  &0.04  &           0.02  &           1.66  &   F  &   e \\
   NGC4623  &   a  & $\setminus$  &     F475W  &          10.18  &          20.06  &           0.75  &           0.10  &           2.12  &           2.06  &0.85  &$<$0.1  &$<$8.43  &   F  &   e \\
   NGC4636$^\ddag$  &   d  &      $\cap$  &     F814W  &          10.00  &          19.75  &           2.00  &           6.45  &           0.26  &           0.26  &0.07  &           2.01  &         139.24  &   S  &   a \\
   NGC4638  &   a  & $\setminus$  &     F475W  &           6.84  &          18.07  &           5.00  &           1.99  &           0.77  &           0.77  &0.04  &$<$0.1  &$<$8.48  &   F  &   e \\
   NGC4649  &   a  &      $\cap$  &     F475W  &           5.05  &          17.16  &           1.66  &           1.60  &           0.14  &           0.14  &0.01  &           2.58  &         216.12  &   F  &   e \\
   NGC4660  &  b1  & $\setminus$  &         -  &           1.76  &          16.34  &           5.61  &           1.50  &           0.91  &           0.91  &   -  &           0.04  &           2.91  &   F  &   e \\
   NGC4697  &   c  & $\setminus$  &     F475W  &           2.35  &          17.35  &           0.07  &           0.00  &           1.47  &           0.82  &0.09  &$<$0.1  &$<$5.52  &   F  &   e \\
   NGC4733  &   d  &    $\wedge$  &     F555W  &           3.94  &          20.56  &           1.82  &           1.95  &           0.35  &           0.35  &0.03  &           1.13  &          79.63  &   F  &   a \\
   NGC4754  &   a  & $\setminus$  &     F475W  &         100.00  &          23.83  &           0.44  &           4.48  &           0.42  &           0.60  &0.03  &           0.01  &           1.09  &   F  &   e \\
   NGC4762  &   a  & $\setminus$  &     F475W  &          75.32  &          24.00  &           0.47  &           5.56  &           0.16  &           0.40  &0.02  &           0.24  &          26.40  &   F  &   e \\
   NGC5173  &   d  & $\setminus$  &     F702W  &           0.79  &          16.25  &           3.61  &           1.35  &           0.52  &           0.52  &0.02  &$<$0.1  &$<$18.6  &   F  &   e \\
   NGC5198  &  b2  &      $\cap$  &         -  &           0.16  &          15.40  &           2.60  &           1.13  &           0.23  &           0.26  &   -  &           0.12  &          22.17  &   S  &   b \\
   NGC5273  &   d  & $\setminus$  &     F606W  &           0.10  &          14.95  &           5.00  &           2.84  &           0.54  &           1.66  &0.02  &$<$0.1  &$<$7.80  &   F  &   e \\
   NGC5308  &   c  & $\setminus$  &     F475W  &           0.56  &          16.15  &           1.34  &           1.19  &           0.78  &           0.82  &0.02  &$<$0.1  &$<$15.2  &   F  &   e \\
   NGC5322  &   d  &      $\cap$  &     F555W  &          13.87  &          21.23  &           0.79  &           5.22  &           0.05  &           0.15  &0.05  &           0.71  &         104.00  &   S  &   c \\
   NGC5422  &   d  &    $\wedge$  &     F814W  &          47.48  &          24.00  &           0.40  &           5.52  &           0.02  &           0.45  &0.04  &           0.13  &          20.04  &   F  &   e \\
   NGC5475  &   c  &    $\wedge$  &     F475W  &           1.50  &          19.86  &           2.29  &          10.00  &           0.38  &           0.40  &0.07  &           0.22  &          30.83  &   F  &   e \\
   NGC5485  &   c  &      $\cap$  &     F814W  &           0.73  &          16.22  &           4.21  &           1.01  &           0.19  &           0.19  &0.03  &           0.65  &          79.24  &   F  &   b \\
   NGC5557  &  b1  &      $\cap$  &         -  &           0.80  &          16.40  &           0.80  &           1.68  &           0.02  &           0.07  &   -  &           0.26  &          48.89  &   S  &   b \\
   NGC5576$^\dag$  &   d  &    $\wedge$  &     F555W  &           6.00  &          18.38  &           0.41  &           3.01  &           0.00  &           0.47  &0.01  &           0.12  &          14.10  &   S  &   b \\
   NGC5813  &  b1  &      $\cap$  &         -  &           1.15  &          16.82  &           1.60  &           1.60  &           0.05  &           0.06  &   -  &           0.66  &          99.82  &   S  &   c \\
   NGC5831  &   d  & $\setminus$  &     F702W  &          36.56  &          24.00  &           0.96  &           7.41  &           0.77  &           0.79  &0.02  &$<$0.1  &$<$12.7  &   S  &   c \\
   NGC5838  &  b3  & $\setminus$  &         -  &           4.35  &          17.69  &           2.60  &           1.87  &           0.93  &           0.93  &   -  &$<$0.1  &$<$10.5  &   F  &   e \\
   NGC5845  &  b4  & $\setminus$  &         -  &           1.38  &          15.86  &           2.10  &           2.18  &           0.51  &           0.52  &   -  &$<$0.1  &$<$12.2  &   F  &   e \\
   NGC5846  &   d  &      $\cap$  &     F555W  &           2.00  &          18.57  &           2.01  &           1.72  &          -0.03  &          -0.02  &0.04  &           1.32  &         154.98  &   S  &   a \\
   NGC5854  &   d  & $\setminus$  &     F814W  &           0.30  &          14.00  &           0.27  &           3.86  &          -1.10  &           1.01  &0.07  &           0.02  &           2.44  &   F  &   e \\
   NGC6278  &  b2  & $\setminus$  &         -  &           0.60  &          16.30  &           0.80  &           1.62  &           0.55  &           0.67  &   -  &$<$0.1  &$<$20.7  &   F  &   e \\
   NGC6703  &   d  & $\setminus$  &     F555W  &           0.07  &          14.86  &           5.00  &           1.12  &           0.20  &           0.95  &0.01  &           0.06  &           7.60  &   S  &   a \\
   NGC7280  &   d  & $\setminus$  &     F606W  &           0.10  &          14.00  &           3.47  &           1.26  &           0.53  &           0.87  &0.04  &$<$0.1  &$<$11.4  &   F  &   e \\
   NGC7457  &  b1  & $\setminus$  &         -  &           0.22  &          16.33  &           1.00  &           1.05  &          -0.10  &           0.61  &   -  &           0.04  &           2.50  &   F  &   e \\
  UGC04551  &  b2  &    $\wedge$  &         -  &           2.26  &          17.67  &           2.20  &           2.16  &           0.49  &           0.49  &   -  &           0.22  &          30.04  &   F  &   e \\
  UGC06062  &  b2  & $\setminus$  &         -  &           2.75  &          18.96  &           0.90  &           1.81  &           0.80  &           0.82  &   -  &$<$0.1  &$<$18.7  &   F  &   e \\
\enddata
\tablecomments{A machine readable version of this table, as well as all other values used in this study, but published in previous papers of the \atlas series, can be found on http://www-astro.physics.ox.ac.uk/atlas3d/.\\
Column (1): Galaxy name. Only galaxies from the \atlas sample for which we were able to perform ``Nuker-law" are included. \\
Column (2): (a) data from the ACSVCS \citep{2006ApJS..164..334F}; (b) data from Lauer et al. (2007) originating in: b1 $-$ Lauer et al. (2005), b2 $-$ Rest et al. (2001), b3 (NICMOS images) $-$ Ravindranath et al. (2001), b4 (NICMOS images) $-$ Quillen et al. (2000); (c) supplementary archival HST/ACS imaging; (d) supplemental archival WFC2/PC1 images.  Details of the observations for the archival images are provided in Table 1. \\
Column (3): {\it core} = $\cap$, {\it intermediate} = $\wedge$,  "{\it power-law} = $\setminus$. Intermediate and power-law galaxies are grouped in {\it core-less} galaxies in the text. \\
Column (4): filter of images analysed in this work. \\
Columns (5$-$10): ``Nuker'' fit parameters. \\
Column (11): root-mean-square of ``Nuker-law" fits done in this work.\\
Column (12 -13): ``cusp radius" (see eq.~\ref{e:cusp})\\
Column (14): classification into fast or slow rotators, from Paper III\\
Column (15): kinematic group from Paper II, where: a -- non-rotating galaxies, b -- featureless non-regular rotators, c -- kinematically distinct cores (and counter-rotating cores), d -- $2\sigma$ peak galaxies, e -- regular-rotators, f -- unclassified. \\
$^\dag$ -- galaxies for which our classification differs from that of \citet{2005AJ....129.2138L}, primarily due to different definition of $\gamma^\prime$\\
$^\ddag$ -- galaxies for which the ``Nuker" core is not also seems as a partially depleted core using core-S\'ersic/Sersic fitting method.
}
\label{tab:results}
\end{deluxetable}
%%%%%%%%%%%%%%%%%%%%%%%%%%%%%%%%%%%%%%%%%%%%%%%%%%%%%%%%%%

\label{lastpage}

\end{document}